\documentclass[superscriptaddress,aps,twocolumn,longbibliography]{revtex4-1}
\usepackage{amsmath,amssymb}
\usepackage{mathtools}
\usepackage{graphicx}
\usepackage{dcolumn}
\usepackage{tcolorbox}
\usepackage{enumitem}
\usepackage{float}
\usepackage{bm,dsfont}
\usepackage{multirow}
\usepackage{simpler-wick}
\usepackage{cuted}
\usepackage{color}
\usepackage{lipsum}
\usepackage{hyperref}
\hypersetup{
colorlinks = true,
linkcolor = [rgb]{0.70,0.13,0.13},
citecolor = [rgb]{0.13,0.55,0.13},
urlcolor = [rgb]{0.25, 0.41, 0.88}}
\newcommand{\bra}[1]{{\langle #1|}}
\newcommand{\ket}[1]{{|#1 \rangle}}

\newcommand{\ii}{\mathrm{i}}
\newcommand{\id}{\mathds{1}}
\newcommand{\Wg}{\mathsf{Wg}}
\newcommand{\U}{\mathrm{U}}

\newcommand{\dsE}{\mathbb{E}}

\newcommand{\dsR}{\mathbb{R}}

\newcommand{\scC}{\mathcal{C}}
\newcommand{\scD}{\mathcal{D}}
\newcommand{\scE}{\mathcal{E}}
\newcommand{\scF}{\mathcal{F}}

\newcommand{\scM}{\mathcal{M}}

\newcommand{\scO}{\mathcal{O}}

\newcommand{\scS}{\mathcal{S}}

\newcommand{\Tr}{\operatorname{Tr}}

\newcommand{\var}{\operatorname{Var}}
\newcommand{\avg}{\operatorname{avg}}
\renewcommand{\Pr}{\operatorname{Pr}}
\newcommand{\E}{\mathop{\dsE}}

\newcommand{\norm}[1]{{\lVert #1\rVert}}
\newcommand{\mat}[1]{\left[\begin{matrix}#1\end{matrix}\right]}

\newcommand{\dia}[3]{\raisebox{#3pt}{\includegraphics[height=#2pt]{dia_#1}}}
\newcommand{\eq}[1]{\begin{equation}#1\end{equation}}
\newcommand{\eqs}[1]{\begin{equation}\begin{split}#1\end{split}\end{equation}}
\newcommand{\eqnref}[1]{Eq.\,\eqref{#1}}
\newcommand{\figref}[1]{Fig.\,\ref{#1}}

\newcommand{\secref}[1]{Sec.\,\ref{#1}}
\newcommand{\appref}[1]{Appendix\,\ref{#1}}
\newcommand{\refcite}[1]{Ref.\,\cite{#1}}

\usepackage[mathlines]{lineno}

\begin{document}


\title{Classical Shadow Tomography with Locally Scrambled Quantum Dynamics}

\author{Hong-Ye Hu}
\affiliation{Department of Physics, University of California San Diego, La Jolla, CA 92093, USA}
\affiliation{Department of Physics, Harvard University, 17 Oxford Street, Cambridge, MA 02138, USA}
\author{Soonwon Choi}
\affiliation{Department of Physics, University of California, Berkeley, California 94720, USA}
\affiliation{Center for Theoretical Physics, Massachusetts Institute of Technology, Cambridge, MA 02139, USA}
\author{Yi-Zhuang You}
\email{yzyou@physics.ucsd.edu}
\affiliation{Department of Physics, University of California San Diego, La Jolla, CA 92093, USA}

\date{\today}

\begin{abstract}
We generalize the classical shadow tomography scheme to a broad class of finite-depth or finite-time local unitary ensembles, known as locally scrambled quantum dynamics, where the unitary ensemble is invariant under local basis transformations. In this case, the reconstruction map for the classical shadow tomography depends only on the average entanglement feature of classical snapshots. We provide an unbiased estimator of the quantum state as a linear combination of reduced classical snapshots in all subsystems, where the combination coefficients are solely determined by the entanglement feature.
We also bound the number of experimental measurements required for the tomography scheme, so-called sample complexity, by formulating the operator shadow norm in the entanglement feature formalism. We numerically demonstrate our approach for finite-depth local unitary circuits and finite-time local-Hamiltonian generated evolutions. The shallow-circuit measurement can achieve a lower tomography complexity compared to the existing method based on Pauli or Clifford measurements. Our approach is also applicable to approximately locally scrambled unitary ensembles with a controllable bias that vanishes quickly.  Surprisingly, we find a single instance of time-dependent local Hamiltonian evolution is sufficient to perform an approximate tomography as we numerically demonstrate it using a paradigmatic spin chain Hamiltonian modeled after trapped ion or Rydberg atom quantum simulators. Our approach significantly broadens the application of classical shadow tomography on near-term quantum devices. 

\end{abstract}

\pacs{Valid PACS appear here}

\maketitle


\section{Introduction}

Quantum state tomography\cite{Vogel1989Determination,James2001Measurement,Caves2002Unknown} is an essential task in many quantum technology applications. It seeks to reconstruct a quantum state from experimental data of repeated measurements. While reconstructing the full density matrix of a many-body system quickly becomes unfeasible with increasing system size due to the curse of dimensionality\cite{ODonnell2015Efficient,Haah2015Sample-optimal}, predicting a collection of (possibly exponentially many) properties of the quantum system can still be efficiently achieved with an only polynomial number of state copies, which was the idea of \emph{shadow tomography} proposed by Aaronson\cite{Aaronson2017Shadow,Aaronson2019Gentle}. The idea is further improved by the recent work\cite{Huang2020Predicting} to propose the \emph{classical shadow tomography}, which significantly reduces the demand on the quantum hardware and enables efficient classical post-processing.

\begin{figure}[htbp]
\begin{center}
\includegraphics[width=0.82\columnwidth]{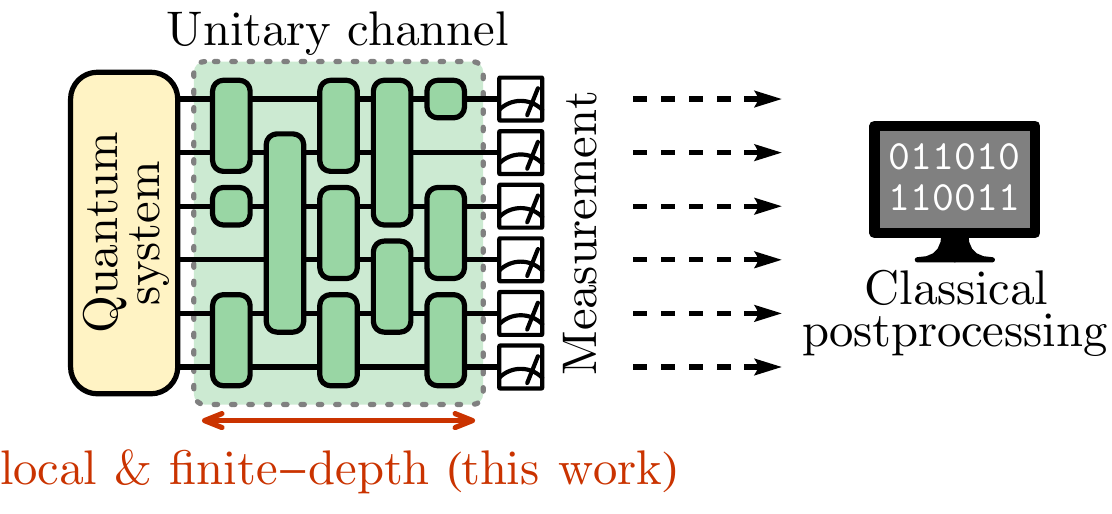}
\caption{Illustration of classical shadow tomography protocol. This work focuses on the case when the unitary channel is of finite depth and respects locality.}
\label{fig:protocol}
\end{center}
\end{figure}

Given a copy of an unknown quantum state $\rho$ of $N$ qubits, the classical shadow tomography protocol (see \figref{fig:protocol}) first transforms the state $\rho\to\rho'=U\rho U^\dagger$ by a unitary $U$, which is randomly sampled (independently each time) from some probability distribution $P(U)$, then measures the transformed state $\rho'$ in the computational basis, $\rho'\to\ket{b}\bra{b}$, which collapses the system to a product state $\ket{b}$ labeled by a bit-string $b\in\{0,1\}^{\times N}$ of measurement outcomes $b=(b_1,\cdots,b_N)$ with the probability $P(b|\rho')=\bra{b}\rho'\ket{b}$. Based on the observed bit-string $b$ and the classical description of the unitary $U$, a classical snapshot $\hat{\sigma}_{U,b}=U^\dagger\ket{b}\bra{b}U$ can be constructed in principle, which essentially encodes the measurement outcomes together with their basis choice (pulled back through the unitary evolution). Repeating such measurements on independent and identical copies of $\rho$ for a few times, a collection of classical snapshots $\scE_{\sigma|\rho}=\{\hat{\sigma}_{U,b}\}$ can be obtained (which correlates with $\rho$). \refcite{Ohliger2013Efficient} showed that as long as the unitary ensemble is expressive enough (i.e.~tomographically complete), there exist a linear reconstruction map $\scM^{-1}$ such that the density matrix $\rho$ can be formally recovered as
$\rho=\E_{\hat{\sigma}\in\scE_{\sigma|\rho}}\scM^{-1}[\hat{\sigma}]$. This also enables the prediction of many properties of $\rho$, like the expectation value of any physical observable $O$ as: $\langle O\rangle=\Tr(O\rho)=\E_{\hat{\sigma}\in\scE_{\sigma|\rho}}\Tr (O\scM^{-1}[\hat{\sigma}])$. The construction of classical snapshots $\hat{\sigma}_{U,b}$ and the computation of their associated properties are performed on a classical computer. 

However, the existing methods\cite{Aaronson2017Shadow,Aaronson2019Gentle,Huang2020Predicting,Hu2021Hamiltonian-Driven} have limitations in applying to near-term quantum devices. First, depending on the type of observables $O$ that we are interested in, one needs to employ different strategies to design the unitary circuit $U$. Two limiting cases have been analyzed in \refcite{Huang2020Predicting}: (i) if the observable is low-rank (such as many-body overlap fidelity), it is most efficient to adopt deep circuits, such that $U$ effectively forms a global Haar random ensemble; (ii) if the observable is high-rank and quasi-local, it would be more efficient to adopt shallow circuits (e.g.~the on-site Haar random). Otherwise the sample complexity will be high. However, the flexibility to interpolate between these two limits has not been available yet, such that the tomography protocol can not adjust to the target observables in a more adaptive manner. Second, more importantly, in existing quantum simulation platforms, applying random unitary circuits is very challenging, because it requires high degrees of sophisticated quantum controls. In particular, for programmable quantum simulators of large systems based on trapped ions or Rydberg atom systems,\cite{2020arXiv201212281E,2020arXiv201212268S,NatureTrapIon} a certain set of entangling unitary evolution is much more favorable to implement than typical random unitaries that require fine-tuned control.
Therefore, it is desirable to develop a method applicable for systems with limited controls.


In this work, we address these challenges by generalizing the classical shadow tomography methods to a broad class of unitary ensembles. In our approach, the specific details of the unitary ensemble is not important as long as the ensemble generates \emph{locally scrambled} quantum dynamics\cite{Kuo2020Markovian}. Rigorously speaking, the probability distribution $P(U)$ of evolution unitaries is invariant under local basis transformations, i.e.~$\forall V\in\U(d)^N: P(U)=P(UV)=P(VU)$ where $V=\prod_iV_i$ is a product of local unitary operator $V_i$ on each qudit. This basically means that the unitary evolution $U$ is efficient in scrambling local quantum information, such that the initial local basis choice is quickly ``forgotten'' under the quantum dynamics. Examples of locally scrambled quantum dynamics includes random unitary circuits (including random Clifford circuit at the 3-design level)\cite{Nahum2017Quantum,Zhou2018Emergent,Nahum2018Operator,Choi2019QECEPTRUCWPM,Bao2019TPTRUCWM,Fan2020Self-Organized} and quantum Brownian dynamics\cite{Lashkari2013TFSC,Xu2018LQFS,Gharibyan2018ORMBSS,Zhou2019ODBQC,Chen2019QCDLPIS}. As the unitary ensemble does not care about local basis choice, the only information that matters will be the quantum entanglement that the unitary dynamics can create in the quantum system. Therefore, for locally scrambled quantum dynamics, the reconstruction map only depends on the entanglement property of the classical snapshots. The density matrix $\rho$ can be reconstructed as a linear superposition of the classical snapshot $\hat{\sigma}$ reduced in different subsystems. The combination coefficient can be calculated from the entanglement feature\cite{You2018Machine,You2018Entanglement} of the classical snapshots, which is simply the collection of average purities of classical snapshots in all possible subregions.

Since our method is applicable to a broad class of quantum dynamics, it is natural to consider an ensemble of realistic Hamiltonian evolutions that are readily available in near-term quantum devices. To this end, we introduce an \emph{approximate} classical shadow tomography (with a non-vanishing but small bias) applicable to an ensemble of time-dependent Hamiltonian evolution that generates approximately locally scrambled dynamics. We numerically demonstrate this idea by using a simple spin chain Hamiltonian modeled after programmable trapped ions or Rydberg atom array systems. We introduce the local frame potential to characterize the bias and we show the bias decreases rapidly for the initial short period of time, and reaches a vanishingly small plateau value for the proposed Hamiltonian. Surprisingly, we find even a single instance from an ensemble of Hamiltonian evolution suffices to perform an approximate tomography, implying that our method is hardware efficient for existing quantum devices\cite{Saffman_2016,RevModPhys.93.025001}.

In the following, we will first establish the general theoretical framework to calculate the reconstruction map in \secref{sec:reconstruction} and to bound the sample complexity in \secref{sec:variance}. We also provide a two-qudit toy model to analytically demonstrate our construction in \secref{sec:twoqudit}. We comment on how to carry out the computation efficiently in \secref{sec:remarks}. Then we apply our construction for local unitary circuits and numerically demonstrates its accuracy in quantum fidelity and Pauli observable estimation tasks in \secref{sec:RUC}, as well as their scaling of sample complexity in \secref{sec:complexity}. Finally, we show in \secref{sec:Hamiltonian} that our approach can be extended to broader classes of unitary ensembles that are approximately locally scrambled. We propose a  frame potential to characterize the level of approximation, which serves as a powerful indicator to design nearly-locally-scrambled unitary ensembles that are available for existing analog quantum simulators\cite{Saffman_2016,RevModPhys.93.025001}. We summarize our classical post-processing protocol and outline a few interesting future applications in \secref{sec:summary}

\section{Theoretical Framework}\label{sec:theory}

\subsection{Reconstruction Map from Entanglement Features}\label{sec:reconstruction}

To be general, we consider a quantum system consists of $N$ qudits, where each qudit has the Hilbert space dimension $d$ (where $d=2$ corresponds to the qubit system). The protocol of classical shadow tomography describes a process that first measures the unknown quantum state $\rho$ in a random basis specified by the unitary transformation $U$ and then prepare the classical snapshot $\hat{\sigma}_{U,b}\equiv U^\dagger\ket{b}\bra{b}U$ based on the measurement outcome $b$. The randomness involved in the process includes (i) sampling $U$ from the distribution $P(U)$ and (ii) obtaining the measurement outcome $b$ conditioned on the evolved state $\rho'=U\rho U^\dagger$ with the probability $P(b|\rho')=\bra{b}\rho'\ket{b}=\Tr (\hat{\sigma}_{U,b}\rho)$.     Inspired by the discussion in \refcite{Acharya2021Informationally}, we define \eq{\label{eq:Sigma_rho}\scE_{\sigma|\rho}=\{\hat{\sigma}_{U,b} \;|\; P(\hat{\sigma}_{U,b}|\rho)= \Tr (\hat{\sigma}_{U,b}\rho) P(U)\}}
as the \emph{posterior snapshot ensemble}, as it is conditioned on the observation of $\rho$.
The posterior snapshot ensemble reduces to the \emph{prior snapshot ensemble}
\eq{\label{eq:Sigma_0}\scE_{\sigma}=\{\hat{\sigma}_{U,b} \;|\; P(\hat{\sigma}_{U,b})= d^{-N}P(U)\},}
when there is no knowledge contained in $\rho$, i.e.~$\rho=d^{-N}\id$. For the prior distribution $P(\hat{\sigma}_{U,b})$, the outcome $b$ is uniformly drawn from all possible outcomes in $\{0,1,\cdots,d-1\}^{\times N}$ (independent of $U,\rho$). The prior snapshot ensemble $\scE_{\sigma}$ only depends on the unitary ensemble $\scE_U=\{U|P(U)\}$. 

With the notation introduced above, the expected classical snapshot $\sigma$ can be expressed as
\eqs{\label{eq:sigma_general}\sigma&\equiv \E_{\hat{\sigma}\in\scE_{\sigma|\rho}}\hat{\sigma}= \E_{\hat{\sigma}\in\scE_{\sigma}}\hat{\sigma}\Tr (\hat{\sigma}\rho)d^N=\scM[\rho],}
which is related to the original state $\rho$ by a quantum channel $\scM$, called the \emph{measurement channel}. It is easy to check that the measurement channel $\scM$ is trace-preserving, completely positive and self-adjoint. It is generally difficult to obtain an explicit expression of $\scM$ for generic unitary ensemble $\scE_U$ (or for generic prior snapshot ensemble $\scE_{\sigma}$). Results of $\scM$ are known for global and on-site 2-design unitaries\cite{Ohliger2013Efficient,Guta2018Fast,Elben2019Statistical} (possibly with noise\cite{Enshan-Koh2020Classical,Chen2020Robust}), fermionic Gaussian unitaries\cite{Zhao2020Fermionic}, and many-body Gaussian unitaries\cite{Hu2021Hamiltonian-Driven}.

We can make progress in computing the measurement channel $\scM$ (and its inverse) for yet another class of unitary ensemble, namely the \emph{locally scrambled unitaries}\cite{Kuo2020Markovian}, for which $P(U)$ obeys the local-basis invariance condition
\eq{\label{eq:LS_unitary}\forall V\in\U(d)^N: P(U)=P(UV)=P(VU),} 
where the local scrambling unitary $V$ is an element in the group $\U(d)^{N}$ (the tensor product of the on-site unitary group $\U(d)$ of each qudit). This condition is sufficient to ensure the prior ensemble $\scE_{\sigma}$ of snapshot states $\hat{\sigma}$ to be invariant under $\hat{\sigma}\to V^\dagger \hat{\sigma} V$,\eq{\label{eq:LS_state}\forall V\in\U(d)^N:P(\hat{\sigma})=P(V^\dagger \hat{\sigma} V).} 
In this case, we say that $\scE_{\sigma}$ is a \emph{locally scrambled ensemble}. In fact, our following derivation only requires the weaker condition \eqnref{eq:LS_state} at the state level, instead of \eqnref{eq:LS_unitary} at the channel level, though it will be practically more straight forward to design unitary circuits that satisfies \eqnref{eq:LS_unitary} by assembling locally scrambled unitary gates. 

Nevertheless, as long as the states $\hat{\sigma}$ are locally scrambled (even if the unitaries $U$ may or may not be locally scrambled), we will be able to insert local basis transformations $V$ in \eqnref{eq:sigma_general}, and average $V$ over any ensemble of our choice,
\eq{\label{eq:sigma_V}\sigma=\E_{V\in\U(d)^N}\E_{\hat{\sigma}\in\scE_{\sigma}}V^\dagger \hat{\sigma}V\Tr (V^\dagger \hat{\sigma}V \rho)d^N.}
We can choose the ensemble of $V=\prod_i V_i$ to be such that every $V_i$ is independently a local 2-design unitary. With this choice, the ensemble average of $V$ can be evaluated by averaging every $V_i$ over the Haar unitary measure following \refcite{Weingarten1978Asymptotic,Collins2006Integration}, and the result can be written as (see \appref{app:EF} for derivation)
\eq{\label{eq:sigma}\sigma=\sum_{B,C\in 2^{\Omega_N}}d^{2N-|B|}\rho_{B}\Wg_{B,C}W_{\scE_{\sigma},C}^{(2)},}
with $B, C$ summing over all possible subregions of the $N$ qudit system, where each subregion is labeled by a subset of $\Omega_N=\{1,\cdots,N\}$ (as an element in the power set $2^{\Omega_N}$). $|B|$ denotes the size (cardinality) of the region $B$. $\rho_{B}=(\Tr_{\bar{B}}\rho)\otimes(\id_{\bar{B}}/d^{|\bar{B}|})$ is the reduced density matrix of $\rho$ in region $B$ embedded back into the total Hilbert space. $\bar{B}$ denotes the complement of region $B$. Note that $B$ and $\bar{B}$ do not need to be consecutive regions in the space, and they can intertwine with each other in general. $\Wg_{B,C}=(d^2-1)^{-N}(-1/d)^{|B\ominus C|}$ is the Weingarten function of regions $B$ and $C$, where $B\ominus C$ denotes the subregions that belong to either $B$ or $C$ but not both. 
\eq{\label{eq:WC0}
W_{\scE_{\sigma},C}^{(2)}\equiv\E_{\hat{\sigma}\in\scE_{\sigma}}\Tr_{C}(\Tr_{\bar{C}}\hat{\sigma})^2=\E_{\hat{\sigma}\in\scE_{\sigma}}e^{-S_{C}^{(2)}(\hat{\sigma})}} is the 2nd \emph{entanglement feature}\cite{You2018Machine,You2018Entanglement} of the prior snapshot ensemble $\scE_{\sigma}$, where $S_C^{(2)}(\hat{\sigma})$ denotes the 2nd R\'enyi entanglement entropy of the state $\hat{\sigma}$ in region $C$. The entanglement feature $W_{\scE_{\sigma},C}^{(2)}$ is merely a property of the unitary ensemble $\scE_U$ (which determines $\scE_{\sigma}$). It describes how the unitary channel entangles a product state in general. It depends on neither the underlying state $\rho$ to be reconstructed nor any particular snapshot state $\hat{\sigma}$ collected in the tomography process.

Given the entanglement feature $W_{\scE_{\sigma},C}^{(2)}$, \eqnref{eq:sigma} spells out how the expected classical snapshot $\sigma$ is written as a linear combination of reduced density matrices $\rho_B$ in all regions, which explicitly specifies the measurement channel $\scM$ as a linear map $\sigma=\scM[\rho]$ from $\rho$ to $\sigma$. Therefore, any reduced classical snapshot $\sigma_A$ must also be a linear combination of reduced density matrices $\rho_B$, which implies that the measurement channel can be represented as a matrix $\scM_{AB}$ such that $\sigma_{A}=\sum_{B}\scM_{AB}\rho_B$. Suppose the map $\scM$ is invertible (i.e.~the unitary ensemble is tomographically complete), the inverse map $\scM^{-1}$ (the reconstruction map) must also be a linear map that combines all reduced classical snapshots $\sigma_A$ to reconstruct $\rho_B=\sum_{A}(\scM^{-1})_{BA}\sigma_A$. In particular, we are most interested to reconstruct the full density matrix $\rho$ (because all reduced density matrices follows from its partial trace), which must also be a linear combination of $\sigma_A$ with some coefficients $r_A\in\dsR$,
\eq{\label{eq:rho}\rho=\scM^{-1}[\sigma]=d^N\sum_{A\in 2^{\Omega_N}}r_{A}\sigma_{A},}
where $\sigma_A=(\Tr_{\bar{A}}\sigma)\otimes(\id_{\bar{A}}/d^{|\bar{A}|})$ follows the same definition as the reduced density matrix. The reconstruction map $\scM^{-1}$ is not a physical channel, because the reconstruction coefficients $r_{A}$ may not be positive definite in general. Nevertheless, $\scM^{-1}$ is still trace-preserving and self-adjoint. Since $\scM^{-1}$ is linear, we have $\rho=\scM^{-1}[\E_{\hat{\sigma}\in\scE_{\sigma|\rho}}\hat{\sigma}]=\E_{\hat{\sigma}\in\scE_{\sigma|\rho}}\scM^{-1}[\hat{\sigma}]$, which enables us to reconstruct the underlying state $\rho$ from the ensemble of classical snapshots. The collection of $\hat{\rho}=\scM^{-1}[\hat{\sigma}]$ is also called the \emph{classical shadow}\cite{Huang2020Predicting} of $\rho$, which can then be used to predict many properties of $\rho$ efficiently.

Now the key problem is to compute $r_A$ from $W_{\scE_{\sigma},C}^{(2)}$. For a system of $N$ qudits, there will be $2^N$ many reconstruction coefficients $r_{A}$. To determine them, we substitute \eqnref{eq:sigma} to \eqnref{eq:rho} and find
\eq{\label{eq:rho_combination}\rho=\sum_{A,B,C\in 2^{\Omega_N}}f_{A,B,C}r_{A}\rho_{B}W_{\scE_{\sigma},C}^{(2)},}
with the fusion coefficient $f_{A,B,C}$ given by
\eqs{&f_{A,B,C}=\sum_{D\in 2^{\Omega_N}}\delta_{B,A\cap D}d^{2N+|A|-|B|+|\bar{A}\cap \bar{D}|}\mathsf{Wg}_{D,C}\\
&=\Big(\frac{d^3}{d^2-1}\Big)^{N}\sum_{D\in 2^{\Omega_N}}\delta_{B,A\cap D}d^{-|D|}\Big(-\frac{1}{d}\Big)^{|C\ominus D|},}
which is universally determined by the qudit dimension $d$. Here $\delta_{A,B}$ denotes the Kronecker delta of two regions $A$ and $B$, s.t.~$\delta_{A,B}=1$ (or $0$) if $A=B$ (or $A\neq B$). \eqnref{eq:rho_combination} will hold for any choice of $\rho$ if and only if
\eq{\label{eq:rA}
\sum_{A,C\in 2^{\Omega_N}} r_{A} f_{A,B,C} W_{\scE_{\sigma},C}^{(2)}=\delta_{B,\Omega_N},}
where $\Omega_N=\{1,\cdots,N\}$ is the full set that labels the full system of $N$ qudits. By solving this linear equation, we can determine the reconstruction coefficients $r_{A}$ in terms of of the entanglement feature $W_{\scE_{\sigma},C}^{(2)}$, such that the reconstruction map $\scM^{-1}$ can be constructed according to \eqnref{eq:rho}. 

In conclusion, we provide a general framework to compute the reconstruction map for the classical shadow tomography with locally scrambled quantum dynamics. The protocol is summarized as:
\begin{tcolorbox}
\begin{enumerate}[leftmargin=*,align=left]
\vspace{-5pt}
\item Given the prior snapshot ensemble $\scE_\sigma$, first calculate its entanglement feature by $$W_{\scE_{\sigma},C}^{(2)}=\E_{\hat{\sigma}\in\scE_{\sigma}}\Tr_{C}(\Tr_{\bar{C}}\hat{\sigma})^2.$$
\vspace{-15pt}
\item Solve for the reconstruction coefficient $r_A$ by
$$\sum_{A,C\in 2^{\Omega_N}} r_{A} f_{A,B,C} W_{\scE_{\sigma},C}^{(2)}=\delta_{B,\Omega_N}.$$
\vspace{-15pt}
\item Then the reconstruction map is given by
$$\rho=\scM^{-1}[\sigma]=d^N\sum_{A\in 2^{\Omega_N}}r_{A}\sigma_{A}.$$
\end{enumerate}
\end{tcolorbox}
All computations are supposed to be carried out on a classical computer in the post-processing procedure. Although solving for $r_A$ may be computationally demanding for large systems, it only needed to be done once and its result can be applied to process all classical snapshots collected from all possible states $\rho$ to be learned.

\subsection{Variance Estimation and Sample Complexity}\label{sec:variance}

Given the ensemble $\scE_{\sigma|\rho}$ of classical snapshots collected from measuring the unknown state $\rho$, we can use the reconstruction map $\scM^{-1}$ to predict properties of $\rho$. For example, let $O$ be a traceless Hermitian operator representing a physical observable. Its expectation value $\langle O\rangle\equiv \Tr(O\rho)$ can be predicted via
\eq{\label{eq:avg_O}\langle O\rangle=\E_{\hat{\sigma}\in\scE_{\sigma|\rho}}\Tr (O\scM^{-1}[\hat{\sigma}])=\E_{\hat{\sigma}\in\scE_{\sigma|\rho}}\Tr (\scM^{-1}[O]\hat{\sigma}),}
where we have used the self-adjoint property of $\scM^{-1}$ to transpose its action from $\hat{\sigma}$ to $O$. We can interpret $\hat{o}(\hat{\sigma})\equiv \Tr (\scM^{-1}[O]\hat{\sigma})$ as the single-shot estimation of the observable (based on a particular classical snapshot $\hat{\sigma}$), such that $\langle O\rangle=\E_{\hat{\sigma}\in\scE_{\sigma|\rho}}\hat{o}(\hat{\sigma})$.

The variance of the single-shot estimation is defined as
$\var\hat{o}\equiv \E_{\hat{\sigma}\in\scE_{\sigma|\rho}}\hat{o}(\hat{\sigma})^2-(\E_{\hat{\sigma}\in\scE_{\sigma|\rho}}\hat{o}(\hat{\sigma}))^2$,
which can be bounded by (the first term in $\var\hat{o}$)
\eqs{\label{eq:VarO}\var\hat{o}\leq \norm{O}_{\scE_{\sigma|\rho}}^2&\equiv \E_{\hat{\sigma}\in\scE_{\sigma|\rho}}\hat{o}(\hat{\sigma})^2\\
&=\E_{\hat{\sigma}\in\scE_{\sigma}}(\Tr \scM^{-1}[O]\hat{\sigma})^2\Tr(\hat{\sigma}\rho)d^N.}
The bound $\norm{O}_{\scE_{\sigma|\rho}}$ can be considered as a generalized $\rho$-dependent notion of the (squared) \emph{shadow norm}\cite{Huang2020Predicting} of an operator $O$ (whereas the shadow norm originally defined in \refcite{Huang2020Predicting} further maximizes over all possible underlying states $\rho$ to remove the dependence on $\rho$). Assuming $\scE_{\sigma}$ is locally scrambled, following the same approach of inserting and averaging local-basis transformations as in \eqnref{eq:sigma_V}, the bound in \eqnref{eq:VarO} becomes
\eqs{\label{eq:normO0}\norm{O}_{\scE_{\sigma|\rho}}^2= \sum_{g,h\in S_3^{N}}\norm{O}_{\rho,g}^2 \Wg_{g,h}W_{\scE_{\sigma},h}^{(3)},}
where $g,h$ are group elements in the $S_3^N$ (product of 3-fold permutation groups over $N$ qudits). $\Wg_{g,h}$ is the Weingarten function of permutations $g$ and $h$, which is equivalent to traditional Weingarten function as $\Wg_{g,h}=\Wg(gh^{-1},d)$, where $d$ is the local Hilbert dimension of qudit. $\norm{O}_{\rho,g}^2$ is a generalized operator norm for $O$, which is defined as
\eq{\label{eq:normO1}\norm{O}_{\rho,g}^2\equiv d^N\Tr((\scM^{-1}[O]^{\otimes 2}\otimes \rho)\chi_g),}
where $\chi_g$ is the representation of the $S_3^N$ permutation $g$ in the 3-fold Hilbert space. $W_{\scE_{\sigma},h}^{(3)}$ is the 3rd \emph{entanglement feature} of the ensemble $\scE_{\sigma}$, defined as
\eq{\label{eq:W3}W_{\scE_{\sigma},h}^{(3)}\equiv \E_{\hat{\sigma}\in\scE_{\sigma}}\Tr(\hat{\sigma}^{\otimes 3}\chi_h).}
Note that the 2nd entanglement feature previously defined in \eqnref{eq:WC0} can be consistently cast into the form of \eqnref{eq:W3} in terms of permutation operators (see \refcite{You2018Entanglement}).

In practice, the expectation value $\langle O\rangle$ is always estimated based on a finite collection of the snapshot states. 
Let $M$ be the number of samples of $\rho$ measured in the data acquisition stage (each sample results in a snapshot state $\hat{\sigma}_k$). The finite average estimation $\bar{o}=\frac{1}{M}\sum_{k=1}^{M}\hat{o}(\hat{\sigma}_k)$ will fluctuate around the true expectation value $\langle O\rangle$ with a variance that scales as $(\var\hat{o})/M$. By the Chebyshev inequality, the probability for $\bar{o}$ to deviate from $\langle O\rangle$ by more than $\epsilon$ amount is bounded by
\eq{\Pr(|\bar{o}-\langle O \rangle|\geq \epsilon)\leq \frac{\var\hat{o}}{\epsilon^2 M}\leq \frac{\norm{O}_{\scE_{\sigma|\rho}}^2}{\epsilon^2 M}.}
Therefore, to control the failure probability within a threshold $\delta$, i.e.~$\Pr(|\bar{o}-\langle O \rangle|\geq \epsilon)\leq \delta$, sufficient number of samples is required
\eq{\label{eq:M0}M\geq \frac{\norm{O}_{\scE_{\sigma|\rho}}^2}{\epsilon^2 \delta}.} 
A larger (smaller) shadow norm $\norm{O}_{\scE_{\sigma|\rho}}^2$ indicates that more (less) samples are needed. 

However, the $\rho$-dependent shadow norm $\norm{O}_{\scE_{\sigma|\rho}}^2$ is generally complicated to evaluate. If we are not interested in the shadow norm for a specific state $\rho$, but rather the expectation of the shadow norm over an ensemble of states $\{V\rho V^\dagger\}$ that are similar to $\rho$ by local basis transformations $V\in \U(d)^N$, we can actually define a $\rho$-independent shadow norm by averaging over $V$,
\eqs{\norm{O}_{\scE_{\sigma}}^2&\equiv\E_{V\in\U(d)^N}\norm{O}_{\scE_{\sigma|{V \rho V^\dagger}}}^2.}
The expected shadow norm can be expressed purely in terms of the entanglement features of $\scE_{\sigma}$ and $\scE_O$ (see \appref{app:var} for derivation),
\eq{\label{eq:normO5}\norm{O}_{\scE_{\sigma}}^2=\sum_{A,B,C,D\in 2^{\Omega_N}}v_{A,B,C,D}W_{\scE_{\sigma},{A\cap B\cap C}}^{(2)}W_{\scE_O,D}^{(2)},}
where the coefficient $v_{A,B,C,D}$ is given by
\eq{v_{A,B,C,D}=r_A r_B \Big(\frac{d^2}{d^2-1}\Big)^N d^{|{A\cap B\cap C}|-|C|}\Big(-\frac{1}{d}\Big)^{|C\ominus D|},}
and $\scE_{O}=\{V^\dagger O V|V\in \U(d)^N\}$ denotes the locally scrambled ensemble (or known as $\U(d)^N$-twirling) associated with the observable $O$ in question.

In conclusion, given a traceless Hermitian operator $O$, its expected shadow norm $\norm{O}_{\scE_{\sigma}}^2$ provides a typical lower bound for the number of samples needed
\eq{\label{eq:M1}M\gtrsim \frac{\norm{O}_{\scE_{\sigma}}^2}{\epsilon^2 \delta},} 
in order to control the error of the prediction $\bar{o}$ given by the classical shadow tomography within the probability bound $\Pr(|\bar{o}-\langle O \rangle|\geq \epsilon)\leq \delta$. Here we have only analyzed the sample complexity for a single linear observable. For the analysis of multiple and/or non-linear observables, we refer to the original paper of \refcite{Huang2020Predicting}. Their result applies to our case simply by replacing the shadow norm with our version.

\subsection{A Toy Example of Two-Qudit System}\label{sec:twoqudit}

\begin{figure}[htbp]
\begin{center}
\includegraphics[width=0.65\columnwidth]{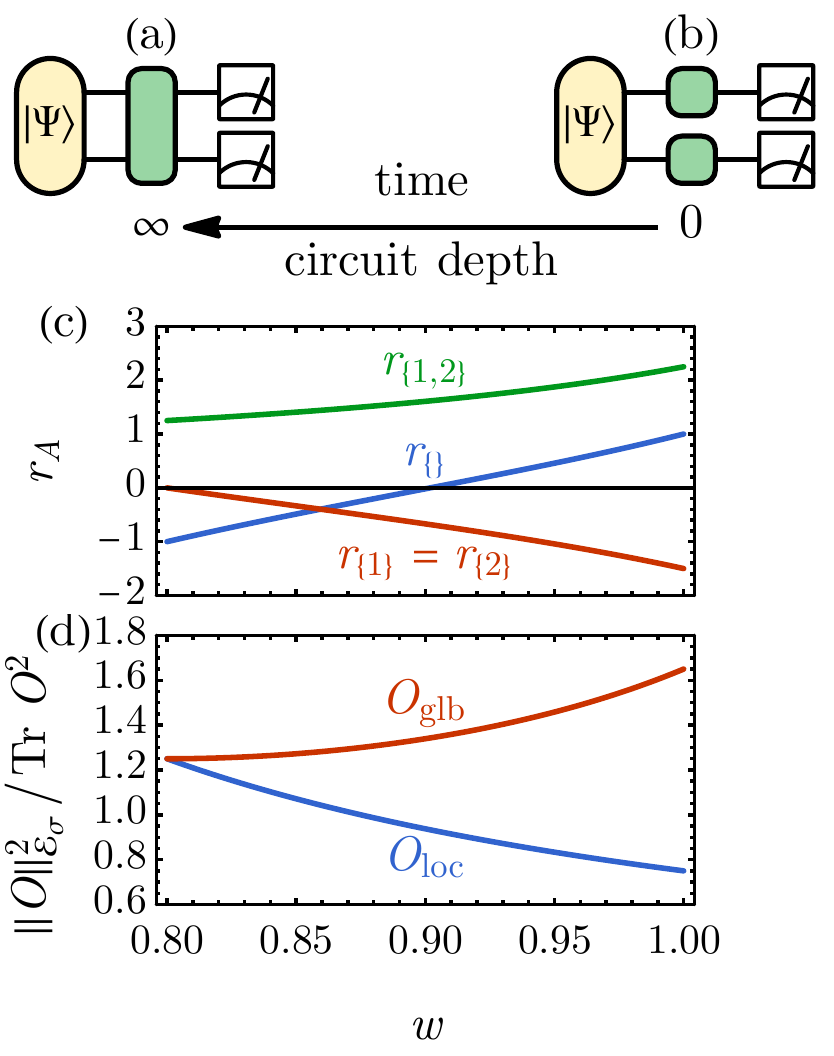}
\caption{Two-qudit unitary channel in (a) the long-time (Page state) limit and (b) the short-time (product state) limit. (c) Reconstruction coefficients $r_A$ and (d) the shadow norm $\norm{O}_{\scE_{\sigma}}^2$ v.s. the single-qudit purity $w$, for $d=2$. $w$ varying from $1$ to $4/5$ effectively models the circuit depth (or evolution time) growing from 0 to $\infty$.}
\label{fig:twoqudit}
\end{center}
\end{figure}

To demonstrate our framework and to gain some analytical intuition, we present a toy example to compute the reconstruction map in a two-qudit ($N=2$) system. We assume that the two-qudit system always evolves under a locally scrambled quantum dynamics, which can be modeled (for example) by a finite-time Brownian evolution\footnote{The Brownian unitary evolution is a product of a sequence of infinitesimal time-evolution $U=\prod_t e^{-\ii H_t\;\delta t}$, but the Hamiltonian $H_t$ at each time step is independent drawn from a random Hamiltonian ensemble (unlike 
the coherent quantum dynamics, where the same Hamiltonian drives the dynamics though all time.)} driven by random Hamiltonians. Every classical snapshot $\hat{\sigma}_{U,b}=U^\dagger \ket{b}\bra{b}U$ is generated by the reversed evolution from the product state $\ket{b}\bra{b}$. In the long-time limit (\figref{fig:twoqudit}(a)), the entanglement feature $W_{\scE_{\sigma}}^{(2)}=(1,\frac{2d}{d^2+1},\frac{2d}{d^2+1},1)$ follows from that of Page states, where the subregion basis are arranged in the order of $\{\},\{1\},\{2\},\{1,2\}$. This is because the evolution of entanglement feature under any locally scrambled quantum dynamics always converges to the Page state, regardless of the initial state, as proven in \refcite{Kuo2020Markovian}. In the short-time limit (\figref{fig:twoqudit}(b)), $\hat{\sigma}$ remains as a product state, therefore the entanglement entropy vanishes for all regions, which translates to $W_{\scE_{\sigma}}^{(2)}=(1,1,1,1)$. In general, for any intermediate time, the entanglement feature should take the form of 
\eq{\label{eq:WC2}W_{
\scE_{\sigma}}^{(2)}=(1,w,w,1),}
with $w$ varies between $\frac{2d}{d^2+1}$ (the long-time limit) and $1$ (the short-time limit). The physical meaning of $w$ is the average single-qudit purity in the snapshot state $\hat{\sigma}$.

Given $W_{\scE_{\sigma}}^{(2)}$ in \eqnref{eq:WC2}, \eqnref{eq:rA} reads
\eq{d^2\mat{
1 & \frac{d(d-w)}{d^2-1} & \frac{d(d-w)}{d^2-1} & \frac{d^2(d^2-2dw+1)}{(d^2-1)^2}\\
0 & \frac{dw-1}{d^2-1} & 0 & \frac{d(d^2w-2d+w)}{(d^2-1)^2}\\
0 & 0 & \frac{dw-1}{d^2-1} & \frac{d(d^2w-2d+w)}{(d^2-1)^2}\\
0 & 0 & 0 & \frac{d^2-2dw+1}{(d^2-1)^2}
}\mat{r_{\{\}}\vspace{5pt}\\r_{\{1\}}\vspace{5pt}\\r_{\{2\}}\vspace{5pt}\\r_{\{1,2\}}} 
=\mat{0\vspace{5pt}\\0\vspace{5pt}\\0\vspace{5pt}\\1}.}
By solving this linear equation, the reconstruction coefficient $r_A$ can be obtained
\eq{\label{eq:rsol}
r=\mat{r_{\{\}}\vspace{5pt}\\r_{\{1\}}\vspace{5pt}\\r_{\{2\}}\vspace{5pt}\\r_{\{1,2\}}} 
=\mat{\frac{d^3w-3d^2+3dw-2w^2+1}{(dw-1)(d^2-2dw+1)}\\\frac{-d^4w+2d^3-2d+w}{d(dw-1)(d^2-2dw+1)}\\\frac{-d^4w+2d^3-2d+w}{d(dw-1)(d^2-2dw+1)}\\\frac{(d^2-1)^2}{d^2(d^2-2dw+1)}}.}
The behavior of $r_A$ as a function of $w$ is shown in \figref{fig:twoqudit}(c), which continuously interpolates the two limits.

In the short-time limit, $w=1$ and \eqnref{eq:rsol} reduces to $r_A=(1,-(d+1)/d,-(d+1)/d,(d+1)^2/d^2)$, corresponding to the reconstruction map
\eq{\scM^{-1}[\sigma]=\bigotimes_{i=1,2}((d+1)\sigma_i-\id_i),}
matching the result of on-site 2-design circuits\cite{Guta2018Fast,Elben2019Statistical}. In the long-time limit, $w=\frac{2d}{d^2+1}$ and \eqnref{eq:rsol} reduces to $r_A=(-1,0,0,(d^2+1)/d^2)$, corresponding to the reconstruction map
\eq{\scM^{-1}[\sigma]=(d^2+1)\sigma-\id,}
matching the result of global 2-design circuits\cite{Guta2018Fast,Elben2019Statistical}. The general result in \eqnref{eq:rsol} provides the reconstruction map that interpolates these two limits, which allows us to perform classical shadow tomography for intermediate unitary channels that are neither on-site nor global 2-design.

To investigate the sample complexity of the tomography scheme in the two-qudit system, we consider a traceless Hermitian operator $O$ (i.e.~$\Tr O =0$) and define two parameters $k_1$ and $k_2$ to parameterize the purity:
\eqs{k_1&=\Tr_{\{1\}}(\Tr_{\{2\}}O)^2/\Tr O^2\\
k_2&=\Tr_{\{2\}}(\Tr_{\{1\}}O)^2/\Tr O^2.}
Then the entanglement feature of the observable $O$ can be arranged as the following vector
\eq{W_{\scE_O}^{(2)}=(0,k_1,k_2,1)\Tr O^2,}
with the same choice of region basis as in \eqnref{eq:WC2}. Given $r$, $W_{\scE_O}^{(2)}$ and $W_{\scE_{\sigma}}^{(2)}$, we have all the information needed to calculate the shadow norm, according to \eqnref{eq:normO5}
\eq{\label{eq:normO6}\norm{O}_{\scE_{\sigma}}^2=\tfrac{d^2-1}{d^3}\big(\tfrac{k_\text{tot}}{d w-1}+\tfrac{(d^2-1)(d-k_\text{tot})}{d^2-2d w+1}\big)\Tr O^2,}
where $k_\text{tot}=k_1+k_2$. 

The operator locality crucially affects $k_\text{tot}$. Consider modeling a local operator $O_\text{loc}$ by a random operator drawn from the Gaussian unitary ensemble (GUE) and acting on the first qudit only, we have \eq{W_{\scE_{O_\text{loc}}}^{(2)}=(0,d,0,1)\Tr O_\text{loc}^2,} hence $k_\text{tot}=d$. On the other hand, for a global operator $O_\text{glb}$ modeled by a global GUE random operator acting on both qudits simultaneously, we have \eq{W_{\scE_{O_\text{glb}}}^{(2)}=(0,\tfrac{d}{d^2+1},\tfrac{d}{d^2+1},1)\Tr O_\text{glb}^2,}
hence $k_\text{tot}=\tfrac{2d}{d^2+1}$. In these two cases, the shadow norm in \eqnref{eq:normO6} becomes
\eqs{\norm{O_\text{loc}}_{\scE_{\sigma}}^2&=\tfrac{d^2-1}{d^2(dw-1)}\Tr O_\text{loc}^2,\\
\norm{O_\text{glb}}_{\scE_{\sigma}}^2&=\tfrac{d^2-1}{d^2(d^2+1)}\big(\tfrac{(d^2-1)^2}{d^2-2dw-1}+\tfrac{2}{dw-1}\big)\Tr O_\text{glb}^2.}
Their dependence in $w$ is plotted in \figref{fig:twoqudit}(d). In the short-time limit ($w=1$), $\norm{O_\text{loc}}_{\scE_{\sigma}}^2<\norm{O_\text{glb}}_{\scE_{\sigma}}^2$, meaning that the shallow circuit is more efficient in predicting local observables. In the long-time limit ($w=\tfrac{2d}{d^2+1}$), $\norm{O_\text{loc}}_{\scE_{\sigma}}^2=\norm{O_\text{glb}}_{\scE_{\sigma}}^2=(1+d^{-2})\Tr O^2$, such that there is no difference in predicting both local and global observables in terms of the sample efficiency, because all operators are equally scrambled in this limit.

\subsection{Additional Remarks on Computational Methods and Future Directions}\label{sec:remarks}

Efficient numerical methods have been developed\cite{Fan2020Self-Organized,Akhtar2020Multiregion} to calculate the evolution of entanglement feature $W_{\scE_{\sigma}}^{(k)}$ under locally scrambled quantum dynamics by solving the corresponding entanglement dynamics equation (without simulating the quantum dynamics using brute force). However, we will leave this approach for future exploration. In this work, we will compute the entanglement feature beforehand based on the definition \eqnref{eq:WC0}, by direct sampling from the prior snapshot ensemble $\scE_\sigma$. For experimentally generated random unitaries whose distribution is a priori unknown, it is also possible to estimate the entanglement feature efficiently from R\'enyi entropy measurements\cite{Elben2018Renyi,Vermersch2018Unitary,Brydges2019Probing} following the definition \eqnref{eq:WC0}. 

 As shown in \refcite{Akhtar2020Multiregion}, for one-dimensional quantum systems, the entanglement feature vector $W_{\scE_{\sigma}}^{(k)}$ admits efficient matrix product state (MPS) representation, even if snapshot states in $\scE_{\sigma}$ are volume-law entangled. Combining the MPS representation of $W_{\scE_{\sigma}}^{(2)}$ with the fact that $f_{A,B,C}$ is factorizable to every qudit, one can develop efficient MPS-based numerical approach to find the solution of $r_{A}$ (also as a MPS). However, we will defer the development of this approach to future work. In the following numerical demonstrations, we will directly solve \eqnref{eq:rA} for small systems as a proof of concept. 

The MPS representations for $r$, $W_{\scE_{\sigma}}^{(2)}$ and $W_{\scE_O}^{(2)}$ also enables us to calculate the shadow norm $\norm{O}_{\scE_{\sigma}}^2$ efficiently by a four-way MPS contraction. The ability to compute the shadow norm efficiently will be particularly useful if we want to design optimal unitary channels to minimize the sample complexity for a given set of designated observables. It is possible to apply machine learning approaches (such as deep reinforcement learning) to perform the circuit structure optimization. Therefore, our construction provides the flexibility to allow the classical shadow tomography to adapt to designated observables, which has not been possible before. We will also leave this promising direction to future research. 

\section{Numerical Demonstrations}

\begin{figure}[htbp]
\begin{center}
\includegraphics[width=0.99\columnwidth]{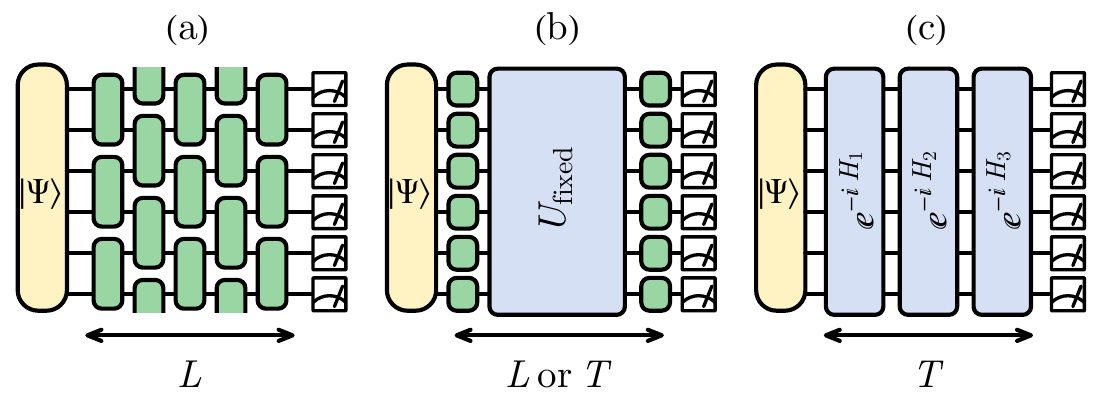}
\caption{Classical shadow tomography with (a) finite-depth random unitary/Clifford circuits (of $L$ layers), (b) a fixed unitary twirled by single qubit random Clifford gates, and (c) discrete-time Hamiltonian dynamics (of $T$ steps).}
\label{fig:circuits}
\end{center}
\end{figure}

To demonstrate the effectiveness of our approach, we consider three types of unitary ensembles for the unitary channel in the data acquisition protocol, as illustrated in \figref{fig:circuits}.

\subsection{Classical Shadow Tomography with Shallow Random Unitary/Clifford Circuits} \label{sec:RUC}

We first consider using random unitary circuits (RUCs) \cite{Nahum2017Quantum} for the unitary channel. As illustrated in \figref{fig:circuits}(a), the unitary circuit consists of two-qubit local unitary gates arranged in the brick-wall pattern with a periodic boundary condition. Each gate in the circuit is independently drawn from the Haar random unitary ensemble. The depth $L$ of the circuit can be adjusted. Obviously, RUCs are locally scrambled, as any local-basis transformation (from both left and right) can be absorbed by the Haar random unitary gates in the circuit. Therefore, we expect our reconstruction map to work perfectly in this case for any choice of the circuit depth $L$.

\begin{figure}[htbp]
\centering
\includegraphics[width=0.8\columnwidth]{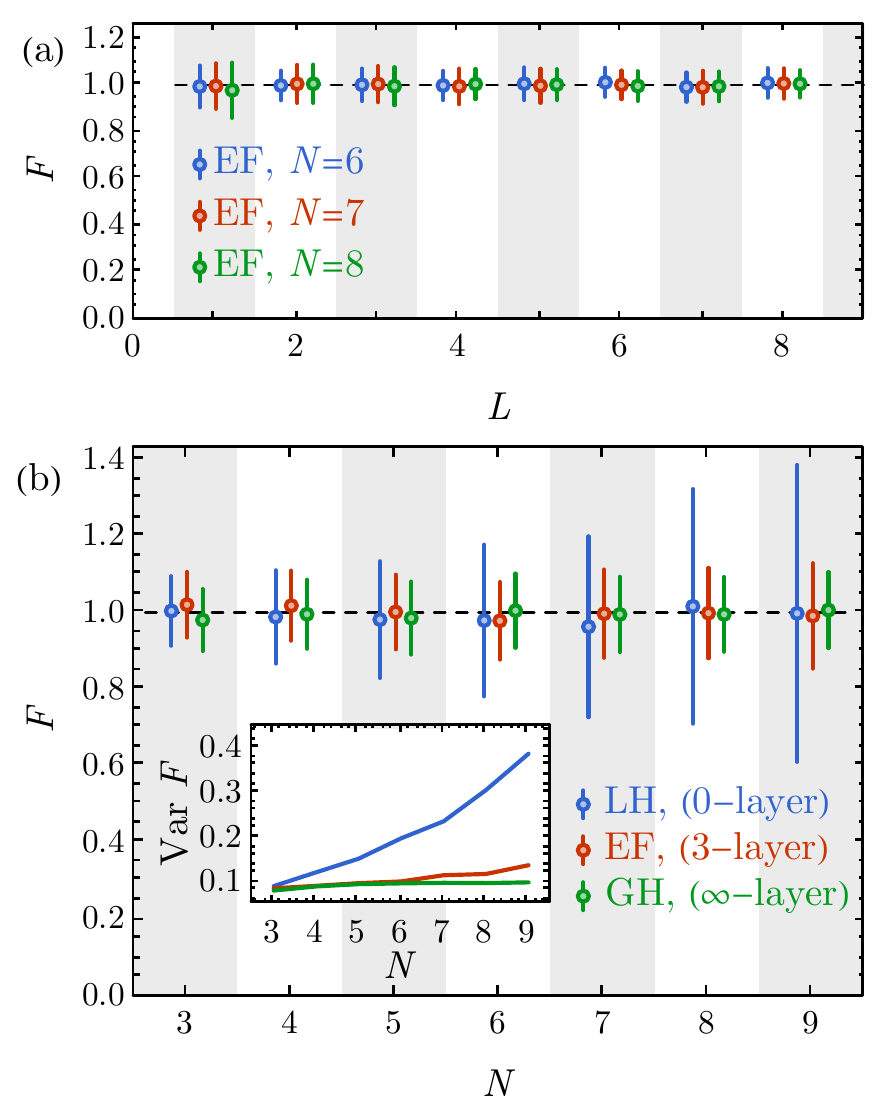}
\caption{(a) Fidelity estimation of GHZ state with RUC of different circuit depth $L$ using entanglement-feature-based reconstruction $\scM^{-1}_\text{EF}$ (denoted by EF)  over different number $N$ of qubits. (b) Fidelity estimation of GHZ state using shallow RUC (3-layer, with $\scM^{-1}_\text{EF}$, denoted by EF), random on-site (local Haar) gates (0-layer, with $\scM^{-1}_\text{LH}$, denoted by LH) and global Haar unitary ($\infty$-layer, with $\scM^{-1}_\text{GH}$, denoted by GH). The inset shows the variance $\var F$ of the predicted fidelity as a function of system size $N$. In both subfigures, the sample size is 5000. Error bar indicates 3-standard-deviation estimated by the bootstrap method. Points are split horizontally to avoid the overlap of markers.}
\label{fig:fidelity_RUC}
\end{figure}

For illustration purpose, we start with a Greenberger-Horne-Zeilinger (GHZ) state $\rho=\ket{\Psi}\bra{\Psi}$, where $\ket{\Psi}=\frac{1}{\sqrt{2}}(\ket{00\cdots 0}+\ket{11\cdots 1})$. For every given circuit depth $L$, we first calculate the entanglement feature $W^{(2)}_{\mathcal{E}_{\sigma},C}$ to determine the reconstruction map $\scM^{-1}$. This calculation is done for once and stored in the classical memory for future reference. In our numerical simulation of the data acquisition process, we sample the RUC, apply it to the GHZ state $\ket{\Psi}$, and perform the computational basis measurement. We generate a collection of classical snapshots $\scE_{\sigma|\rho}=\{\hat{\sigma}\}$ of size $M$ by repeated measurements. We then estimate the fidelity $F$ of the reconstructed state by
\eq{\label{eq:F} F=\sqrt{\frac{1}{M}\sum_{\hat{\sigma}\in\scE_{\sigma|\rho}}\bra{\Psi}\scM^{-1}[\hat{\sigma}]\ket{\Psi}}.}
Following the philosophy of classical shadow tomography, one should view \eqnref{eq:F} as a prediction task. If the shadow tomography is successful, then the estimated  fidelity should converge to $F=1$. This estimation can be achieved accurately by a few measurements, even though the full density matrix estimation $\avg_{\hat{\sigma}\in\scE_{\sigma|\rho}}\scM^{-1}[\hat{\sigma}]$ may still have large fluctuations. In addition, when the reconstruction is biased, for example the experimental channel $\scM$ doesn't match the theoretical assumption of the unitary ensemble, then the fidelity estimation will deviate from one.
\figref{fig:fidelity_RUC} (a) shows that the entanglement-feature-based reconstruction map $\scM^{-1}_\text{EF}$ indeed gives unbiased estimation of fidelity $F$ for different circuit depths $L$ and for different system sizes $N$. Furthermore, when the GHZ state is prepared with $Z$ errors, our method can give the correct fidelity estimation that decreases linearly with the probability of $Z$ error, which is challenging for the current state-of-art machine-learning quantum state tomography method\cite{Carrasquilla2019Reconstructing,Huang2020Predicting} (see \appref{app:mixed_state} for more discussions).

\begin{figure}[htbp]
\begin{center}
\includegraphics[width=0.92\columnwidth]{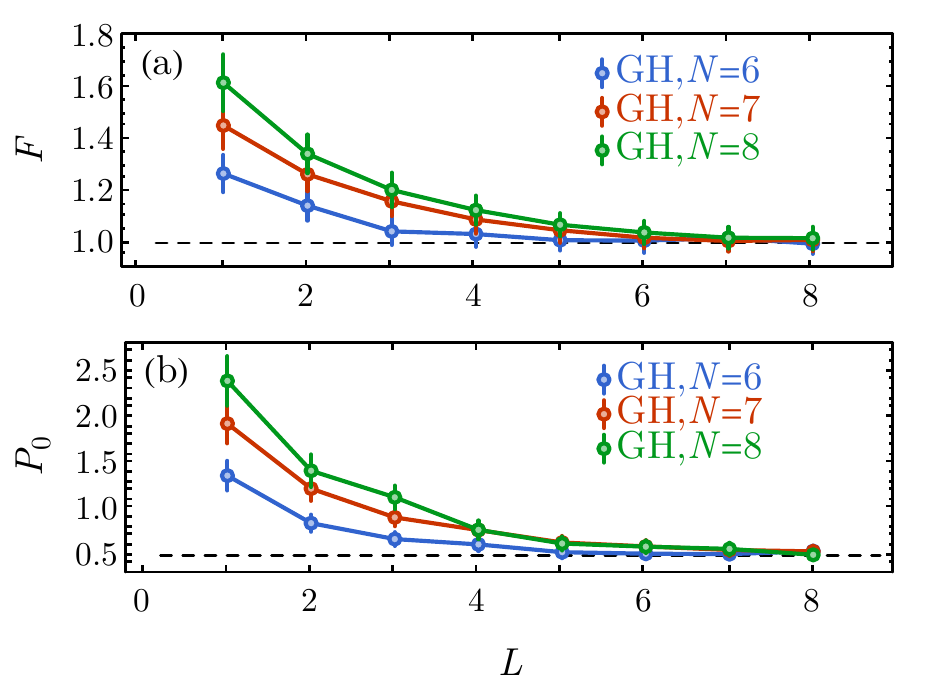}
\caption{(a) Fidelity estimation of the reconstructed GHZ state with RUC of finite depth $L$. (b) Estimation of observable $P_0=|\langle \Psi|00\cdots 0\rangle|^2$ (the projection operator to the $\ket{00\cdots 0}$ state) on the reconstructed GHZ state with RUC of finite depth $L$. In both cases, the reconstruction uses the global Haar reconstruction map. The sample size is 5000. Error bar indicates 3-standard-deviation estimated by the bootstrap method.}
\label{fig:bias}
\end{center}
\end{figure}

To compare with the existing classical shadow tomography method\cite{Huang2020Predicting}, we consider the reconstruction maps $\scM_\text{GH}^{-1}[\sigma]=(d^{N}+1)\sigma-\id$ and $\scM_\text{LH}^{-1}[\sigma]=\bigotimes_{i}((d+1)\sigma_i-\id_i)$, where $\scM_\text{GH}^{-1}$ (or $\scM_\text{LH}^{-1}$) assumes the unitary ensemble is global (or on-site local) Haar random. They can be viewed as special limits where circuit depth $L$ tends to infinity and zero respectively. Although they can also achieve an unbiased estimation of quantum fidelity, the tomography efficiency differs. In \figref{fig:fidelity_RUC}(b), the error bar shows how the (3-times) standard deviation of the estimated fidelity scales with the number of qubits $N$ at 5000 sample size. As we can see (both from the error bar and from the inset of \figref{fig:fidelity_RUC}(b)), the variance of (on-site) local Haar estimation increases drastically as $N$ increases, which implies an increasingly high sample complexity for large systems. 

In the other limit, the variance of global Haar estimation is independent of system size, achieving the optimal sample complexity as advocated in \refcite{Huang2020Predicting}. However, to realize the global Haar ensemble, the circuit depth needs to be at least of order $\scO(N)$, which is quite demanding for quantum devices. If we approximate the global Haar ensemble with finite-depth circuits and use the reconstruction map $\scM_\text{GH}^{-1}$ on data collected from finite-depth circuit measurements,\cite{Ohliger2013Efficient} this will yield systematically biased predictions for physical quantities when the circuit is not deep enough. In \figref{fig:bias}(a), we show that the biased prediction tends to over-estimate the fidelity, leading to the unphysical result of $F>1$ (the correct behavior is $F=1$). 
This occurs because, when the measurement channel $\scM$ in data acquisition protocol disagrees with the reconstruction channel $\scM^{-1}$ in classical post-processing protocol, the reconstructed density matrix $\frac{1}{M}\sum_{\hat{\sigma}\in\scE_{\sigma|\rho}}\scM^{-1}[\hat{\sigma}]$ may not be positive-definite (see \appref{app:purification} for detailed discussions), resulting in the unphysical fidelity estimation. This bias gets worse for larger system size. In \figref{fig:bias}(b), we also show the estimation of $P_0=|\langle \Psi|00\cdots 0\rangle|^2$. For GHZ state, the correct behavior is $P_0=0.5$, and we still see significant bias when applying $\scM^{-1}_{\text{GH}}$ for shallow circuits.


However, with the entanglement-feature-based reconstruction map $\scM_\text{EF}^{-1}$, as demonstrated in \figref{fig:fidelity_RUC} (b), we are able to achieve an \emph{unbiased} fidelity estimation with a 3-layer shallow circuit, approaching \emph{similar variance level} (i.e.~similar sample efficiency) as global Haar ensemble while keeping a \emph{low circuit complexity}. This clearly demonstrates the advantage of our approach.


\subsection{Scaling of Variance and Tomography Complexity}\label{sec:complexity}

The above discussion motivate us to define the \emph{tomography complexity} as $\scC=(L+1)M$, where $L$ is the \emph{circuit complexity} (the number of layers in the quantum circuit), and $M$ is the \emph{sample complexity} (the number of sample needed). $M$ will be proportional to the single-shot variance $\var \hat{o}$. Suppose applying each layer of quantum gates and performing measurements both take a unit of time on the quantum device, then $\scC$ is roughly the total amount of time needed to collect the classical shadow from $M$ copies of the quantum state, which characterizes the complexity of the data acquisition protocol. This notion of complexity is consistent with the Quantum Algorithmic Measurement (QUALM) complexity introduced in \refcite{Aharonov2021Quantum} (with $LM$ and $M$ being their gate and query complexities respectively). In the following, we will investigate the scaling of single-shot variance $\var \hat{o}$ as a function of circuit depth $L$ and system size $N$ for both low-rank operators (such as fidelity) and full-rank operators (such as Pauli operators), and show how the tomography complexity $\mathcal{C}$ can guide us to find the optimal circuit depth $L_*$.

\begin{figure}[htbp]
\centering
\includegraphics[width=\columnwidth]{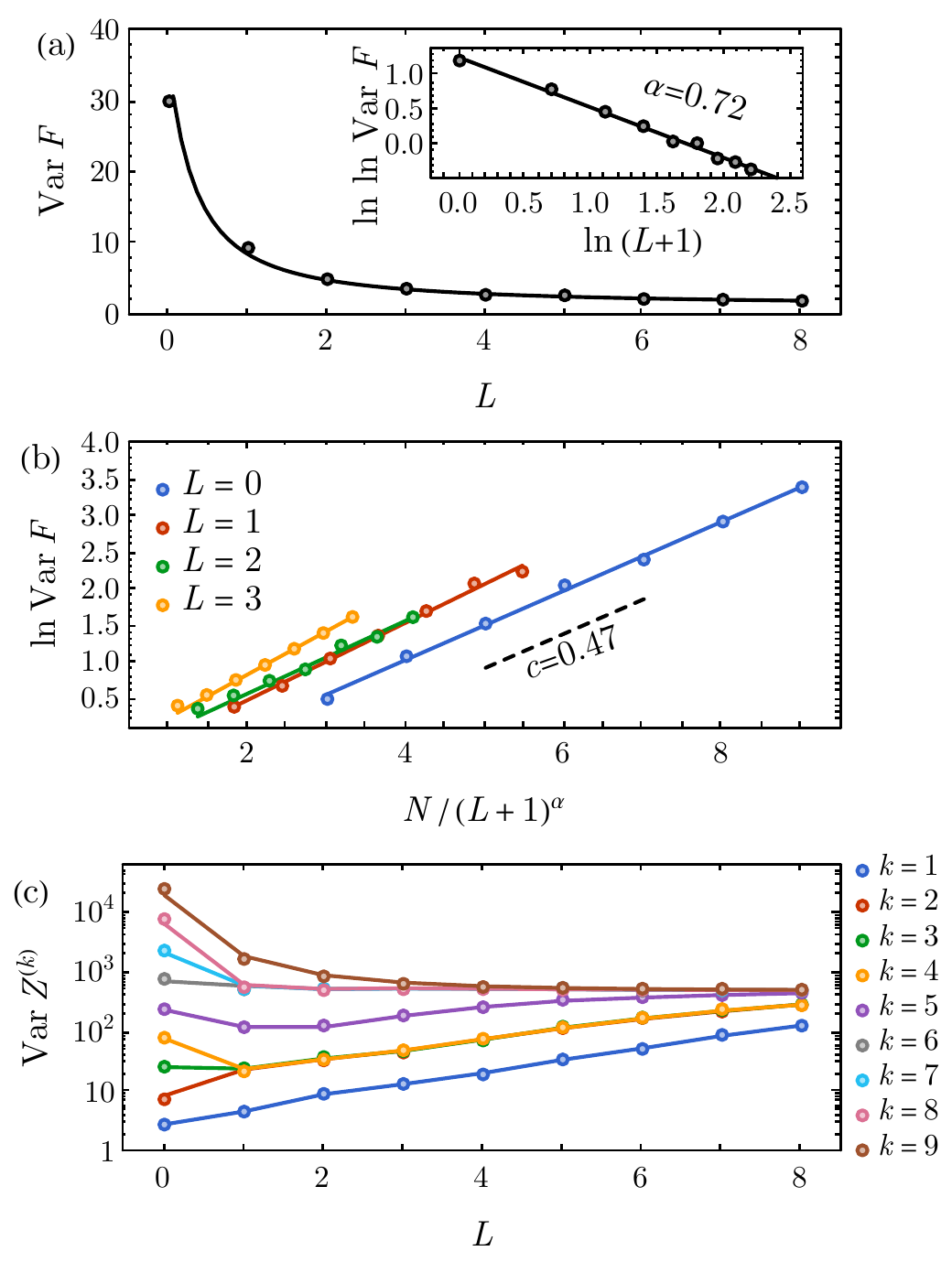}
\caption{(a) Single-shot variance of estimated fidelity v.s. circuit depth $L$ for a 9-qubits GHZ state. (b) Single-shot variance of estimated fidelity as a function of the effective system size $N_\text{eff}$. The best fit for $\var F\propto \exp\big({c\frac{N}{(L+1)^\alpha}}\big)$ gives $c=0.47\pm 0.08$ and $\alpha=0.72\pm 0.1$. (c) Variance of full rank operator estimation on 9-qubit GHZ state. The full rank operators are Pauli-$Z$ operators of the form: $Z^{(k)}=Z^{\otimes k}I^{\otimes(N-k)}$ with different support $k$. The dots are experimental results from simulation, and the lines are theoretical prediction using operator shallow norm by \eqnref{eq:normO5}. They match perfectly.}
\label{fig:variance_scaling}
\end{figure}

For low-rank operators, we will focus on quantum fidelity, which is important in many quantum information applications, such as (variational) state preparation. We will define the zero-depth limit ($L\rightarrow 0$) of the RUC to be a single layer of on-site Haar-random gates because even if there is no two-qubit gate in the ``zero-depth'' circuit, we still assume that the unitary ensemble is locally scrambled such that on-site scrambling unitaries continue to persist. In this limit, the single-shot variance $\var F$ of fidelity estimation scales exponentially with the number of qubits $N$. On the other hand, in the deep circuit limit ($L\rightarrow\infty$), RUCs will approach the global Haar unitary ensemble, and the variance $\var F$ will be independent of system size. We are interested to investigate how $\var F$ behaves in the shallow circuit regime. In \figref{fig:variance_scaling}(a), we calculated $\var F$ numerically using the bootstrap method for the 9-qubit GHZ state. It shows the variance $\var F$ will decrease quickly in the shallow circuit regime. Interestingly, we found that an empirical formula $\var F\propto \exp\big(c\frac{N}{(L+1)^{\alpha}}\big)$ fits the data well in the shallow circuit regime, with $\alpha=0.7\pm 0.1$. In \figref{fig:variance_scaling}(b), we plot $\var F$ as a function of $\frac{N}{(L+1)^{\alpha}}$ for different fixed circuit depth $L$. We find curves with different choices of circuit depth $L$ all collapse together with the same coefficient $c=0.47\pm 0.08$. 

The physical intuition behind the empirical formula has to do with the operator growth in RUCs. If the quantum circuit is very shallow, then the computational basis measurement will only probe local information in the original basis. If the circuit becomes deeper, computational basis measurement can probe information in larger regions in the original basis, because the measurement operator has grown under the (backward) circuit evolution. Suppose the size of the measurement operator grows in a power-law manner  $\sim(L+1)^{\alpha}$\footnote{In the $L\to 0$ limit, the measurement operator is still of at least size 1, which motivates the ``$+1$'' regularization in $(L+1)^{\alpha}$.} with respect to the depth $L$ of the RUC, the relative size of the system will effectively shrink to $N_{\text{eff}}=\frac{N}{(L+1)^{\alpha}}$, such that $\var F$ should scale universally with $N_{\text{eff}}$, as proposed in the empirical formula. We might expect $\alpha=1/2$ (or $\alpha=1$), if the operator grew diffusively (or ballisticaly). However, the best fit of our numerical result seems to indicate an effective operator growth between the diffusive and ballistic limits. Due to the limited system size in this study, we are unable to determine whether our observation persists to the thermodynamic limit. We will leave this intriguing scaling behavior for further investigation in the future. Nevertheless, for any $\alpha$, the variance decreases faster than exponential with $L$ in the shallow circuit regime, which already speaks for the advantage of applying shallow circuits in classical shadow tomography. 

For full-rank operators, we mainly focus on consecutive strings of Pauli operators of the form\eq{Z^{(k)}=Z^{\otimes k}I^{\otimes(N-k)}=\underbrace{ZZ\cdots Z}_{k}\underbrace{II\cdots I}_{N-k},} where $Z$ is the Pauli-$Z$ operator, and $I$ is the identity operator. We define the locality of the Pauli string operator by its length $k$. In the shallow circuit limit ($L\rightarrow 0$), the variance of estimation for $Z^{(k)}$ scales $\var {Z^{(k)}}\propto 4^{k}$. So shallow circuit is only efficient for predicting the local observables, and becomes inefficient for non-local observables. In the deep circuit limit ($L\rightarrow \infty$), as the unitary ensemble becomes globally Haar, there is no difference between local and non-local operators in this limit, and $\var {Z^{(k)}}\propto 2^N$. A simple comparison indicates: when $k\gtrsim N/2$, $\var Z^{(k)}$ will decrease with $L$, thus deep circuits will have lower sample complexity; when $k\lesssim N/2$, $\var Z^{(k)}$ will increase with $L$, thus shallow circuits will have lower sample complexity. In \figref{fig:variance_scaling}(c), the dots shows the variance $\var {Z^{(k)}}$ as a function of circuit depth $L$ for different support $k$. The trend agrees with our simple argument. For non-local operators, their variance will quickly decrease with the circuit depth $L$, while the variance for local operators will mildly increase with $L$. The behavior is theoretically described by how the operator shadow norm $\norm{O}_{\scE_{\sigma}}^{2}$ depends on both the circuit depth $L$ and the operator locality $k$, which are separately encoded in the entanglement features of $\scE_\sigma$ and $\scE_O$. We calculate the shadow norm $\norm{Z^{(k)}}_{\scE_{\sigma}}^{2}$ based on the entanglement feature formalism using \eqnref{eq:normO5}, and plot the result as lines in \figref{fig:variance_scaling}(c). The theoretical calculation agrees perfectly with the numerical results, which also indicates that the shadow norm bounds the single-shot variance (and hence the sample complexity) quite tightly.

\begin{figure}[htbp]
\centering
\includegraphics[width=0.9\columnwidth]{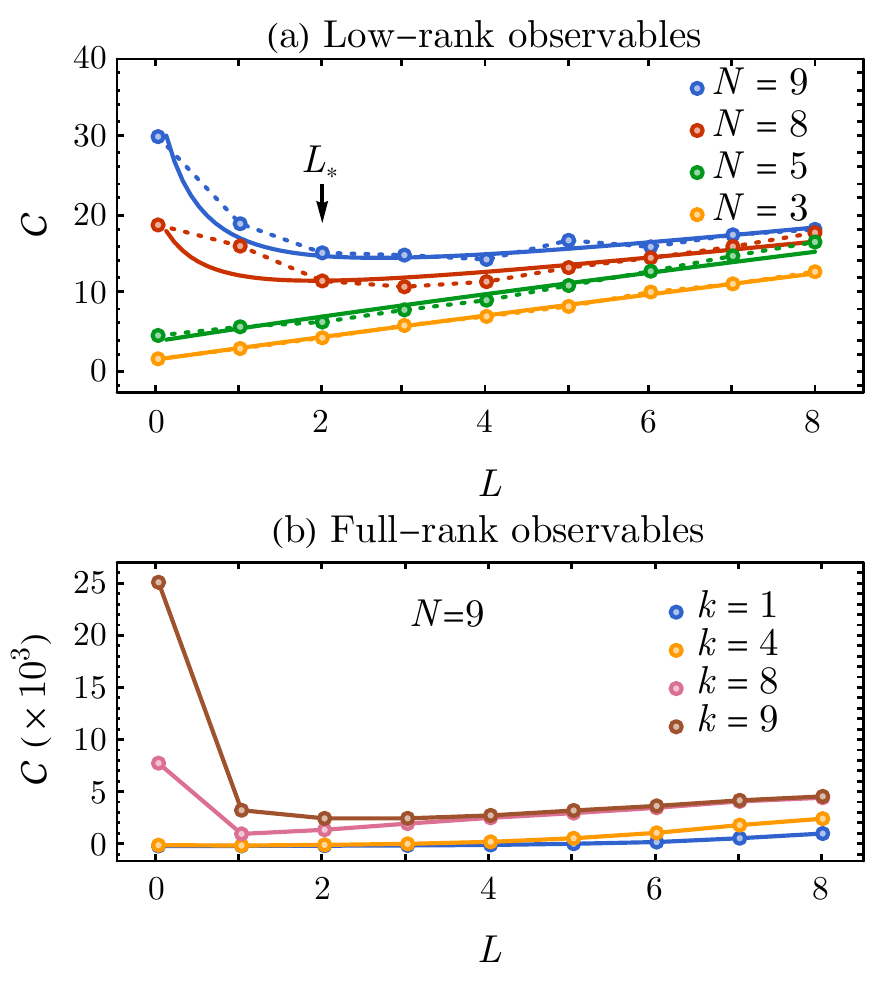}
\caption{(a) Tomography complexity $\mathcal{C}\propto (L+1) \var F$ as a function of circuit depth $L$ for the fidelity (low-rank observable) estimation task. Dots are tomography complexities for GHZ states of qubit number $N$ by our numerical simulation. Solid curves are best fits based on the empirical formula \eqnref{eq:C}. (b) Tomography complexity $\mathcal{C}\propto (L+1) \var Z^{(k)}$ for the Pauli string (full-rank observable) estimation task. Dots are numerical simulation results. Solid curves are analytic calculations using the operator shadow norm formula \eqnref{eq:normO5}.}
\label{fig:complexity}
\end{figure}

Based on the discussion in \secref{sec:variance}, the sample complexity $M$ is proportional to the single-shot variance. Given the scaling of variance with the the circuit depth $L$, we can study the scaling of the sample complexity, as well as that of the tomography complexity $\scC$. For fidelity estimation task, $\scC$ scales as
\eqs{\label{eq:C}\scC&=(L+1) M\propto (L+1) \var F\\
&\propto (L+1) \exp\Big(\frac{cN}{(L+1)^{\alpha}}\Big).}
For sufficiently large systems, the complexity $\scC$ can have a non-trivial minimum at a finite circuit depth $L_*\simeq (\alpha cN)^{1/\alpha}-1$. Our simulation result in \figref{fig:complexity}(a) verifies such behavior. For small systems ($N\lesssim 5$), random single-qubit measurements can efficiently benchmark the quantum state, so we do not need to use a finite-depth circuit for data acquisition. However, as the system size $N$ gets larger, to maintain the prediction accuracy, single-qubit measurements will require more and more samples that have to grow exponentially with $N$. As shown in \figref{fig:variance_scaling}(a), applying a few layers of quantum circuits before the measurement can quickly bring down the single-shot variance (and hence reduce the sample complexity). However, we also do not want to go too far in the circuit depth, because that would increase the circuit complexity. Therefore, we expect an optimal circuit depth $L_*$ where the sample complexity and the circuit complexity reach a balance, and the total tomography complexity is minimized. This explains the advantage of shallow circuits in classical shadow tomography, as compared to the existing method that requires either on-site Haar random ($L\to 0$) or global Haar random ($L\to\infty$) unitaries. 

We also study the tomography complexity $\scC\propto (L+1) \var Z^{(k)}$ for the full-rank observables, such as Pauli strings $Z^{(k)}$, as shown in \figref{fig:variance_scaling}(b). In this case, 
what matters is the locality $k$ of the full-rank operator (the length $k$ of the Pauli string). For local operators (small $k$), on-site measurement will be most efficient. However, for non-local operators (large $k$), we observe that the tomography complexity is minimized at some finite circuit depth, again demonstrating the advantage of employing shallow circuits in classical shadow tomography.  For different classes of physical observables, we can use the tomography complexity $\scC$ as an objective function to guide the design of the optimal circuit structure. We will leave this promising direction for future investigation.

\subsection{Classical Shadow Tomography with Fixed Quantum Circuits or Hamiltonian Dynamics}

Compared to other classical shadow tomography protocols, our method can be applied to a large family of unitary ensembles that only requires the local scrambling condition, which is more appealing to near term quantum devices. One of the biggest challenges in realizing the original proposal of global Clifford classical shadow tomography is that the realization of global Clifford unitary requires $\sim N^2$ many local Clifford gates (for a $N$-qubit system), which remains challenging for near term quantum devices. Even though global Clifford shadow tomography is very efficient in predicting non-local properties, it has not been implemented even for few-qubit systems as far as we know. 

As we have seen in \secref{sec:RUC}, the \emph{quantum entanglement} created by the unitary channel plays an important role in reducing the sample complexity. With the quantum entanglement generated by the unitary channel, the classical shadow tomography is essentially an \emph{entanglement-assisted} non-local measurement protocol. To circumvent the difficulty of sampling  (fully-scrambled) deep random unitaries but still harness the power of entanglement, we can 
use the idea of locally scrambled unitaries to design randomized measurement protocols that have sandwich structures like \figref{fig:circuits} (b), where random single-qubit Clifford gates (green boxes) are introduced at the beginning and the end of the unitary channel, and a fixed unitary circuit/quantum dynamics (the blue box) is sandwiched in between to provide entanglement generation. This sandwiched protocol satisfies the local scrambling condition rigorously, therefore the reconstruction map in \eqnref{eq:rA} can be applied. 

\begin{figure}[htbp]
\centering
\includegraphics[width=0.9\columnwidth]{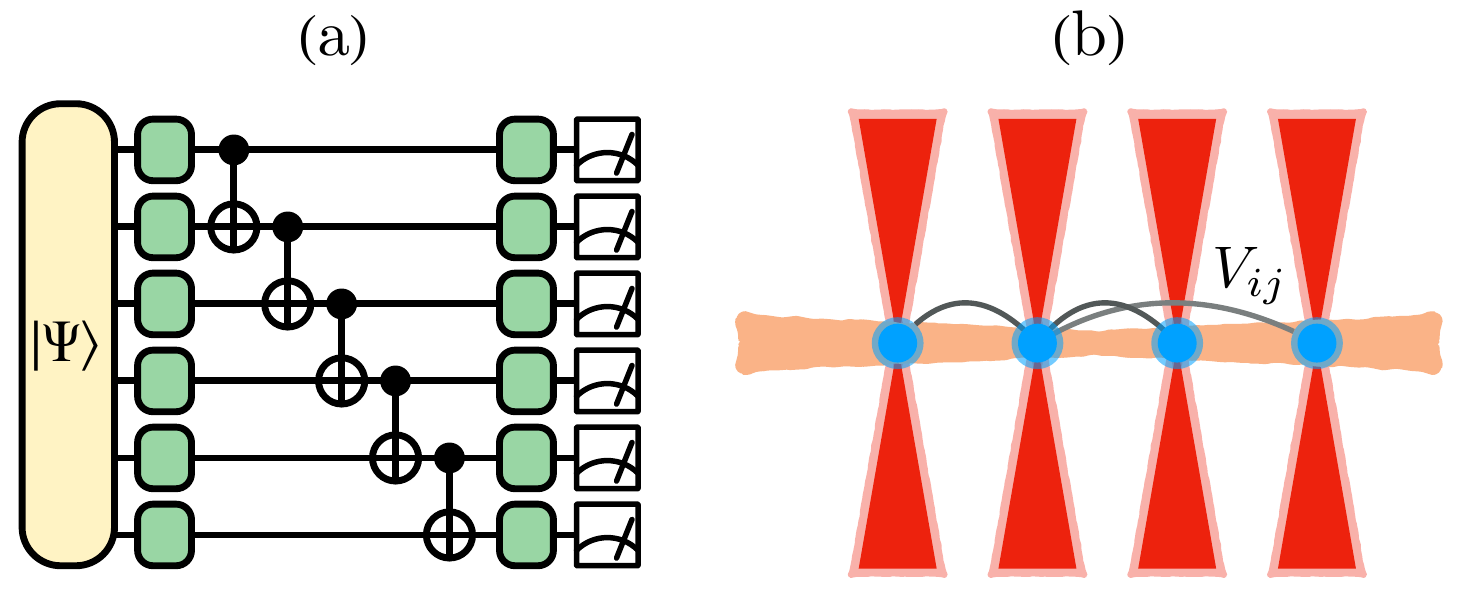}
\caption{Classical shadow tomography with a fixed unitary and onsite random Clifford gates. The fixed unitary can be generated with single quantum circuit, such as (a) CNOT gates, or fixed quantum dynamics, such as (b) Rydberg Hamiltonian dynamics.}
\label{fig:circuit}
\end{figure}

We will give two examples to demonstrate this sandwiched protocol. In the first example, as shown in \figref{fig:circuit}(a), the fixed unitary is taken to be a fixed Clifford circuit consist of a sequence of controlled-NOT (CNOT) gates that generates entanglement. In the second example, as illustrated in \figref{fig:circuit}(b), the fixed unitary is generated by the time evolution of a Rydberg atom Hamiltonian \footnote{For simulation of Rydberg Hamiltonian, we choose parameters $\Omega=2.75, \Delta=1, R_b=1$.}:
\eqs{
H = \dfrac{\Omega}{2}\sum_{i}X_i-\Delta \sum_{i}Z_i+\Omega \sum_{i<j}\left(\dfrac{R_b}{a|i-j|}\right)^6 Z_i Z_j.\label{eq:rydberg}
}
Both cases are ready to be implemented with near term quantum devices, such as trapped ion based quantum simulator or Rydberg based quantum simulator, given the fact that single qubit Clifford gates can be efficiently implemented, and randomized Pauli measurements have been demonstrated.\cite{Ma2016Pure}

\begin{figure}[htbp]
\centering
\includegraphics[width=0.8\columnwidth]{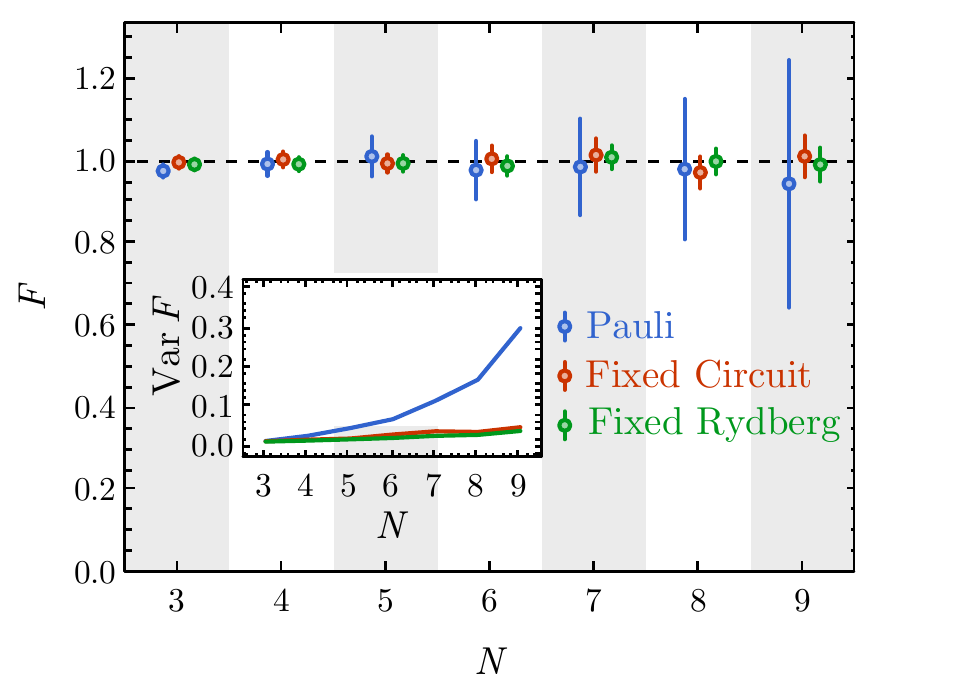}
\caption{Fidelity estimation of GHZ state using randomized Pauli measurements (denoted as Pauli), classical shadow tomography with fixed CNOT gates as in \figref{fig:circuit} (a) and classical shadow tomography with fixed Rydberg Hamiltonian dynamics as in \figref{fig:circuit} (b). The inset shows the variance $\var F$ of the predicted fidelity as a function of system size $N$. The sample size is 10000. Error bar indicates 3-standard-deviation estimated by the bootstrap method. Points are split horizontally to avoid the overlap of markers.}
\label{fig:fixed}
\end{figure}

For comparison, we use both our proposed sandwiched protocol and the standard randomized Pauli measurement to perform the classical shadow tomography on a GHZ state and to evaluate the fidelity of the reconstructed state. The results are shown in \figref{fig:fixed}. As we can see, the variance of the fidelity estimation based on randomized Pauli measurements grows exponentially with increasing system size. As expected, the variance (or the sample complexity) reduces dramatically if one adds a fixed unitary generated by the CNOT circuit or the Rydberg Hamiltonian dynamics. More specifically, the variance of prediction is reduced by one order of magnitude even for a small system of $N=9$ qubits. In both cases, the quantum entanglement generated by the locally scrambled quantum dynamics helps to improve the tomography efficiency. This result demonstrates the power of our protocol: it is both \emph{very flexible} in terms of the design and \emph{very efficient} in terms of the sample complexity.

\subsection{Approximate Classical Shadow Tomography with Local Hamiltonian Dynamics}\label{sec:Hamiltonian}

Requiring an unitary ensemble to be strictly locally scrambled could be restrictive.
To this end, we would like to explore a broader class of unitary ensembles that are only approximately locally scrambled. In particular, we study unitary evolutions $U=e^{-\ii H T}$ generated by a local Hamiltonian $H$ for finite amount of time $T$, as depicted in \figref{fig:circuits}(b). Two classes of Hamiltonians are of particular interest. In the first class, we consider a model of random local Hamiltonians
\eq{H=\sum_{i} H_{i,i+1},}
where each term $H_{i,i+1}$ is independently sampled as 2-local GUE random matrices. We dub this class the \emph{GUE2 ensemble} to remind ourselves that the Hamiltonian is 2-local. The local Hamiltonian describes a disordered one-dimensional quantum system in general. Once every $H_{i,i+1}$ term is sampled, we will use the Hamiltonian $H$ to drive the time evolution without changing $H$ during the evolution. The unitary GUE2 ensemble is only invariant under $U\to V^\dagger U V$ (not $U\to UV$) for $V\in U(d)^N$, such that that its corresponding prior snapshot ensemble $\scE_{\sigma}$ will transform as $\hat{\sigma}_{U,b}=U^\dagger \ket{b}\bra{b}U\to V^\dagger U^\dagger V\ket{b}\bra{b}V^\dagger U V\neq V^\dagger \hat{\sigma}_{U,b} V$, which does not satisfy the locally scrambling condition at the state level (i.e.~the invariance under $\hat{\sigma}\to V^\dagger \hat{\sigma} V$). However, we anticipate that under a sufficient amount of time evolution, the original local basis choice (of $\ket{b}$) will be quickly randomized given the chaotic nature of the local Hamiltonian, such that the initial choice of $V\ket{b}\bra{b}V^\dagger$ or $\ket{b}\bra{b}$ will not make a substantial difference statistically, so the GUE2 ensemble will become approximately locally scrambled after some local thermalization (scrambling) time $T_\text{Th}$.

Another more realistic class of random Hamiltonians to be considered is based on the quantum Ising model with both disordered coupling in space and random fields in time
\eq{\label{eq:DQIM}H_t= \sum_{\langle ij\rangle}J_{ij}X_iX_j+ h\sum_{i} (\cos\theta_t X_i+\sin\theta_t Y_i),}
where the local coupling $J_{ij}\sim \text{Uni}[J-\frac{J}{2},J+\frac{J}{2}]$ is drawn from a uniform distribution, and the angle of magnetic field $\theta_t\sim \text{Uni}[0,2\pi]$ is also random. We use this Hamiltonian to drive the quantum dynamics in discrete time steps. In each period of time, the magnetic field $h$ will be applied along a different random direction $\theta_t$ in the $x$-$y$ plane for all spins uniformly. However, $J_{ij}$ will remain the same throughout the time evolution. The ensemble of unitary consists of 
\eq{U=\prod_{t=1}^{T} e^{-iH_t}.} 
which we name as the \emph{Disordered Quantum Ising Model} or DQIM for short. The DQIM ensemble is friendly for quantum technology such as Rydberg-atom-based\cite{Saffman_2016} or trapped-ion-based\cite{RevModPhys.93.025001} quantum simulators. Similar construction of approximate unitary designs by Hamiltonian evolution with random quenches in time was also proposed in \refcite{Elben2018Renyi,Vermersch2018Unitary}. We would like to investigate how well our framework applies to these two cases.


Each approximately locally scrambled unitary ensembles $\scE_U$ leads to a prior snapshot ensemble $\scE_\sigma=\{\hat{\sigma}_{U,b}|b\in\{0,1\}^{\times N},U\in\scE_U\}$ that is also approximately locally scrambled. We propose to characterize how close the prior snapshot ensemble $\scE_{\sigma}$ is towards its local-basis invariant limit by the following frame potential
\eq{\scF_{\scE_{\sigma}}^{(k)}=\E_{\hat{\sigma},\hat{\sigma}'\in\scE_{\sigma}}(\Tr\hat{\sigma}\hat{\sigma}')^k.}
Recall that in deriving \eqnref{eq:sigma_V} from \eqnref{eq:sigma_general}, we only require the 2nd moment to match, i.e.~\eq{\E_{\hat{\sigma}\in\scE_{\sigma}}\hat{\sigma}^{\otimes 2}=\E_{V\in U(d)^N}\E_{\hat{\sigma}\in\scE_{\sigma}}(V^\dagger \hat{\sigma} V)^{\otimes 2},} therefore we will be most interested in the 2nd frame potential $\scF_{\scE_{\sigma}}^{(2)}$. The frame potential $\scF_{\scE_{\sigma}}^{(2)}$ for any ensemble $\scE_{\sigma}$ is lower bounded by its locally-scrambled ($\U(d)^N$-twirled) limit $\scF_{\scE_{\sigma}^\text{LS}}^{(2)}$ as
\eq{\scF_{\scE_{\sigma}}^{(2)}\geq \scF_{\scE_{\sigma}^\text{LS}}^{(2)}=\sum_{A,B}W_{\scE_{\sigma},A}^{(2)}\Wg_{A,B}W_{\scE_{\sigma},B}^{(2)}.}
The fact that $\scF_{\scE_{\sigma}^\text{LS}}^{(2)}$ is expressed purely in terms of the entanglement feature of $\scE_{\sigma}$ indicates that it is indeed free of any local-basis-dependent information. We can define the gap between the frame potential and its locally-scrambled limit as \eqs{\Delta^{(2)}_{\scE_\sigma}&=\scF_{\scE_{\sigma}}^{(2)}- \scF_{\scE_{\sigma}^\text{LS}}^{(2)}\\ &=\Tr\Big(\E_{\hat{\sigma}\in\scE_{\sigma}}\big(\hat{\sigma}^{\otimes 2}-\E_{V\in\U(d)^N}(V^\dagger \hat{\sigma} V)^{\otimes 2}\big)\Big)^2,}
which turns out to match the trace-square-difference between the 2nd moment $\E_{\hat{\sigma}}\hat{\sigma}^{\otimes 2}$ and its local twirling $\E_{V,\hat{\sigma}}(V^\dagger \hat{\sigma} V)^{\otimes 2}$. The frame potential gap $\Delta_{\scE_\sigma}^{(2)}$ serves as an indicator of the validity of our approach, as it vanishes if $\scE_\sigma$ is locally scrambled such that our construction becomes exact. 

\begin{figure}[htpb]
\centering
\includegraphics[width=0.85\columnwidth]{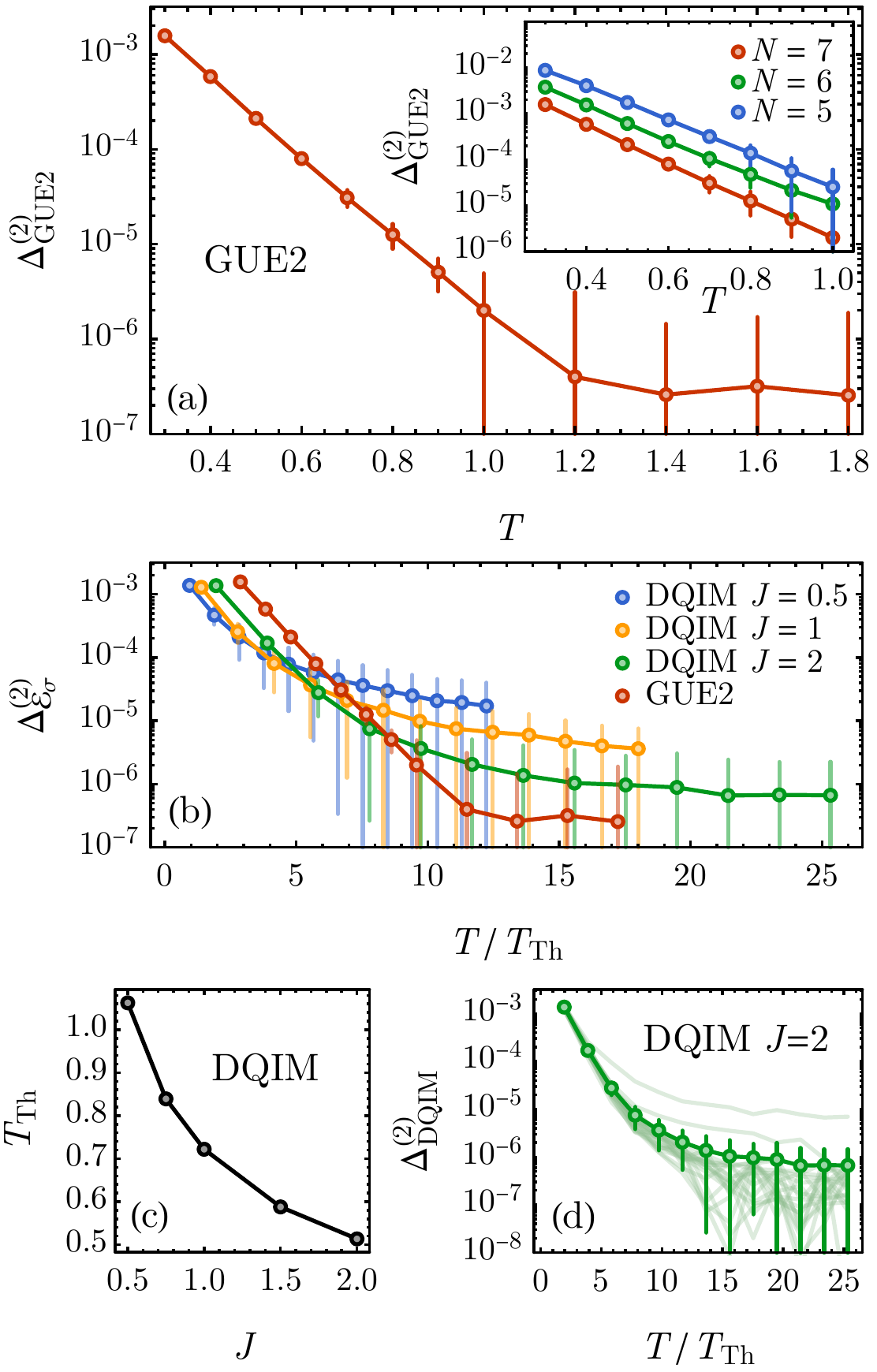}\caption{(a) Frame potential gap $\Delta^{(2)}_{\scE_\sigma}$ of the GUE2 ensemble as a function of evolution time $T$. The inset shows the decay behavior for different system sizes $N$. (b) Frame potential gap of DQIM ensemble at different coupling strength $J$, in comparison with that of the GUE2 ensemble. (c) The dependence of the local scrambling time $T_\text{Th}$ on the coupling strength $J$. (d) Frame potential gap for single instances in the DQIM ensemble. Each instance corresponds to a light-green curve in the background.}
\label{fig:frame_potential}
\end{figure}

Different unitary ensembles can lead to different frame potential gaps of $\scE_{\sigma}$, which can be used to evaluate the quality of the unitary ensemble in obeying the local scrambling condition. In \figref{fig:frame_potential}(a), we first focus on the frame potential gap $\Delta^{(2)}$ for the GUE2 ensemble. We find the gap will first decay exponentially and then saturate to a plateau at a very low level. The quickly vanishing gap implies that the GUE2 ensemble quickly becomes approximately locally scrambled as time evolves. We define the characteristic time associated with the exponential decay as $T_\text{Th}$, i.e. $\Delta^{(2)}(T)\propto \exp(-T/T_\text{Th})$, which can be considered as the local scrambling (thermalization) time. Such an exponential decaying behavior in the early time regime is generally expected for non-critical quantum dynamics, which admit typical local energy scales (or time scales). In addition, the inset plot in \figref{fig:frame_potential}(a) shows that $T_\text{Th}$ is independent of the system size $N$, as the slope remains the same for different $N$ within error bar. Unlike global scrambling (global thermalization) which requires a long time $(\sim N)$ to achieve, achieving local scrambling only requires a fixed amount of time set by the ultra-violet energy scale that is independent of the system size $N$. This is another advantage of using locally scrambled quantum dynamics for classical shadow tomography in practice. 

As for the DQIM ensemble, we fix the strength of the magnetic field at $h=\pi/4$, since this value produces the fastest on-site scrambling of a single qubit. According to the definition Eq.\eqref{eq:DQIM}, the only tuning parameter will be the mean value $J$ of Ising couplings (which also sets their disorder strength). We calculate the frame potential gap $\Delta^{(2)}$ for DQIM ensemble with different $J$. We observe that the frame potential gap always decays exponentially in the early time regime, in correspondence to the local thermalization process. Then it will typically crossover to a plateau (i.e.~saturate to a finite constant) in the late time. The early-time exponential decay region is larger for larger $J$, and we use the exponential decay regime to define the local scrambling time $T_\text{Th}$. The result is shown in \figref{fig:frame_potential}(b). The DQIM ensemble also approaches local scrambling as time evolves, although the final saturation plateau is not as low as the GUE2 ensemble. Larger Ising coupling $J$ will lower the saturation plateau and shorter the local scrambling time $T_\text{Th}$, as shown in \figref{fig:frame_potential}(c). In addition, as shown in \figref{fig:frame_potential}(d), we find the frame potential gap for a single realization quenched-disorder Hamiltonian does not deviate significantly from the ensemble mean value. This indicates that a single fixed disordered Ising chain under a randomly rotating uniform magnetic field is already good to generate an approximately locally scrambled ensemble that can be used for classical shadow tomography.

\begin{figure}[htbp]
\centering
\includegraphics[width=0.99\columnwidth]{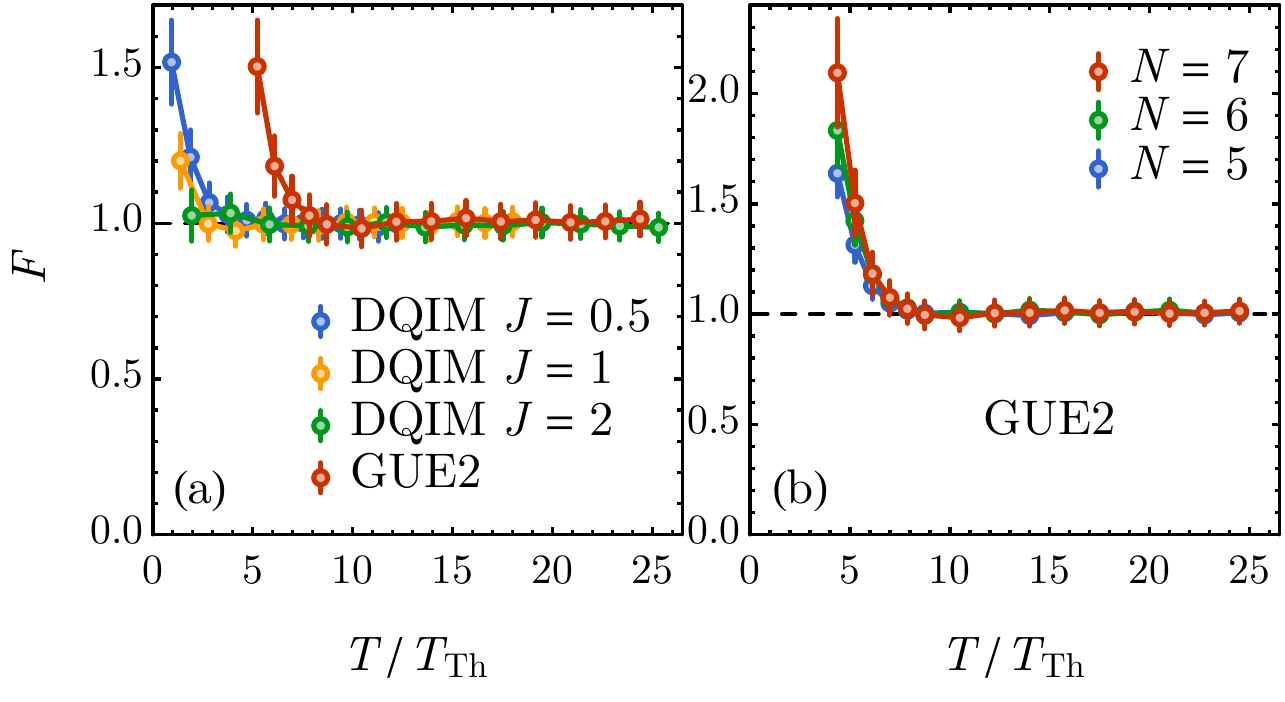}
\caption{Fidelity prediction by (a) different approximated locally scrambled ensembles, and (b) the GUE ensemble at different system sizes $N$. Sample number is 10000 and error bar indicates 3-standard deviation.}
\label{fig:practical_fidelity}
\end{figure}

In practice, we use the two proposed approximated ensembles, (i) the GUE2 ensemble and (ii) a single instance of the DQIM ensemble, to perform the tomography task and predict the fidelity of a 7-qubit GHZ state. In \figref{fig:practical_fidelity}(a), we see the predicted fidelity will be biased in the beginning (the biased fidelity can be greater than one, see \appref{app:purification} for more discussions), due to the fact that the quantum dynamics is still on its way to establish local scrambling. After around $T\sim 10 T_\text{Th}$, the local scrambling condition is approximately established, then the entanglement-feature-based reconstruction map $\scM_\text{EF}^{-1}$ can provide a good reconstruction of the quantum state, as indicated by the convergence of the quantum fidelity to identity. In \appref{app:purification}, we further investigate the quantum fidelity of $\tilde{\rho}$ projected to the physical space (to tame the unphysical $F>1$ behavior) and show that the reconstruction is nearly perfect after around $T\sim 10 T_\text{Th}$. In addition, \figref{fig:practical_fidelity}(b) also shows the local scrambling time for GUE2 is independent of system size, which is consistent with the same behavior in \figref{fig:frame_potential}(a). The results in \figref{fig:practical_fidelity} suggest that the entanglement-feature-based approach could be applicable for approximately locally scrambled unitary ensembles. The reconstruction bias vanishes as the frame potential gap decays. As long as the frame potential gap is low enough, the bias is also expected to be vanishingly small for all predictions. This significantly broadens the application of classical shadow tomography to a large class of quantum dynamics that can be achieved on NISQ devices.

\section{Summary and Discussions}\label{sec:summary}

Our result can be further extended to more general measurement channels, which can involve ancilla qubits and partial measurements. The unitary channel can be noisy and the measurements can be weak. Under generalized measurements, the state $\rho$ collapses to $\rho\to K_a\rho K_a^\dagger/(\Tr K_a\rho K_a^\dagger)$, where $K_a$ is the Kraus operators\cite{Kraus1971General} associated with the measurement outcome $a$. We can define the measurement operator $\hat{\sigma}_a=K_a^\dagger K_a$ (with the standard normalization $\sum_a\hat{\sigma}_a=\id$), which forms the prior snapshot ensemble $\scE_\sigma=\{\hat{\sigma}_a|P(\hat{\sigma}_a)=d^{-N}\}$, and the posterior snapshot ensemble will be $\scE_{\sigma|\rho}=\{\hat{\sigma}_a|P(\hat{\sigma}_a|\rho)=\Tr \hat{\sigma}_a\rho\}$ correspondingly. As long as the generalized prior snapshot ensemble $\scE_{\sigma}$ is locally scrambled, i.e.~$\forall V\in\U(d)^N:P(\hat{\sigma})=P(V^\dagger \hat{\sigma} V)$, our theoretical framework automatically applies, and all formulations in this work remain valid in the same form. This enables us to consider classical shadow tomography with very general data acquisition protocols.

\begin{figure}[htbp]
\begin{center}
\includegraphics[width=0.65\columnwidth]{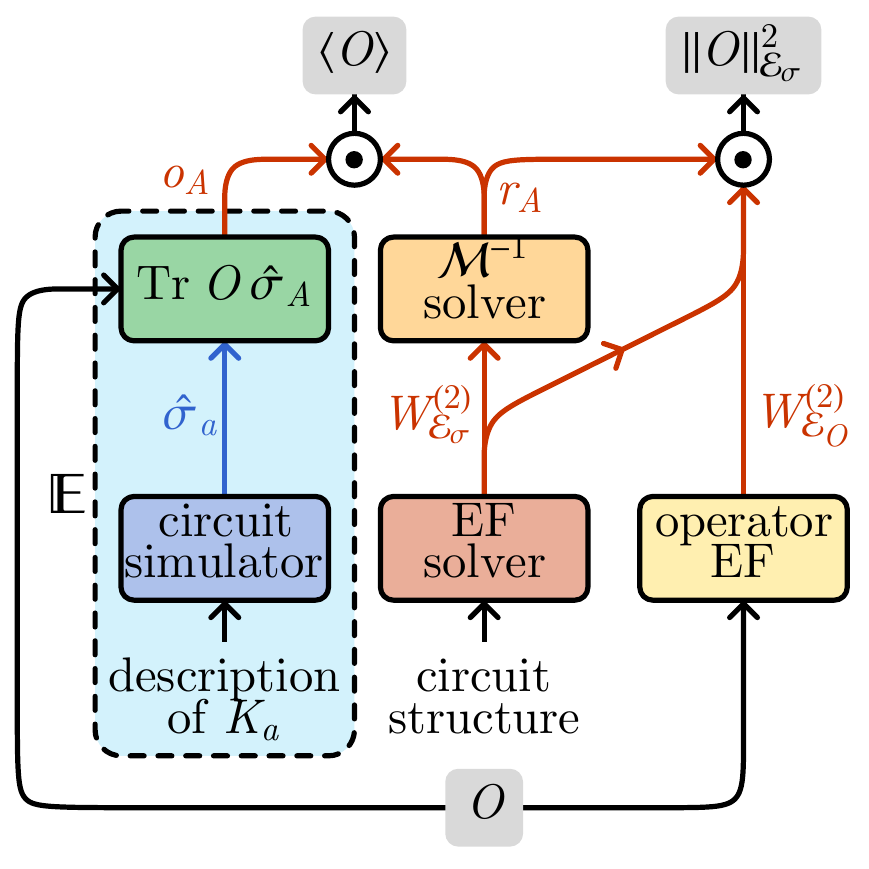}
\caption{Classical post-processing protocol to estimate the operator expectation value and shadow norm.}
\label{fig:flow}
\end{center}
\end{figure}

The entanglement feature formalism plays a central role in our approach. \figref{fig:flow} summarizes the proposed classical post-processing protocol to predict the expectation value $\langle O\rangle$ of a physical observable $O$, together with its estimated variance (given by the shadow norm $\norm{O}_{\scE_\sigma}^2$ divided by the sample size $M$). Given the circuit structure, the entanglement feature (EF) solver calculates the entanglement feature $W_{\scE_\sigma}^{(2)}$ of the prior snapshot ensemble as defined in \eqnref{eq:WC0} (the algorithm is developed in previous works\cite{Kuo2020Markovian,Fan2020Self-Organized,Akhtar2020Multiregion}). The result is passed to the inverse channel solver to calculate the reconstruction coefficients $r_A$ by solving \eqnref{eq:rA}. With $r_A$, we can predict any physical observable $O$ by $\langle O\rangle=d^{N}\sum_{A}r_A o_A$ where $o_A=\E_{\hat{\sigma} \in\scE_{\sigma|\rho}}\Tr O\hat{\sigma}_A$ (the median-of-means trick\cite{Huang2020Predicting} can be used here if multiple observables are to be predicted). For every sample of classical description of the Kraus operator $K$, a quantum circuit simulator (running on a classical computer) is needed to construct the (efficient representation of) measurement operator $\hat{\sigma}=K^\dagger K$. The classical simulation could be efficient if the circuit is Clifford\cite{Gottesman1998The-Heisenberg} (our formalism applies to random Clifford circuits with no problem). The part of computation in the dashed box of \figref{fig:flow} should be repeated for every sample to evaluate the ensemble average. Finally, given the reconstruction coefficient $r$ and the entanglement features $W_{\scE_\sigma}^{(2)}$ and $W_{\scE_O}^{(2)}$, the shadow norm $\norm{O}_{\scE_\sigma}$ can be calculated, which provides an estimation for variance of the predicted observable. Although it takes some effort to process the entanglement feature data and to calculate the reconstruction coefficients, such computation (everything outside the dashed box in \figref{fig:flow}) only occurs once for a given circuit structure, therefore this computational effort is usually affordable (especially when efficient tensor-network approaches are developed and employed)\cite{2022arXiv220902093A}. 

\begin{figure}[htbp]
\begin{center}
\includegraphics[width=0.6\columnwidth]{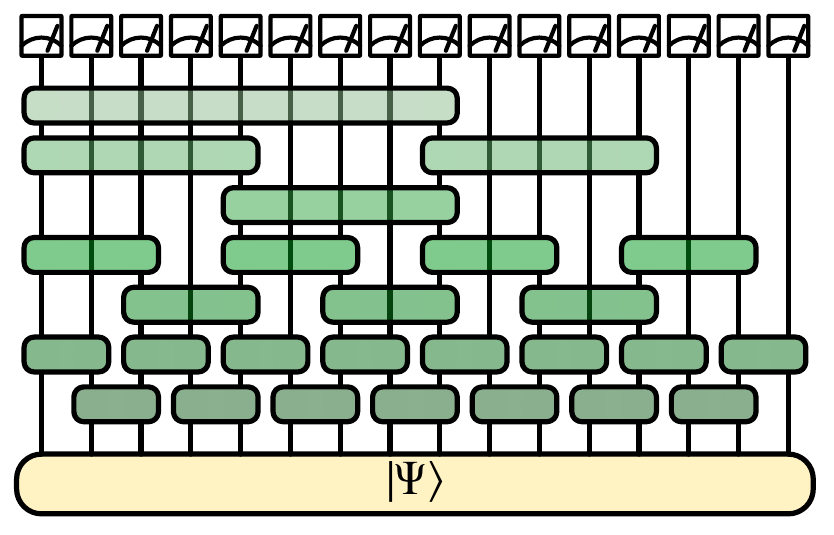}
\caption{Illustration of holographic classical shadow tomography scheme, where the quantum circuit is arranged in a hierarchical structure (forming the hyperbolic bulk space).}
\label{fig:holographic}
\end{center}
\end{figure}

The theoretical framework established in this work extends the classical shadow tomography to general quantum circuits, which opens up many possible applications. As one interesting example, we consider performing the classical shadow tomography in the ``holographic bulk'' by transforming the original state by a random Clifford circuit arranged in a hierarchical structure (see \figref{fig:holographic}), similar to the multi-scale entanglement renormalization ansatz (MERA) network\cite{Vidal2007Entanglement,Vidal2008Class} or the holographic quantum error-correcting code\cite{Pastawski2015Holographic}. Following the idea of holographic duality, local measurements in the holographic bulk translate to measurements at all different scales on the holographic boundary. Therefore it is conceivable that the holographic classical shadow tomography could achieve high sample efficiency for operators of all scales, potentially evading the dichotomy between sample complexity and circuit complexity. 

Another interesting application is to consider random circuits hybrid with random measurements inserted into the circuit at a fixed rate\cite{Li2018QZEMET,Li2019METHQC,Szyniszewski2019ETFVWM,Chan2019UED,Skinner2019MPTDE}. Conditioned on the intermediate measurement outcomes, the hybrid quantum circuit forms a quantum channel that transmits quantum information from end to end. Driven by the measurement rate, the final state can undergo an entanglement transition\cite{Vasseur2018Entanglement,Jian2019MCRQC,Bao2019TPTRUCWM} (or the quantum channel can undergo a purification transition\cite{Gullans2019Dynamical} equivalently). When the measurement rate is high, the quantum information in the initial state can be efficiently extracted by intermediate measurements (eavesdroppers), such that the channel has zero transmission capacity. When the measurement rate is lower than a critical threshold, the channel will have a finite capacity and can transmit quantum information in an error-correcting manner.\cite{Choi2019QECEPTRUCWPM,Fan2020Self-Organized,Gullans2020Quantum} However, it is unclear how to take advantage of the self-organized quantum error correction in these hybrid quantum circuits. We anticipate that the classical shadow tomography with a flexible measurement scheme can help decoding the measurement-induced quantum error-correcting code. We will leave these interesting applications for future explorations.

Finally, the classical shadow tomography provides an efficient interface that converts quantum states to classical shadow data, which enables us to exploit the power of classical computation, especially data-driven and machine learning approaches, to advance our understanding of complex quantum systems and to solve challenging quantum many-body problems. As shown in \refcite{Huang2021Provably}, classical algorithms that learns from the classical shadow data has provable performance advantages over conventional numerical approaches that do not learn of data. Our work further adds to this promising direction by providing a more flexible classical shadow tomography scheme that works with very general measurement protocols (beyond on-site Pauli measurements), which could lead to potentially more efficient classical-shadow-based learning algorithms.

\section{Acknowledgment}

We thank the insightful discussion with Hsin-Yuan Huang, Xun Gao,  Daniel Mark, Andreas Elben, and Jens Eisert. HYH and YZY are supported by a UC Hellman
fellowship. SC acknowledges support from the Miller Institute for Basic Research in Science.

\bibliography{QST}

\begin{thebibliography}{72}%
\makeatletter
\providecommand \@ifxundefined [1]{%
 \@ifx{#1\undefined}
}%
\providecommand \@ifnum [1]{%
 \ifnum #1\expandafter \@firstoftwo
 \else \expandafter \@secondoftwo
 \fi
}%
\providecommand \@ifx [1]{%
 \ifx #1\expandafter \@firstoftwo
 \else \expandafter \@secondoftwo
 \fi
}%
\providecommand \natexlab [1]{#1}%
\providecommand \enquote  [1]{``#1''}%
\providecommand \bibnamefont  [1]{#1}%
\providecommand \bibfnamefont [1]{#1}%
\providecommand \citenamefont [1]{#1}%
\providecommand \href@noop [0]{\@secondoftwo}%
\providecommand \href [0]{\begingroup \@sanitize@url \@href}%
\providecommand \@href[1]{\@@startlink{#1}\@@href}%
\providecommand \@@href[1]{\endgroup#1\@@endlink}%
\providecommand \@sanitize@url [0]{\catcode `\\12\catcode `\$12\catcode
  `\&12\catcode `\#12\catcode `\^12\catcode `\_12\catcode `\%12\relax}%
\providecommand \@@startlink[1]{}%
\providecommand \@@endlink[0]{}%
\providecommand \url  [0]{\begingroup\@sanitize@url \@url }%
\providecommand \@url [1]{\endgroup\@href {#1}{\urlprefix }}%
\providecommand \urlprefix  [0]{URL }%
\providecommand \Eprint [0]{\href }%
\providecommand \doibase [0]{http://dx.doi.org/}%
\providecommand \selectlanguage [0]{\@gobble}%
\providecommand \bibinfo  [0]{\@secondoftwo}%
\providecommand \bibfield  [0]{\@secondoftwo}%
\providecommand \translation [1]{[#1]}%
\providecommand \BibitemOpen [0]{}%
\providecommand \bibitemStop [0]{}%
\providecommand \bibitemNoStop [0]{.\EOS\space}%
\providecommand \EOS [0]{\spacefactor3000\relax}%
\providecommand \BibitemShut  [1]{\csname bibitem#1\endcsname}%
\let\auto@bib@innerbib\@empty
\bibitem [{\citenamefont {Vogel}\ and\ \citenamefont
  {Risken}(1989)}]{Vogel1989Determination}%
  \BibitemOpen
  \bibfield  {author} {\bibinfo {author} {\bibfnamefont {K.}~\bibnamefont
  {Vogel}}\ and\ \bibinfo {author} {\bibfnamefont {H.}~\bibnamefont {Risken}},\
  }\bibfield  {title} {\enquote {\bibinfo {title} {Determination of
  quasiprobability distributions in terms of probability distributions for the
  rotated quadrature phase},}\ }\href {\doibase 10.1103/PhysRevA.40.2847}
  {\bibfield  {journal} {\bibinfo  {journal} {Phys. Rev. A}\ }\textbf {\bibinfo
  {volume} {40}},\ \bibinfo {pages} {2847--2849} (\bibinfo {year}
  {1989})}\BibitemShut {NoStop}%
\bibitem [{\citenamefont {James}\ \emph {et~al.}(2001)\citenamefont {James},
  \citenamefont {Kwiat}, \citenamefont {Munro},\ and\ \citenamefont
  {White}}]{James2001Measurement}%
  \BibitemOpen
  \bibfield  {author} {\bibinfo {author} {\bibfnamefont {Daniel F.~V.}\
  \bibnamefont {James}}, \bibinfo {author} {\bibfnamefont {Paul~G.}\
  \bibnamefont {Kwiat}}, \bibinfo {author} {\bibfnamefont {William~J.}\
  \bibnamefont {Munro}}, \ and\ \bibinfo {author} {\bibfnamefont {Andrew~G.}\
  \bibnamefont {White}},\ }\bibfield  {title} {\enquote {\bibinfo {title}
  {Measurement of qubits},}\ }\href {\doibase 10.1103/PhysRevA.64.052312}
  {\bibfield  {journal} {\bibinfo  {journal} {Phys. Rev. A}\ }\textbf {\bibinfo
  {volume} {64}},\ \bibinfo {pages} {052312} (\bibinfo {year}
  {2001})}\BibitemShut {NoStop}%
\bibitem [{\citenamefont {{Caves}}\ \emph {et~al.}(2002)\citenamefont
  {{Caves}}, \citenamefont {{Fuchs}},\ and\ \citenamefont
  {{Schack}}}]{Caves2002Unknown}%
  \BibitemOpen
  \bibfield  {author} {\bibinfo {author} {\bibfnamefont {Carlton~M.}\
  \bibnamefont {{Caves}}}, \bibinfo {author} {\bibfnamefont {Christopher~A.}\
  \bibnamefont {{Fuchs}}}, \ and\ \bibinfo {author} {\bibfnamefont
  {R{\"u}diger}\ \bibnamefont {{Schack}}},\ }\bibfield  {title} {\enquote
  {\bibinfo {title} {{Unknown quantum states: The quantum de Finetti
  representation}},}\ }\href {\doibase 10.1063/1.1494475} {\bibfield  {journal}
  {\bibinfo  {journal} {Journal of Mathematical Physics}\ }\textbf {\bibinfo
  {volume} {43}},\ \bibinfo {pages} {4537--4559} (\bibinfo {year} {2002})},\
  \Eprint {http://arxiv.org/abs/quant-ph/0104088} {arXiv:quant-ph/0104088
  [quant-ph]} \BibitemShut {NoStop}%
\bibitem [{\citenamefont {O'Donnell}\ and\ \citenamefont
  {Wright}(2016)}]{ODonnell2015Efficient}%
  \BibitemOpen
  \bibfield  {author} {\bibinfo {author} {\bibfnamefont {Ryan}\ \bibnamefont
  {O'Donnell}}\ and\ \bibinfo {author} {\bibfnamefont {John}\ \bibnamefont
  {Wright}},\ }\bibfield  {title} {\enquote {\bibinfo {title} {Efficient
  quantum tomography},}\ }in\ \href {\doibase 10.1145/2897518.2897544} {\emph
  {\bibinfo {booktitle} {Proceedings of the Forty-Eighth Annual ACM Symposium
  on Theory of Computing}}},\ \bibinfo {series and number} {STOC '16}\
  (\bibinfo  {publisher} {Association for Computing Machinery},\ \bibinfo
  {address} {New York, NY, USA},\ \bibinfo {year} {2016})\ pp.\ \bibinfo
  {pages} {899--912}\BibitemShut {NoStop}%
\bibitem [{\citenamefont {Haah}\ \emph {et~al.}(2017)\citenamefont {Haah},
  \citenamefont {Harrow}, \citenamefont {Ji}, \citenamefont {Wu},\ and\
  \citenamefont {Yu}}]{Haah2015Sample-optimal}%
  \BibitemOpen
  \bibfield  {author} {\bibinfo {author} {\bibfnamefont {Jeongwan}\
  \bibnamefont {Haah}}, \bibinfo {author} {\bibfnamefont {Aram~W.}\
  \bibnamefont {Harrow}}, \bibinfo {author} {\bibfnamefont {Zhengfeng}\
  \bibnamefont {Ji}}, \bibinfo {author} {\bibfnamefont {Xiaodi}\ \bibnamefont
  {Wu}}, \ and\ \bibinfo {author} {\bibfnamefont {Nengkun}\ \bibnamefont
  {Yu}},\ }\bibfield  {title} {\enquote {\bibinfo {title} {Sample-optimal
  tomography of quantum states},}\ }\href {\doibase 10.1109/TIT.2017.2719044}
  {\bibfield  {journal} {\bibinfo  {journal} {IEEE Transactions on Information
  Theory}\ }\textbf {\bibinfo {volume} {63}},\ \bibinfo {pages} {5628--5641}
  (\bibinfo {year} {2017})}\BibitemShut {NoStop}%
\bibitem [{\citenamefont {Aaronson}(2018)}]{Aaronson2017Shadow}%
  \BibitemOpen
  \bibfield  {author} {\bibinfo {author} {\bibfnamefont {Scott}\ \bibnamefont
  {Aaronson}},\ }\bibfield  {title} {\enquote {\bibinfo {title} {Shadow
  tomography of quantum states},}\ }in\ \href {\doibase
  10.1145/3188745.3188802} {\emph {\bibinfo {booktitle} {Proceedings of the
  50th Annual ACM SIGACT Symposium on Theory of Computing}}},\ \bibinfo {series
  and number} {STOC 2018}\ (\bibinfo  {publisher} {Association for Computing
  Machinery},\ \bibinfo {address} {New York, NY, USA},\ \bibinfo {year}
  {2018})\ pp.\ \bibinfo {pages} {325--338}\BibitemShut {NoStop}%
\bibitem [{\citenamefont {Aaronson}\ and\ \citenamefont
  {Rothblum}(2019)}]{Aaronson2019Gentle}%
  \BibitemOpen
  \bibfield  {author} {\bibinfo {author} {\bibfnamefont {Scott}\ \bibnamefont
  {Aaronson}}\ and\ \bibinfo {author} {\bibfnamefont {Guy~N.}\ \bibnamefont
  {Rothblum}},\ }\bibfield  {title} {\enquote {\bibinfo {title} {Gentle
  measurement of quantum states and differential privacy},}\ }in\ \href
  {\doibase 10.1145/3313276.3316378} {\emph {\bibinfo {booktitle} {Proceedings
  of the 51st Annual ACM SIGACT Symposium on Theory of Computing}}},\ \bibinfo
  {series and number} {STOC 2019}\ (\bibinfo  {publisher} {Association for
  Computing Machinery},\ \bibinfo {address} {New York, NY, USA},\ \bibinfo
  {year} {2019})\ pp.\ \bibinfo {pages} {322--333}\BibitemShut {NoStop}%
\bibitem [{\citenamefont {{Huang}}\ \emph {et~al.}(2020)\citenamefont
  {{Huang}}, \citenamefont {{Kueng}},\ and\ \citenamefont
  {{Preskill}}}]{Huang2020Predicting}%
  \BibitemOpen
  \bibfield  {author} {\bibinfo {author} {\bibfnamefont {Hsin-Yuan}\
  \bibnamefont {{Huang}}}, \bibinfo {author} {\bibfnamefont {Richard}\
  \bibnamefont {{Kueng}}}, \ and\ \bibinfo {author} {\bibfnamefont {John}\
  \bibnamefont {{Preskill}}},\ }\bibfield  {title} {\enquote {\bibinfo {title}
  {{Predicting many properties of a quantum system from very few
  measurements}},}\ }\href {\doibase 10.1038/s41567-020-0932-7} {\bibfield
  {journal} {\bibinfo  {journal} {Nature Physics}\ }\textbf {\bibinfo {volume}
  {16}},\ \bibinfo {pages} {1050--1057} (\bibinfo {year} {2020})},\ \Eprint
  {http://arxiv.org/abs/2002.08953} {arXiv:2002.08953 [quant-ph]} \BibitemShut
  {NoStop}%
\bibitem [{\citenamefont {{Ohliger}}\ \emph {et~al.}(2013)\citenamefont
  {{Ohliger}}, \citenamefont {{Nesme}},\ and\ \citenamefont
  {{Eisert}}}]{Ohliger2013Efficient}%
  \BibitemOpen
  \bibfield  {author} {\bibinfo {author} {\bibfnamefont {M.}~\bibnamefont
  {{Ohliger}}}, \bibinfo {author} {\bibfnamefont {V.}~\bibnamefont {{Nesme}}},
  \ and\ \bibinfo {author} {\bibfnamefont {J.}~\bibnamefont {{Eisert}}},\
  }\bibfield  {title} {\enquote {\bibinfo {title} {{Efficient and feasible
  state tomography of quantum many-body systems}},}\ }\href {\doibase
  10.1088/1367-2630/15/1/015024} {\bibfield  {journal} {\bibinfo  {journal}
  {New Journal of Physics}\ }\textbf {\bibinfo {volume} {15}},\ \bibinfo {eid}
  {015024} (\bibinfo {year} {2013})},\ \Eprint {http://arxiv.org/abs/1204.5735}
  {arXiv:1204.5735 [quant-ph]} \BibitemShut {NoStop}%
\bibitem [{\citenamefont {{Hu}}\ and\ \citenamefont
  {{You}}(2021)}]{Hu2021Hamiltonian-Driven}%
  \BibitemOpen
  \bibfield  {author} {\bibinfo {author} {\bibfnamefont {Hong-Ye}\ \bibnamefont
  {{Hu}}}\ and\ \bibinfo {author} {\bibfnamefont {Yi-Zhuang}\ \bibnamefont
  {{You}}},\ }\bibfield  {title} {\enquote {\bibinfo {title}
  {{Hamiltonian-Driven Shadow Tomography of Quantum States}},}\ }\href@noop {}
  {\bibfield  {journal} {\bibinfo  {journal} {arXiv e-prints}\ ,\ \bibinfo
  {eid} {arXiv:2102.10132}} (\bibinfo {year} {2021})},\ \Eprint
  {http://arxiv.org/abs/2102.10132} {arXiv:2102.10132 [quant-ph]} \BibitemShut
  {NoStop}%
\bibitem [{\citenamefont {{Ebadi}}\ \emph {et~al.}(2020)\citenamefont
  {{Ebadi}}, \citenamefont {{Wang}}, \citenamefont {{Levine}}, \citenamefont
  {{Keesling}}, \citenamefont {{Semeghini}}, \citenamefont {{Omran}},
  \citenamefont {{Bluvstein}}, \citenamefont {{Samajdar}}, \citenamefont
  {{Pichler}}, \citenamefont {{Ho}}, \citenamefont {{Choi}}, \citenamefont
  {{Sachdev}}, \citenamefont {{Greiner}}, \citenamefont {{Vuletic}},\ and\
  \citenamefont {{Lukin}}}]{2020arXiv201212281E}%
  \BibitemOpen
  \bibfield  {author} {\bibinfo {author} {\bibfnamefont {Sepehr}\ \bibnamefont
  {{Ebadi}}}, \bibinfo {author} {\bibfnamefont {Tout~T.}\ \bibnamefont
  {{Wang}}}, \bibinfo {author} {\bibfnamefont {Harry}\ \bibnamefont
  {{Levine}}}, \bibinfo {author} {\bibfnamefont {Alexander}\ \bibnamefont
  {{Keesling}}}, \bibinfo {author} {\bibfnamefont {Giulia}\ \bibnamefont
  {{Semeghini}}}, \bibinfo {author} {\bibfnamefont {Ahmed}\ \bibnamefont
  {{Omran}}}, \bibinfo {author} {\bibfnamefont {Dolev}\ \bibnamefont
  {{Bluvstein}}}, \bibinfo {author} {\bibfnamefont {Rhine}\ \bibnamefont
  {{Samajdar}}}, \bibinfo {author} {\bibfnamefont {Hannes}\ \bibnamefont
  {{Pichler}}}, \bibinfo {author} {\bibfnamefont {Wen~Wei}\ \bibnamefont
  {{Ho}}}, \bibinfo {author} {\bibfnamefont {Soonwon}\ \bibnamefont {{Choi}}},
  \bibinfo {author} {\bibfnamefont {Subir}\ \bibnamefont {{Sachdev}}}, \bibinfo
  {author} {\bibfnamefont {Markus}\ \bibnamefont {{Greiner}}}, \bibinfo
  {author} {\bibfnamefont {Vladan}\ \bibnamefont {{Vuletic}}}, \ and\ \bibinfo
  {author} {\bibfnamefont {Mikhail~D.}\ \bibnamefont {{Lukin}}},\ }\bibfield
  {title} {\enquote {\bibinfo {title} {{Quantum Phases of Matter on a 256-Atom
  Programmable Quantum Simulator}},}\ }\href@noop {} {\bibfield  {journal}
  {\bibinfo  {journal} {arXiv e-prints}\ ,\ \bibinfo {eid} {arXiv:2012.12281}}
  (\bibinfo {year} {2020})},\ \Eprint {http://arxiv.org/abs/2012.12281}
  {arXiv:2012.12281 [quant-ph]} \BibitemShut {NoStop}%
\bibitem [{\citenamefont {{Scholl}}\ \emph {et~al.}(2020)\citenamefont
  {{Scholl}}, \citenamefont {{Schuler}}, \citenamefont {{Williams}},
  \citenamefont {{Eberharter}}, \citenamefont {{Barredo}}, \citenamefont
  {{Schymik}}, \citenamefont {{Lienhard}}, \citenamefont {{Henry}},
  \citenamefont {{Lang}}, \citenamefont {{Lahaye}}, \citenamefont
  {{L{\"a}uchli}},\ and\ \citenamefont {{Browaeys}}}]{2020arXiv201212268S}%
  \BibitemOpen
  \bibfield  {author} {\bibinfo {author} {\bibfnamefont {Pascal}\ \bibnamefont
  {{Scholl}}}, \bibinfo {author} {\bibfnamefont {Michael}\ \bibnamefont
  {{Schuler}}}, \bibinfo {author} {\bibfnamefont {Hannah~J.}\ \bibnamefont
  {{Williams}}}, \bibinfo {author} {\bibfnamefont {Alexander~A.}\ \bibnamefont
  {{Eberharter}}}, \bibinfo {author} {\bibfnamefont {Daniel}\ \bibnamefont
  {{Barredo}}}, \bibinfo {author} {\bibfnamefont {Kai-Niklas}\ \bibnamefont
  {{Schymik}}}, \bibinfo {author} {\bibfnamefont {Vincent}\ \bibnamefont
  {{Lienhard}}}, \bibinfo {author} {\bibfnamefont {Louis-Paul}\ \bibnamefont
  {{Henry}}}, \bibinfo {author} {\bibfnamefont {Thomas~C.}\ \bibnamefont
  {{Lang}}}, \bibinfo {author} {\bibfnamefont {Thierry}\ \bibnamefont
  {{Lahaye}}}, \bibinfo {author} {\bibfnamefont {Andreas~M.}\ \bibnamefont
  {{L{\"a}uchli}}}, \ and\ \bibinfo {author} {\bibfnamefont {Antoine}\
  \bibnamefont {{Browaeys}}},\ }\bibfield  {title} {\enquote {\bibinfo {title}
  {{Programmable quantum simulation of 2D antiferromagnets with hundreds of
  Rydberg atoms}},}\ }\href@noop {} {\bibfield  {journal} {\bibinfo  {journal}
  {arXiv e-prints}\ ,\ \bibinfo {eid} {arXiv:2012.12268}} (\bibinfo {year}
  {2020})},\ \Eprint {http://arxiv.org/abs/2012.12268} {arXiv:2012.12268
  [quant-ph]} \BibitemShut {NoStop}%
\bibitem [{\citenamefont {Zhang}\ \emph {et~al.}(2017)\citenamefont {Zhang},
  \citenamefont {Pagano}, \citenamefont {Hess}, \citenamefont {Kyprianidis},
  \citenamefont {Becker}, \citenamefont {Kaplan}, \citenamefont {Gorshkov},
  \citenamefont {Gong},\ and\ \citenamefont {Monroe}}]{NatureTrapIon}%
  \BibitemOpen
  \bibfield  {author} {\bibinfo {author} {\bibfnamefont {J.}~\bibnamefont
  {Zhang}}, \bibinfo {author} {\bibfnamefont {G.}~\bibnamefont {Pagano}},
  \bibinfo {author} {\bibfnamefont {P.~W.}\ \bibnamefont {Hess}}, \bibinfo
  {author} {\bibfnamefont {A.}~\bibnamefont {Kyprianidis}}, \bibinfo {author}
  {\bibfnamefont {P.}~\bibnamefont {Becker}}, \bibinfo {author} {\bibfnamefont
  {H.}~\bibnamefont {Kaplan}}, \bibinfo {author} {\bibfnamefont {A.~V.}\
  \bibnamefont {Gorshkov}}, \bibinfo {author} {\bibfnamefont {Z.~X.}\
  \bibnamefont {Gong}}, \ and\ \bibinfo {author} {\bibfnamefont
  {C.}~\bibnamefont {Monroe}},\ }\bibfield  {title} {\enquote {\bibinfo {title}
  {Observation of a many-body dynamical phase transition with a 53-qubit
  quantum simulator},}\ }\href {\doibase 10.1038/nature24654} {\bibfield
  {journal} {\bibinfo  {journal} {Nature}\ }\textbf {\bibinfo {volume} {551}},\
  \bibinfo {pages} {601--604} (\bibinfo {year} {2017})}\BibitemShut {NoStop}%
\bibitem [{\citenamefont {{Kuo}}\ \emph {et~al.}(2020)\citenamefont {{Kuo}},
  \citenamefont {{Akhtar}}, \citenamefont {{Arovas}},\ and\ \citenamefont
  {{You}}}]{Kuo2020Markovian}%
  \BibitemOpen
  \bibfield  {author} {\bibinfo {author} {\bibfnamefont {Wei-Ting}\
  \bibnamefont {{Kuo}}}, \bibinfo {author} {\bibfnamefont {A.~A.}\ \bibnamefont
  {{Akhtar}}}, \bibinfo {author} {\bibfnamefont {Daniel~P.}\ \bibnamefont
  {{Arovas}}}, \ and\ \bibinfo {author} {\bibfnamefont {Yi-Zhuang}\
  \bibnamefont {{You}}},\ }\bibfield  {title} {\enquote {\bibinfo {title}
  {{Markovian entanglement dynamics under locally scrambled quantum
  evolution}},}\ }\href {\doibase 10.1103/PhysRevB.101.224202} {\bibfield
  {journal} {\bibinfo  {journal} {\prb}\ }\textbf {\bibinfo {volume} {101}},\
  \bibinfo {eid} {224202} (\bibinfo {year} {2020})},\ \Eprint
  {http://arxiv.org/abs/1910.11351} {arXiv:1910.11351 [cond-mat.dis-nn]}
  \BibitemShut {NoStop}%
\bibitem [{\citenamefont {Nahum}\ \emph {et~al.}(2017)\citenamefont {Nahum},
  \citenamefont {Ruhman}, \citenamefont {Vijay},\ and\ \citenamefont
  {Haah}}]{Nahum2017Quantum}%
  \BibitemOpen
  \bibfield  {author} {\bibinfo {author} {\bibfnamefont {Adam}\ \bibnamefont
  {Nahum}}, \bibinfo {author} {\bibfnamefont {Jonathan}\ \bibnamefont
  {Ruhman}}, \bibinfo {author} {\bibfnamefont {Sagar}\ \bibnamefont {Vijay}}, \
  and\ \bibinfo {author} {\bibfnamefont {Jeongwan}\ \bibnamefont {Haah}},\
  }\bibfield  {title} {\enquote {\bibinfo {title} {Quantum entanglement growth
  under random unitary dynamics},}\ }\href {\doibase 10.1103/PhysRevX.7.031016}
  {\bibfield  {journal} {\bibinfo  {journal} {Phys. Rev. X}\ }\textbf {\bibinfo
  {volume} {7}},\ \bibinfo {pages} {031016} (\bibinfo {year}
  {2017})}\BibitemShut {NoStop}%
\bibitem [{\citenamefont {Zhou}\ and\ \citenamefont
  {Nahum}(2019)}]{Zhou2018Emergent}%
  \BibitemOpen
  \bibfield  {author} {\bibinfo {author} {\bibfnamefont {Tianci}\ \bibnamefont
  {Zhou}}\ and\ \bibinfo {author} {\bibfnamefont {Adam}\ \bibnamefont
  {Nahum}},\ }\bibfield  {title} {\enquote {\bibinfo {title} {Emergent
  statistical mechanics of entanglement in random unitary circuits},}\ }\href
  {\doibase 10.1103/PhysRevB.99.174205} {\bibfield  {journal} {\bibinfo
  {journal} {Phys. Rev. B}\ }\textbf {\bibinfo {volume} {99}},\ \bibinfo
  {pages} {174205} (\bibinfo {year} {2019})}\BibitemShut {NoStop}%
\bibitem [{\citenamefont {Nahum}\ \emph {et~al.}(2018)\citenamefont {Nahum},
  \citenamefont {Vijay},\ and\ \citenamefont {Haah}}]{Nahum2018Operator}%
  \BibitemOpen
  \bibfield  {author} {\bibinfo {author} {\bibfnamefont {Adam}\ \bibnamefont
  {Nahum}}, \bibinfo {author} {\bibfnamefont {Sagar}\ \bibnamefont {Vijay}}, \
  and\ \bibinfo {author} {\bibfnamefont {Jeongwan}\ \bibnamefont {Haah}},\
  }\bibfield  {title} {\enquote {\bibinfo {title} {Operator spreading in random
  unitary circuits},}\ }\href {\doibase 10.1103/PhysRevX.8.021014} {\bibfield
  {journal} {\bibinfo  {journal} {Phys. Rev. X}\ }\textbf {\bibinfo {volume}
  {8}},\ \bibinfo {pages} {021014} (\bibinfo {year} {2018})}\BibitemShut
  {NoStop}%
\bibitem [{\citenamefont {Choi}\ \emph {et~al.}(2020)\citenamefont {Choi},
  \citenamefont {Bao}, \citenamefont {Qi},\ and\ \citenamefont
  {Altman}}]{Choi2019QECEPTRUCWPM}%
  \BibitemOpen
  \bibfield  {author} {\bibinfo {author} {\bibfnamefont {Soonwon}\ \bibnamefont
  {Choi}}, \bibinfo {author} {\bibfnamefont {Yimu}\ \bibnamefont {Bao}},
  \bibinfo {author} {\bibfnamefont {Xiao-Liang}\ \bibnamefont {Qi}}, \ and\
  \bibinfo {author} {\bibfnamefont {Ehud}\ \bibnamefont {Altman}},\ }\bibfield
  {title} {\enquote {\bibinfo {title} {Quantum error correction in scrambling
  dynamics and measurement-induced phase transition},}\ }\href {\doibase
  10.1103/PhysRevLett.125.030505} {\bibfield  {journal} {\bibinfo  {journal}
  {Phys. Rev. Lett.}\ }\textbf {\bibinfo {volume} {125}},\ \bibinfo {pages}
  {030505} (\bibinfo {year} {2020})}\BibitemShut {NoStop}%
\bibitem [{\citenamefont {Bao}\ \emph {et~al.}(2020)\citenamefont {Bao},
  \citenamefont {Choi},\ and\ \citenamefont {Altman}}]{Bao2019TPTRUCWM}%
  \BibitemOpen
  \bibfield  {author} {\bibinfo {author} {\bibfnamefont {Yimu}\ \bibnamefont
  {Bao}}, \bibinfo {author} {\bibfnamefont {Soonwon}\ \bibnamefont {Choi}}, \
  and\ \bibinfo {author} {\bibfnamefont {Ehud}\ \bibnamefont {Altman}},\
  }\bibfield  {title} {\enquote {\bibinfo {title} {Theory of the phase
  transition in random unitary circuits with measurements},}\ }\href {\doibase
  10.1103/PhysRevB.101.104301} {\bibfield  {journal} {\bibinfo  {journal}
  {Phys. Rev. B}\ }\textbf {\bibinfo {volume} {101}},\ \bibinfo {pages}
  {104301} (\bibinfo {year} {2020})}\BibitemShut {NoStop}%
\bibitem [{\citenamefont {{Fan}}\ \emph {et~al.}(2021)\citenamefont {{Fan}},
  \citenamefont {{Vijay}}, \citenamefont {{Vishwanath}},\ and\ \citenamefont
  {{You}}}]{Fan2020Self-Organized}%
  \BibitemOpen
  \bibfield  {author} {\bibinfo {author} {\bibfnamefont {Ruihua}\ \bibnamefont
  {{Fan}}}, \bibinfo {author} {\bibfnamefont {Sagar}\ \bibnamefont {{Vijay}}},
  \bibinfo {author} {\bibfnamefont {Ashvin}\ \bibnamefont {{Vishwanath}}}, \
  and\ \bibinfo {author} {\bibfnamefont {Yi-Zhuang}\ \bibnamefont {{You}}},\
  }\bibfield  {title} {\enquote {\bibinfo {title} {{Self-organized error
  correction in random unitary circuits with measurement}},}\ }\href {\doibase
  10.1103/PhysRevB.103.174309} {\bibfield  {journal} {\bibinfo  {journal}
  {\prb}\ }\textbf {\bibinfo {volume} {103}},\ \bibinfo {eid} {174309}
  (\bibinfo {year} {2021})},\ \Eprint {http://arxiv.org/abs/2002.12385}
  {arXiv:2002.12385 [cond-mat.stat-mech]} \BibitemShut {NoStop}%
\bibitem [{\citenamefont {{Lashkari}}\ \emph {et~al.}(2013)\citenamefont
  {{Lashkari}}, \citenamefont {{Stanford}}, \citenamefont {{Hastings}},
  \citenamefont {{Osborne}},\ and\ \citenamefont
  {{Hayden}}}]{Lashkari2013TFSC}%
  \BibitemOpen
  \bibfield  {author} {\bibinfo {author} {\bibfnamefont {Nima}\ \bibnamefont
  {{Lashkari}}}, \bibinfo {author} {\bibfnamefont {Douglas}\ \bibnamefont
  {{Stanford}}}, \bibinfo {author} {\bibfnamefont {Matthew}\ \bibnamefont
  {{Hastings}}}, \bibinfo {author} {\bibfnamefont {Tobias}\ \bibnamefont
  {{Osborne}}}, \ and\ \bibinfo {author} {\bibfnamefont {Patrick}\ \bibnamefont
  {{Hayden}}},\ }\bibfield  {title} {\enquote {\bibinfo {title} {{Towards the
  fast scrambling conjecture}},}\ }\href {\doibase 10.1007/JHEP04(2013)022}
  {\bibfield  {journal} {\bibinfo  {journal} {Journal of High Energy Physics}\
  }\textbf {\bibinfo {volume} {2013}},\ \bibinfo {eid} {22} (\bibinfo {year}
  {2013})},\ \Eprint {http://arxiv.org/abs/1111.6580} {arXiv:1111.6580
  [hep-th]} \BibitemShut {NoStop}%
\bibitem [{\citenamefont {Xu}\ and\ \citenamefont
  {Swingle}(2019)}]{Xu2018LQFS}%
  \BibitemOpen
  \bibfield  {author} {\bibinfo {author} {\bibfnamefont {Shenglong}\
  \bibnamefont {Xu}}\ and\ \bibinfo {author} {\bibfnamefont {Brian}\
  \bibnamefont {Swingle}},\ }\bibfield  {title} {\enquote {\bibinfo {title}
  {Locality, quantum fluctuations, and scrambling},}\ }\href {\doibase
  10.1103/PhysRevX.9.031048} {\bibfield  {journal} {\bibinfo  {journal} {Phys.
  Rev. X}\ }\textbf {\bibinfo {volume} {9}},\ \bibinfo {pages} {031048}
  (\bibinfo {year} {2019})}\BibitemShut {NoStop}%
\bibitem [{\citenamefont {{Gharibyan}}\ \emph {et~al.}(2018)\citenamefont
  {{Gharibyan}}, \citenamefont {{Hanada}}, \citenamefont {{Shenker}},\ and\
  \citenamefont {{Tezuka}}}]{Gharibyan2018ORMBSS}%
  \BibitemOpen
  \bibfield  {author} {\bibinfo {author} {\bibfnamefont {Hrant}\ \bibnamefont
  {{Gharibyan}}}, \bibinfo {author} {\bibfnamefont {Masanori}\ \bibnamefont
  {{Hanada}}}, \bibinfo {author} {\bibfnamefont {Stephen~H.}\ \bibnamefont
  {{Shenker}}}, \ and\ \bibinfo {author} {\bibfnamefont {Masaki}\ \bibnamefont
  {{Tezuka}}},\ }\bibfield  {title} {\enquote {\bibinfo {title} {{Onset of
  random matrix behavior in scrambling systems}},}\ }\href {\doibase
  10.1007/JHEP07(2018)124} {\bibfield  {journal} {\bibinfo  {journal} {Journal
  of High Energy Physics}\ }\textbf {\bibinfo {volume} {2018}},\ \bibinfo {eid}
  {124} (\bibinfo {year} {2018})},\ \Eprint {http://arxiv.org/abs/1803.08050}
  {arXiv:1803.08050 [hep-th]} \BibitemShut {NoStop}%
\bibitem [{\citenamefont {{Zhou}}\ and\ \citenamefont
  {{Chen}}(2019)}]{Zhou2019ODBQC}%
  \BibitemOpen
  \bibfield  {author} {\bibinfo {author} {\bibfnamefont {Tianci}\ \bibnamefont
  {{Zhou}}}\ and\ \bibinfo {author} {\bibfnamefont {Xiao}\ \bibnamefont
  {{Chen}}},\ }\bibfield  {title} {\enquote {\bibinfo {title} {{Operator
  dynamics in a Brownian quantum circuit}},}\ }\href {\doibase
  10.1103/PhysRevE.99.052212} {\bibfield  {journal} {\bibinfo  {journal}
  {\pre}\ }\textbf {\bibinfo {volume} {99}},\ \bibinfo {eid} {052212} (\bibinfo
  {year} {2019})},\ \Eprint {http://arxiv.org/abs/1805.09307} {arXiv:1805.09307
  [cond-mat.str-el]} \BibitemShut {NoStop}%
\bibitem [{\citenamefont {{Chen}}\ and\ \citenamefont
  {{Zhou}}(2019)}]{Chen2019QCDLPIS}%
  \BibitemOpen
  \bibfield  {author} {\bibinfo {author} {\bibfnamefont {Xiao}\ \bibnamefont
  {{Chen}}}\ and\ \bibinfo {author} {\bibfnamefont {Tianci}\ \bibnamefont
  {{Zhou}}},\ }\bibfield  {title} {\enquote {\bibinfo {title} {{Quantum chaos
  dynamics in long-range power law interaction systems}},}\ }\href {\doibase
  10.1103/PhysRevB.100.064305} {\bibfield  {journal} {\bibinfo  {journal}
  {\prb}\ }\textbf {\bibinfo {volume} {100}},\ \bibinfo {eid} {064305}
  (\bibinfo {year} {2019})},\ \Eprint {http://arxiv.org/abs/1808.09812}
  {arXiv:1808.09812 [cond-mat.stat-mech]} \BibitemShut {NoStop}%
\bibitem [{\citenamefont {You}\ \emph {et~al.}(2018)\citenamefont {You},
  \citenamefont {Yang},\ and\ \citenamefont {Qi}}]{You2018Machine}%
  \BibitemOpen
  \bibfield  {author} {\bibinfo {author} {\bibfnamefont {Yi-Zhuang}\
  \bibnamefont {You}}, \bibinfo {author} {\bibfnamefont {Zhao}\ \bibnamefont
  {Yang}}, \ and\ \bibinfo {author} {\bibfnamefont {Xiao-Liang}\ \bibnamefont
  {Qi}},\ }\bibfield  {title} {\enquote {\bibinfo {title} {Machine learning
  spatial geometry from entanglement features},}\ }\href {\doibase
  10.1103/PhysRevB.97.045153} {\bibfield  {journal} {\bibinfo  {journal} {Phys.
  Rev. B}\ }\textbf {\bibinfo {volume} {97}},\ \bibinfo {pages} {045153}
  (\bibinfo {year} {2018})}\BibitemShut {NoStop}%
\bibitem [{\citenamefont {{You}}\ and\ \citenamefont
  {{Gu}}(2018)}]{You2018Entanglement}%
  \BibitemOpen
  \bibfield  {author} {\bibinfo {author} {\bibfnamefont {Yi-Zhuang}\
  \bibnamefont {{You}}}\ and\ \bibinfo {author} {\bibfnamefont {Yingfei}\
  \bibnamefont {{Gu}}},\ }\bibfield  {title} {\enquote {\bibinfo {title}
  {{Entanglement features of random Hamiltonian dynamics}},}\ }\href {\doibase
  10.1103/PhysRevB.98.014309} {\bibfield  {journal} {\bibinfo  {journal}
  {\prb}\ }\textbf {\bibinfo {volume} {98}},\ \bibinfo {eid} {014309} (\bibinfo
  {year} {2018})},\ \Eprint {http://arxiv.org/abs/1803.10425} {arXiv:1803.10425
  [quant-ph]} \BibitemShut {NoStop}%
\bibitem [{\citenamefont {Saffman}(2016)}]{Saffman_2016}%
  \BibitemOpen
  \bibfield  {author} {\bibinfo {author} {\bibfnamefont {M}~\bibnamefont
  {Saffman}},\ }\bibfield  {title} {\enquote {\bibinfo {title} {Quantum
  computing with atomic qubits and rydberg interactions: progress and
  challenges},}\ }\href {\doibase 10.1088/0953-4075/49/20/202001} {\bibfield
  {journal} {\bibinfo  {journal} {Journal of Physics B: Atomic, Molecular and
  Optical Physics}\ }\textbf {\bibinfo {volume} {49}},\ \bibinfo {pages}
  {202001} (\bibinfo {year} {2016})}\BibitemShut {NoStop}%
\bibitem [{\citenamefont {Monroe}\ \emph {et~al.}(2021)\citenamefont {Monroe},
  \citenamefont {Campbell}, \citenamefont {Duan}, \citenamefont {Gong},
  \citenamefont {Gorshkov}, \citenamefont {Hess}, \citenamefont {Islam},
  \citenamefont {Kim}, \citenamefont {Linke}, \citenamefont {Pagano},
  \citenamefont {Richerme}, \citenamefont {Senko},\ and\ \citenamefont
  {Yao}}]{RevModPhys.93.025001}%
  \BibitemOpen
  \bibfield  {author} {\bibinfo {author} {\bibfnamefont {C.}~\bibnamefont
  {Monroe}}, \bibinfo {author} {\bibfnamefont {W.~C.}\ \bibnamefont
  {Campbell}}, \bibinfo {author} {\bibfnamefont {L.-M.}\ \bibnamefont {Duan}},
  \bibinfo {author} {\bibfnamefont {Z.-X.}\ \bibnamefont {Gong}}, \bibinfo
  {author} {\bibfnamefont {A.~V.}\ \bibnamefont {Gorshkov}}, \bibinfo {author}
  {\bibfnamefont {P.~W.}\ \bibnamefont {Hess}}, \bibinfo {author}
  {\bibfnamefont {R.}~\bibnamefont {Islam}}, \bibinfo {author} {\bibfnamefont
  {K.}~\bibnamefont {Kim}}, \bibinfo {author} {\bibfnamefont {N.~M.}\
  \bibnamefont {Linke}}, \bibinfo {author} {\bibfnamefont {G.}~\bibnamefont
  {Pagano}}, \bibinfo {author} {\bibfnamefont {P.}~\bibnamefont {Richerme}},
  \bibinfo {author} {\bibfnamefont {C.}~\bibnamefont {Senko}}, \ and\ \bibinfo
  {author} {\bibfnamefont {N.~Y.}\ \bibnamefont {Yao}},\ }\bibfield  {title}
  {\enquote {\bibinfo {title} {Programmable quantum simulations of spin systems
  with trapped ions},}\ }\href {\doibase 10.1103/RevModPhys.93.025001}
  {\bibfield  {journal} {\bibinfo  {journal} {Rev. Mod. Phys.}\ }\textbf
  {\bibinfo {volume} {93}},\ \bibinfo {pages} {025001} (\bibinfo {year}
  {2021})}\BibitemShut {NoStop}%
\bibitem [{\citenamefont {{Acharya}}\ \emph {et~al.}(2021)\citenamefont
  {{Acharya}}, \citenamefont {{Saha}},\ and\ \citenamefont
  {{Sengupta}}}]{Acharya2021Informationally}%
  \BibitemOpen
  \bibfield  {author} {\bibinfo {author} {\bibfnamefont {Atithi}\ \bibnamefont
  {{Acharya}}}, \bibinfo {author} {\bibfnamefont {Siddhartha}\ \bibnamefont
  {{Saha}}}, \ and\ \bibinfo {author} {\bibfnamefont {Anirvan~M.}\ \bibnamefont
  {{Sengupta}}},\ }\bibfield  {title} {\enquote {\bibinfo {title}
  {{Informationally complete POVM-based shadow tomography}},}\ }\href@noop {}
  {\bibfield  {journal} {\bibinfo  {journal} {arXiv e-prints}\ ,\ \bibinfo
  {eid} {arXiv:2105.05992}} (\bibinfo {year} {2021})},\ \Eprint
  {http://arxiv.org/abs/2105.05992} {arXiv:2105.05992 [quant-ph]} \BibitemShut
  {NoStop}%
\bibitem [{\citenamefont {Gu{\c{t}}{\u{a}}}\ \emph {et~al.}(2020)\citenamefont
  {Gu{\c{t}}{\u{a}}}, \citenamefont {Kahn}, \citenamefont {Kueng},\ and\
  \citenamefont {Tropp}}]{Guta2018Fast}%
  \BibitemOpen
  \bibfield  {author} {\bibinfo {author} {\bibfnamefont {M}~\bibnamefont
  {Gu{\c{t}}{\u{a}}}}, \bibinfo {author} {\bibfnamefont {J}~\bibnamefont
  {Kahn}}, \bibinfo {author} {\bibfnamefont {R}~\bibnamefont {Kueng}}, \ and\
  \bibinfo {author} {\bibfnamefont {J~A}\ \bibnamefont {Tropp}},\ }\bibfield
  {title} {\enquote {\bibinfo {title} {Fast state tomography with optimal error
  bounds},}\ }\href {\doibase 10.1088/1751-8121/ab8111} {\bibfield  {journal}
  {\bibinfo  {journal} {Journal of Physics A: Mathematical and Theoretical}\
  }\textbf {\bibinfo {volume} {53}},\ \bibinfo {pages} {204001} (\bibinfo
  {year} {2020})}\BibitemShut {NoStop}%
\bibitem [{\citenamefont {{Elben}}\ \emph {et~al.}(2019)\citenamefont
  {{Elben}}, \citenamefont {{Vermersch}}, \citenamefont {{Roos}},\ and\
  \citenamefont {{Zoller}}}]{Elben2019Statistical}%
  \BibitemOpen
  \bibfield  {author} {\bibinfo {author} {\bibfnamefont {A.}~\bibnamefont
  {{Elben}}}, \bibinfo {author} {\bibfnamefont {B.}~\bibnamefont
  {{Vermersch}}}, \bibinfo {author} {\bibfnamefont {C.~F.}\ \bibnamefont
  {{Roos}}}, \ and\ \bibinfo {author} {\bibfnamefont {P.}~\bibnamefont
  {{Zoller}}},\ }\bibfield  {title} {\enquote {\bibinfo {title} {{Statistical
  correlations between locally randomized measurements: A toolbox for probing
  entanglement in many-body quantum states}},}\ }\href {\doibase
  10.1103/PhysRevA.99.052323} {\bibfield  {journal} {\bibinfo  {journal}
  {\pra}\ }\textbf {\bibinfo {volume} {99}},\ \bibinfo {eid} {052323} (\bibinfo
  {year} {2019})},\ \Eprint {http://arxiv.org/abs/1812.02624} {arXiv:1812.02624
  [quant-ph]} \BibitemShut {NoStop}%
\bibitem [{\citenamefont {{Enshan Koh}}\ and\ \citenamefont
  {{Grewal}}(2020)}]{Enshan-Koh2020Classical}%
  \BibitemOpen
  \bibfield  {author} {\bibinfo {author} {\bibfnamefont {Dax}\ \bibnamefont
  {{Enshan Koh}}}\ and\ \bibinfo {author} {\bibfnamefont {Sabee}\ \bibnamefont
  {{Grewal}}},\ }\bibfield  {title} {\enquote {\bibinfo {title} {{Classical
  Shadows with Noise}},}\ }\href@noop {} {\bibfield  {journal} {\bibinfo
  {journal} {arXiv e-prints}\ ,\ \bibinfo {eid} {arXiv:2011.11580}} (\bibinfo
  {year} {2020})},\ \Eprint {http://arxiv.org/abs/2011.11580} {arXiv:2011.11580
  [quant-ph]} \BibitemShut {NoStop}%
\bibitem [{\citenamefont {{Chen}}\ \emph {et~al.}(2020)\citenamefont {{Chen}},
  \citenamefont {{Yu}}, \citenamefont {{Zeng}},\ and\ \citenamefont
  {{Flammia}}}]{Chen2020Robust}%
  \BibitemOpen
  \bibfield  {author} {\bibinfo {author} {\bibfnamefont {Senrui}\ \bibnamefont
  {{Chen}}}, \bibinfo {author} {\bibfnamefont {Wenjun}\ \bibnamefont {{Yu}}},
  \bibinfo {author} {\bibfnamefont {Pei}\ \bibnamefont {{Zeng}}}, \ and\
  \bibinfo {author} {\bibfnamefont {Steven~T.}\ \bibnamefont {{Flammia}}},\
  }\bibfield  {title} {\enquote {\bibinfo {title} {{Robust shadow
  estimation}},}\ }\href@noop {} {\bibfield  {journal} {\bibinfo  {journal}
  {arXiv e-prints}\ ,\ \bibinfo {eid} {arXiv:2011.09636}} (\bibinfo {year}
  {2020})},\ \Eprint {http://arxiv.org/abs/2011.09636} {arXiv:2011.09636
  [quant-ph]} \BibitemShut {NoStop}%
\bibitem [{\citenamefont {{Zhao}}\ \emph {et~al.}(2020)\citenamefont {{Zhao}},
  \citenamefont {{Rubin}},\ and\ \citenamefont {{Miyake}}}]{Zhao2020Fermionic}%
  \BibitemOpen
  \bibfield  {author} {\bibinfo {author} {\bibfnamefont {Andrew}\ \bibnamefont
  {{Zhao}}}, \bibinfo {author} {\bibfnamefont {Nicholas~C.}\ \bibnamefont
  {{Rubin}}}, \ and\ \bibinfo {author} {\bibfnamefont {Akimasa}\ \bibnamefont
  {{Miyake}}},\ }\bibfield  {title} {\enquote {\bibinfo {title} {{Fermionic
  partial tomography via classical shadows}},}\ }\href@noop {} {\bibfield
  {journal} {\bibinfo  {journal} {arXiv e-prints}\ ,\ \bibinfo {eid}
  {arXiv:2010.16094}} (\bibinfo {year} {2020})},\ \Eprint
  {http://arxiv.org/abs/2010.16094} {arXiv:2010.16094 [quant-ph]} \BibitemShut
  {NoStop}%
\bibitem [{\citenamefont {Weingarten}(1978)}]{Weingarten1978Asymptotic}%
  \BibitemOpen
  \bibfield  {author} {\bibinfo {author} {\bibfnamefont {Don}\ \bibnamefont
  {Weingarten}},\ }\bibfield  {title} {\enquote {\bibinfo {title} {Asymptotic
  behavior of group integrals in the limit of infinite rank},}\ }\href@noop {}
  {\bibfield  {journal} {\bibinfo  {journal} {Journal of Mathematical Physics}\
  }\textbf {\bibinfo {volume} {19}},\ \bibinfo {pages} {999--1001} (\bibinfo
  {year} {1978})}\BibitemShut {NoStop}%
\bibitem [{\citenamefont {{Collins}}\ and\ \citenamefont
  {{{\'S}niady}}(2006)}]{Collins2006Integration}%
  \BibitemOpen
  \bibfield  {author} {\bibinfo {author} {\bibfnamefont {Beno{\^\i}t}\
  \bibnamefont {{Collins}}}\ and\ \bibinfo {author} {\bibfnamefont {Piotr}\
  \bibnamefont {{{\'S}niady}}},\ }\bibfield  {title} {\enquote {\bibinfo
  {title} {{Integration with Respect to the Haar Measure on Unitary, Orthogonal
  and Symplectic Group}},}\ }\href {\doibase 10.1007/s00220-006-1554-3}
  {\bibfield  {journal} {\bibinfo  {journal} {Communications in Mathematical
  Physics}\ }\textbf {\bibinfo {volume} {264}},\ \bibinfo {pages} {773--795}
  (\bibinfo {year} {2006})},\ \Eprint {http://arxiv.org/abs/math-ph/0402073}
  {arXiv:math-ph/0402073 [math-ph]} \BibitemShut {NoStop}%
\bibitem [{Note1()}]{Note1}%
  \BibitemOpen
  \bibinfo {note} {The Brownian unitary evolution is a product of a sequence of
  infinitesimal time-evolution $U=\DOTSB \prod@ \slimits@ _t e^{-\protect
  \mathrm {i}H_t\protect \tmspace +\thickmuskip {.2777em}\delta t}$, but the
  Hamiltonian $H_t$ at each time step is independent drawn from a random
  Hamiltonian ensemble (unlike the coherent quantum dynamics, where the same
  Hamiltonian drives the dynamics though all time.)}\BibitemShut {NoStop}%
\bibitem [{\citenamefont {{Akhtar}}\ and\ \citenamefont
  {{You}}(2020)}]{Akhtar2020Multiregion}%
  \BibitemOpen
  \bibfield  {author} {\bibinfo {author} {\bibfnamefont {A.~A.}\ \bibnamefont
  {{Akhtar}}}\ and\ \bibinfo {author} {\bibfnamefont {Yi-Zhuang}\ \bibnamefont
  {{You}}},\ }\bibfield  {title} {\enquote {\bibinfo {title} {{Multiregion
  entanglement in locally scrambled quantum dynamics}},}\ }\href {\doibase
  10.1103/PhysRevB.102.134203} {\bibfield  {journal} {\bibinfo  {journal}
  {\prb}\ }\textbf {\bibinfo {volume} {102}},\ \bibinfo {eid} {134203}
  (\bibinfo {year} {2020})},\ \Eprint {http://arxiv.org/abs/2006.08797}
  {arXiv:2006.08797 [cond-mat.dis-nn]} \BibitemShut {NoStop}%
\bibitem [{\citenamefont {Elben}\ \emph {et~al.}(2018)\citenamefont {Elben},
  \citenamefont {Vermersch}, \citenamefont {Dalmonte}, \citenamefont {Cirac},\
  and\ \citenamefont {Zoller}}]{Elben2018Renyi}%
  \BibitemOpen
  \bibfield  {author} {\bibinfo {author} {\bibfnamefont {A.}~\bibnamefont
  {Elben}}, \bibinfo {author} {\bibfnamefont {B.}~\bibnamefont {Vermersch}},
  \bibinfo {author} {\bibfnamefont {M.}~\bibnamefont {Dalmonte}}, \bibinfo
  {author} {\bibfnamefont {J.~I.}\ \bibnamefont {Cirac}}, \ and\ \bibinfo
  {author} {\bibfnamefont {P.}~\bibnamefont {Zoller}},\ }\bibfield  {title}
  {\enquote {\bibinfo {title} {R\'enyi entropies from random quenches in atomic
  hubbard and spin models},}\ }\href {\doibase 10.1103/PhysRevLett.120.050406}
  {\bibfield  {journal} {\bibinfo  {journal} {Phys. Rev. Lett.}\ }\textbf
  {\bibinfo {volume} {120}},\ \bibinfo {pages} {050406} (\bibinfo {year}
  {2018})}\BibitemShut {NoStop}%
\bibitem [{\citenamefont {Vermersch}\ \emph {et~al.}(2018)\citenamefont
  {Vermersch}, \citenamefont {Elben}, \citenamefont {Dalmonte}, \citenamefont
  {Cirac},\ and\ \citenamefont {Zoller}}]{Vermersch2018Unitary}%
  \BibitemOpen
  \bibfield  {author} {\bibinfo {author} {\bibfnamefont {B.}~\bibnamefont
  {Vermersch}}, \bibinfo {author} {\bibfnamefont {A.}~\bibnamefont {Elben}},
  \bibinfo {author} {\bibfnamefont {M.}~\bibnamefont {Dalmonte}}, \bibinfo
  {author} {\bibfnamefont {J.~I.}\ \bibnamefont {Cirac}}, \ and\ \bibinfo
  {author} {\bibfnamefont {P.}~\bibnamefont {Zoller}},\ }\bibfield  {title}
  {\enquote {\bibinfo {title} {Unitary $n$-designs via random quenches in
  atomic hubbard and spin models: Application to the measurement of r\'enyi
  entropies},}\ }\href {\doibase 10.1103/PhysRevA.97.023604} {\bibfield
  {journal} {\bibinfo  {journal} {Phys. Rev. A}\ }\textbf {\bibinfo {volume}
  {97}},\ \bibinfo {pages} {023604} (\bibinfo {year} {2018})}\BibitemShut
  {NoStop}%
\bibitem [{\citenamefont {{Brydges}}\ \emph {et~al.}(2019)\citenamefont
  {{Brydges}}, \citenamefont {{Elben}}, \citenamefont {{Jurcevic}},
  \citenamefont {{Vermersch}}, \citenamefont {{Maier}}, \citenamefont
  {{Lanyon}}, \citenamefont {{Zoller}}, \citenamefont {{Blatt}},\ and\
  \citenamefont {{Roos}}}]{Brydges2019Probing}%
  \BibitemOpen
  \bibfield  {author} {\bibinfo {author} {\bibfnamefont {Tiff}\ \bibnamefont
  {{Brydges}}}, \bibinfo {author} {\bibfnamefont {Andreas}\ \bibnamefont
  {{Elben}}}, \bibinfo {author} {\bibfnamefont {Petar}\ \bibnamefont
  {{Jurcevic}}}, \bibinfo {author} {\bibfnamefont {Beno{\^\i}t}\ \bibnamefont
  {{Vermersch}}}, \bibinfo {author} {\bibfnamefont {Christine}\ \bibnamefont
  {{Maier}}}, \bibinfo {author} {\bibfnamefont {Ben~P.}\ \bibnamefont
  {{Lanyon}}}, \bibinfo {author} {\bibfnamefont {Peter}\ \bibnamefont
  {{Zoller}}}, \bibinfo {author} {\bibfnamefont {Rainer}\ \bibnamefont
  {{Blatt}}}, \ and\ \bibinfo {author} {\bibfnamefont {Christian~F.}\
  \bibnamefont {{Roos}}},\ }\bibfield  {title} {\enquote {\bibinfo {title}
  {{Probing R{\'e}nyi entanglement entropy via randomized measurements}},}\
  }\href {\doibase 10.1126/science.aau4963} {\bibfield  {journal} {\bibinfo
  {journal} {Science}\ }\textbf {\bibinfo {volume} {364}},\ \bibinfo {pages}
  {260--263} (\bibinfo {year} {2019})},\ \Eprint
  {http://arxiv.org/abs/1806.05747} {arXiv:1806.05747 [quant-ph]} \BibitemShut
  {NoStop}%
\bibitem [{\citenamefont {Carrasquilla}\ \emph {et~al.}(2019)\citenamefont
  {Carrasquilla}, \citenamefont {Torlai}, \citenamefont {Melko},\ and\
  \citenamefont {Aolita}}]{Carrasquilla2019Reconstructing}%
  \BibitemOpen
  \bibfield  {author} {\bibinfo {author} {\bibfnamefont {Juan}\ \bibnamefont
  {Carrasquilla}}, \bibinfo {author} {\bibfnamefont {Giacomo}\ \bibnamefont
  {Torlai}}, \bibinfo {author} {\bibfnamefont {Roger~G.}\ \bibnamefont
  {Melko}}, \ and\ \bibinfo {author} {\bibfnamefont {Leandro}\ \bibnamefont
  {Aolita}},\ }\bibfield  {title} {\enquote {\bibinfo {title} {Reconstructing
  quantum states with generative models},}\ }\href {\doibase
  10.1038/s42256-019-0028-1} {\bibfield  {journal} {\bibinfo  {journal} {Nature
  Machine Intelligence}\ }\textbf {\bibinfo {volume} {1}},\ \bibinfo {pages}
  {155--161} (\bibinfo {year} {2019})}\BibitemShut {NoStop}%
\bibitem [{\citenamefont {{Aharonov}}\ \emph {et~al.}(2021)\citenamefont
  {{Aharonov}}, \citenamefont {{Cotler}},\ and\ \citenamefont
  {{Qi}}}]{Aharonov2021Quantum}%
  \BibitemOpen
  \bibfield  {author} {\bibinfo {author} {\bibfnamefont {Dorit}\ \bibnamefont
  {{Aharonov}}}, \bibinfo {author} {\bibfnamefont {Jordan}\ \bibnamefont
  {{Cotler}}}, \ and\ \bibinfo {author} {\bibfnamefont {Xiao-Liang}\
  \bibnamefont {{Qi}}},\ }\bibfield  {title} {\enquote {\bibinfo {title}
  {{Quantum Algorithmic Measurement}},}\ }\href@noop {} {\bibfield  {journal}
  {\bibinfo  {journal} {arXiv e-prints}\ ,\ \bibinfo {eid} {arXiv:2101.04634}}
  (\bibinfo {year} {2021})},\ \Eprint {http://arxiv.org/abs/2101.04634}
  {arXiv:2101.04634 [quant-ph]} \BibitemShut {NoStop}%
\bibitem [{Note2()}]{Note2}%
  \BibitemOpen
  \bibinfo {note} {In the $L\to 0$ limit, the measurement operator is still of
  at least size 1, which motivates the ``$+1$'' regularization in
  $(L+1)^{\alpha }$.}\BibitemShut {Stop}%
\bibitem [{Note3()}]{Note3}%
  \BibitemOpen
  \bibinfo {note} {For simulation of Rydberg Hamiltonian, we choose parameters
  $\Omega =2.75, \Delta =1, R_b=1$.}\BibitemShut {Stop}%
\bibitem [{\citenamefont {{Ma}}\ \emph {et~al.}(2016)\citenamefont {{Ma}},
  \citenamefont {{Jackson}}, \citenamefont {{Zhou}}, \citenamefont {{Chen}},
  \citenamefont {{Lu}}, \citenamefont {{Mazurek}}, \citenamefont {{Fisher}},
  \citenamefont {{Peng}}, \citenamefont {{Kribs}}, \citenamefont {{Resch}},
  \citenamefont {{Ji}}, \citenamefont {{Zeng}},\ and\ \citenamefont
  {{Laflamme}}}]{Ma2016Pure}%
  \BibitemOpen
  \bibfield  {author} {\bibinfo {author} {\bibfnamefont {Xian}\ \bibnamefont
  {{Ma}}}, \bibinfo {author} {\bibfnamefont {Tyler}\ \bibnamefont {{Jackson}}},
  \bibinfo {author} {\bibfnamefont {Hui}\ \bibnamefont {{Zhou}}}, \bibinfo
  {author} {\bibfnamefont {Jianxin}\ \bibnamefont {{Chen}}}, \bibinfo {author}
  {\bibfnamefont {Dawei}\ \bibnamefont {{Lu}}}, \bibinfo {author}
  {\bibfnamefont {Michael~D.}\ \bibnamefont {{Mazurek}}}, \bibinfo {author}
  {\bibfnamefont {Kent A.~G.}\ \bibnamefont {{Fisher}}}, \bibinfo {author}
  {\bibfnamefont {Xinhua}\ \bibnamefont {{Peng}}}, \bibinfo {author}
  {\bibfnamefont {David}\ \bibnamefont {{Kribs}}}, \bibinfo {author}
  {\bibfnamefont {Kevin~J.}\ \bibnamefont {{Resch}}}, \bibinfo {author}
  {\bibfnamefont {Zhengfeng}\ \bibnamefont {{Ji}}}, \bibinfo {author}
  {\bibfnamefont {Bei}\ \bibnamefont {{Zeng}}}, \ and\ \bibinfo {author}
  {\bibfnamefont {Raymond}\ \bibnamefont {{Laflamme}}},\ }\bibfield  {title}
  {\enquote {\bibinfo {title} {{Pure State Tomography with Pauli
  Measurements}},}\ }\href@noop {} {\bibfield  {journal} {\bibinfo  {journal}
  {arXiv e-prints}\ ,\ \bibinfo {eid} {arXiv:1601.05379}} (\bibinfo {year}
  {2016})},\ \Eprint {http://arxiv.org/abs/1601.05379} {arXiv:1601.05379
  [quant-ph]} \BibitemShut {NoStop}%
\bibitem [{\citenamefont {{Kraus}}(1971)}]{Kraus1971General}%
  \BibitemOpen
  \bibfield  {author} {\bibinfo {author} {\bibfnamefont {K.}~\bibnamefont
  {{Kraus}}},\ }\bibfield  {title} {\enquote {\bibinfo {title} {{General state
  changes in quantum theory}},}\ }\href {\doibase 10.1016/0003-4916(71)90108-4}
  {\bibfield  {journal} {\bibinfo  {journal} {Annals of Physics}\ }\textbf
  {\bibinfo {volume} {64}},\ \bibinfo {pages} {311--335} (\bibinfo {year}
  {1971})}\BibitemShut {NoStop}%
\bibitem [{\citenamefont {{Gottesman}}(1998)}]{Gottesman1998The-Heisenberg}%
  \BibitemOpen
  \bibfield  {author} {\bibinfo {author} {\bibfnamefont {D.}~\bibnamefont
  {{Gottesman}}},\ }\bibfield  {title} {\enquote {\bibinfo {title} {{The
  Heisenberg Representation of Quantum Computers}},}\ }\href@noop {} {\bibfield
   {journal} {\bibinfo  {journal} {eprint arXiv:quant-ph/9807006}\ } (\bibinfo
  {year} {1998})},\ \Eprint {http://arxiv.org/abs/quant-ph/9807006}
  {quant-ph/9807006} \BibitemShut {NoStop}%
\bibitem [{\citenamefont {{Akhtar}}\ \emph {et~al.}(2022)\citenamefont
  {{Akhtar}}, \citenamefont {{Hu}},\ and\ \citenamefont
  {{You}}}]{2022arXiv220902093A}%
  \BibitemOpen
  \bibfield  {author} {\bibinfo {author} {\bibfnamefont {Ahmed~A.}\
  \bibnamefont {{Akhtar}}}, \bibinfo {author} {\bibfnamefont {Hong-Ye}\
  \bibnamefont {{Hu}}}, \ and\ \bibinfo {author} {\bibfnamefont {Yi-Zhuang}\
  \bibnamefont {{You}}},\ }\bibfield  {title} {\enquote {\bibinfo {title}
  {{Scalable and Flexible Classical Shadow Tomography with Tensor Networks}},}\
  }\href@noop {} {\bibfield  {journal} {\bibinfo  {journal} {arXiv e-prints}\
  ,\ \bibinfo {eid} {arXiv:2209.02093}} (\bibinfo {year} {2022})},\ \Eprint
  {http://arxiv.org/abs/2209.02093} {arXiv:2209.02093 [quant-ph]} \BibitemShut
  {NoStop}%
\bibitem [{\citenamefont {{Vidal}}(2007)}]{Vidal2007Entanglement}%
  \BibitemOpen
  \bibfield  {author} {\bibinfo {author} {\bibfnamefont {G.}~\bibnamefont
  {{Vidal}}},\ }\bibfield  {title} {\enquote {\bibinfo {title} {{Entanglement
  Renormalization}},}\ }\href {\doibase 10.1103/PhysRevLett.99.220405}
  {\bibfield  {journal} {\bibinfo  {journal} {Physical Review Letters}\
  }\textbf {\bibinfo {volume} {99}},\ \bibinfo {eid} {220405} (\bibinfo {year}
  {2007})},\ \Eprint {http://arxiv.org/abs/cond-mat/0512165} {cond-mat/0512165}
  \BibitemShut {NoStop}%
\bibitem [{\citenamefont {{Vidal}}(2008)}]{Vidal2008Class}%
  \BibitemOpen
  \bibfield  {author} {\bibinfo {author} {\bibfnamefont {G.}~\bibnamefont
  {{Vidal}}},\ }\bibfield  {title} {\enquote {\bibinfo {title} {{Class of
  Quantum Many-Body States That Can Be Efficiently Simulated}},}\ }\href
  {\doibase 10.1103/PhysRevLett.101.110501} {\bibfield  {journal} {\bibinfo
  {journal} {Physical Review Letters}\ }\textbf {\bibinfo {volume} {101}},\
  \bibinfo {eid} {110501} (\bibinfo {year} {2008})},\ \Eprint
  {http://arxiv.org/abs/quant-ph/0610099} {quant-ph/0610099} \BibitemShut
  {NoStop}%
\bibitem [{\citenamefont {{Pastawski}}\ \emph {et~al.}(2015)\citenamefont
  {{Pastawski}}, \citenamefont {{Yoshida}}, \citenamefont {{Harlow}},\ and\
  \citenamefont {{Preskill}}}]{Pastawski2015Holographic}%
  \BibitemOpen
  \bibfield  {author} {\bibinfo {author} {\bibfnamefont {Fernando}\
  \bibnamefont {{Pastawski}}}, \bibinfo {author} {\bibfnamefont {Beni}\
  \bibnamefont {{Yoshida}}}, \bibinfo {author} {\bibfnamefont {Daniel}\
  \bibnamefont {{Harlow}}}, \ and\ \bibinfo {author} {\bibfnamefont {John}\
  \bibnamefont {{Preskill}}},\ }\bibfield  {title} {\enquote {\bibinfo {title}
  {{Holographic quantum error-correcting codes: toy models for the
  bulk/boundary correspondence}},}\ }\href {\doibase 10.1007/JHEP06(2015)149}
  {\bibfield  {journal} {\bibinfo  {journal} {Journal of High Energy Physics}\
  }\textbf {\bibinfo {volume} {2015}},\ \bibinfo {eid} {149} (\bibinfo {year}
  {2015})},\ \Eprint {http://arxiv.org/abs/1503.06237} {arXiv:1503.06237
  [hep-th]} \BibitemShut {NoStop}%
\bibitem [{\citenamefont {{Li}}\ \emph {et~al.}(2018)\citenamefont {{Li}},
  \citenamefont {{Chen}},\ and\ \citenamefont {{Fisher}}}]{Li2018QZEMET}%
  \BibitemOpen
  \bibfield  {author} {\bibinfo {author} {\bibfnamefont {Yaodong}\ \bibnamefont
  {{Li}}}, \bibinfo {author} {\bibfnamefont {Xiao}\ \bibnamefont {{Chen}}}, \
  and\ \bibinfo {author} {\bibfnamefont {Matthew P.~A.}\ \bibnamefont
  {{Fisher}}},\ }\bibfield  {title} {\enquote {\bibinfo {title} {{Quantum Zeno
  effect and the many-body entanglement transition}},}\ }\href {\doibase
  10.1103/PhysRevB.98.205136} {\bibfield  {journal} {\bibinfo  {journal}
  {\prb}\ }\textbf {\bibinfo {volume} {98}},\ \bibinfo {eid} {205136} (\bibinfo
  {year} {2018})},\ \Eprint {http://arxiv.org/abs/1808.06134} {arXiv:1808.06134
  [quant-ph]} \BibitemShut {NoStop}%
\bibitem [{\citenamefont {Li}\ \emph {et~al.}(2019)\citenamefont {Li},
  \citenamefont {Chen},\ and\ \citenamefont {Fisher}}]{Li2019METHQC}%
  \BibitemOpen
  \bibfield  {author} {\bibinfo {author} {\bibfnamefont {Yaodong}\ \bibnamefont
  {Li}}, \bibinfo {author} {\bibfnamefont {Xiao}\ \bibnamefont {Chen}}, \ and\
  \bibinfo {author} {\bibfnamefont {Matthew P.~A.}\ \bibnamefont {Fisher}},\
  }\bibfield  {title} {\enquote {\bibinfo {title} {Measurement-driven
  entanglement transition in hybrid quantum circuits},}\ }\href {\doibase
  10.1103/PhysRevB.100.134306} {\bibfield  {journal} {\bibinfo  {journal}
  {Phys. Rev. B}\ }\textbf {\bibinfo {volume} {100}},\ \bibinfo {pages}
  {134306} (\bibinfo {year} {2019})}\BibitemShut {NoStop}%
\bibitem [{\citenamefont {Szyniszewski}\ \emph {et~al.}(2019)\citenamefont
  {Szyniszewski}, \citenamefont {Romito},\ and\ \citenamefont
  {Schomerus}}]{Szyniszewski2019ETFVWM}%
  \BibitemOpen
  \bibfield  {author} {\bibinfo {author} {\bibfnamefont {M.}~\bibnamefont
  {Szyniszewski}}, \bibinfo {author} {\bibfnamefont {A.}~\bibnamefont
  {Romito}}, \ and\ \bibinfo {author} {\bibfnamefont {H.}~\bibnamefont
  {Schomerus}},\ }\bibfield  {title} {\enquote {\bibinfo {title} {Entanglement
  transition from variable-strength weak measurements},}\ }\href {\doibase
  10.1103/PhysRevB.100.064204} {\bibfield  {journal} {\bibinfo  {journal}
  {Phys. Rev. B}\ }\textbf {\bibinfo {volume} {100}},\ \bibinfo {pages}
  {064204} (\bibinfo {year} {2019})}\BibitemShut {NoStop}%
\bibitem [{\citenamefont {{Chan}}\ \emph {et~al.}(2019)\citenamefont {{Chan}},
  \citenamefont {{Nandkishore}}, \citenamefont {{Pretko}},\ and\ \citenamefont
  {{Smith}}}]{Chan2019UED}%
  \BibitemOpen
  \bibfield  {author} {\bibinfo {author} {\bibfnamefont {Amos}\ \bibnamefont
  {{Chan}}}, \bibinfo {author} {\bibfnamefont {Rahul~M.}\ \bibnamefont
  {{Nandkishore}}}, \bibinfo {author} {\bibfnamefont {Michael}\ \bibnamefont
  {{Pretko}}}, \ and\ \bibinfo {author} {\bibfnamefont {Graeme}\ \bibnamefont
  {{Smith}}},\ }\bibfield  {title} {\enquote {\bibinfo {title}
  {{Unitary-projective entanglement dynamics}},}\ }\href {\doibase
  10.1103/PhysRevB.99.224307} {\bibfield  {journal} {\bibinfo  {journal}
  {\prb}\ }\textbf {\bibinfo {volume} {99}},\ \bibinfo {eid} {224307} (\bibinfo
  {year} {2019})},\ \Eprint {http://arxiv.org/abs/1808.05949} {arXiv:1808.05949
  [cond-mat.stat-mech]} \BibitemShut {NoStop}%
\bibitem [{\citenamefont {{Skinner}}\ \emph {et~al.}(2019)\citenamefont
  {{Skinner}}, \citenamefont {{Ruhman}},\ and\ \citenamefont
  {{Nahum}}}]{Skinner2019MPTDE}%
  \BibitemOpen
  \bibfield  {author} {\bibinfo {author} {\bibfnamefont {Brian}\ \bibnamefont
  {{Skinner}}}, \bibinfo {author} {\bibfnamefont {Jonathan}\ \bibnamefont
  {{Ruhman}}}, \ and\ \bibinfo {author} {\bibfnamefont {Adam}\ \bibnamefont
  {{Nahum}}},\ }\bibfield  {title} {\enquote {\bibinfo {title}
  {{Measurement-Induced Phase Transitions in the Dynamics of Entanglement}},}\
  }\href {\doibase 10.1103/PhysRevX.9.031009} {\bibfield  {journal} {\bibinfo
  {journal} {Physical Review X}\ }\textbf {\bibinfo {volume} {9}},\ \bibinfo
  {eid} {031009} (\bibinfo {year} {2019})},\ \Eprint
  {http://arxiv.org/abs/1808.05953} {arXiv:1808.05953 [cond-mat.stat-mech]}
  \BibitemShut {NoStop}%
\bibitem [{\citenamefont {Vasseur}\ \emph {et~al.}(2019)\citenamefont
  {Vasseur}, \citenamefont {Potter}, \citenamefont {You},\ and\ \citenamefont
  {Ludwig}}]{Vasseur2018Entanglement}%
  \BibitemOpen
  \bibfield  {author} {\bibinfo {author} {\bibfnamefont {Romain}\ \bibnamefont
  {Vasseur}}, \bibinfo {author} {\bibfnamefont {Andrew~C.}\ \bibnamefont
  {Potter}}, \bibinfo {author} {\bibfnamefont {Yi-Zhuang}\ \bibnamefont {You}},
  \ and\ \bibinfo {author} {\bibfnamefont {Andreas W.~W.}\ \bibnamefont
  {Ludwig}},\ }\bibfield  {title} {\enquote {\bibinfo {title} {Entanglement
  transitions from holographic random tensor networks},}\ }\href {\doibase
  10.1103/PhysRevB.100.134203} {\bibfield  {journal} {\bibinfo  {journal}
  {Phys. Rev. B}\ }\textbf {\bibinfo {volume} {100}},\ \bibinfo {pages}
  {134203} (\bibinfo {year} {2019})}\BibitemShut {NoStop}%
\bibitem [{\citenamefont {Jian}\ \emph {et~al.}(2020)\citenamefont {Jian},
  \citenamefont {You}, \citenamefont {Vasseur},\ and\ \citenamefont
  {Ludwig}}]{Jian2019MCRQC}%
  \BibitemOpen
  \bibfield  {author} {\bibinfo {author} {\bibfnamefont {Chao-Ming}\
  \bibnamefont {Jian}}, \bibinfo {author} {\bibfnamefont {Yi-Zhuang}\
  \bibnamefont {You}}, \bibinfo {author} {\bibfnamefont {Romain}\ \bibnamefont
  {Vasseur}}, \ and\ \bibinfo {author} {\bibfnamefont {Andreas W.~W.}\
  \bibnamefont {Ludwig}},\ }\bibfield  {title} {\enquote {\bibinfo {title}
  {Measurement-induced criticality in random quantum circuits},}\ }\href
  {\doibase 10.1103/PhysRevB.101.104302} {\bibfield  {journal} {\bibinfo
  {journal} {Phys. Rev. B}\ }\textbf {\bibinfo {volume} {101}},\ \bibinfo
  {pages} {104302} (\bibinfo {year} {2020})}\BibitemShut {NoStop}%
\bibitem [{\citenamefont {Gullans}\ and\ \citenamefont
  {Huse}(2020)}]{Gullans2019Dynamical}%
  \BibitemOpen
  \bibfield  {author} {\bibinfo {author} {\bibfnamefont {Michael~J.}\
  \bibnamefont {Gullans}}\ and\ \bibinfo {author} {\bibfnamefont {David~A.}\
  \bibnamefont {Huse}},\ }\bibfield  {title} {\enquote {\bibinfo {title}
  {Dynamical purification phase transition induced by quantum measurements},}\
  }\href {\doibase 10.1103/PhysRevX.10.041020} {\bibfield  {journal} {\bibinfo
  {journal} {Phys. Rev. X}\ }\textbf {\bibinfo {volume} {10}},\ \bibinfo
  {pages} {041020} (\bibinfo {year} {2020})}\BibitemShut {NoStop}%
\bibitem [{\citenamefont {{Gullans}}\ \emph {et~al.}(2020)\citenamefont
  {{Gullans}}, \citenamefont {{Krastanov}}, \citenamefont {{Huse}},
  \citenamefont {{Jiang}},\ and\ \citenamefont
  {{Flammia}}}]{Gullans2020Quantum}%
  \BibitemOpen
  \bibfield  {author} {\bibinfo {author} {\bibfnamefont {Michael~J.}\
  \bibnamefont {{Gullans}}}, \bibinfo {author} {\bibfnamefont {Stefan}\
  \bibnamefont {{Krastanov}}}, \bibinfo {author} {\bibfnamefont {David~A.}\
  \bibnamefont {{Huse}}}, \bibinfo {author} {\bibfnamefont {Liang}\
  \bibnamefont {{Jiang}}}, \ and\ \bibinfo {author} {\bibfnamefont {Steven~T.}\
  \bibnamefont {{Flammia}}},\ }\bibfield  {title} {\enquote {\bibinfo {title}
  {{Quantum coding with low-depth random circuits}},}\ }\href@noop {}
  {\bibfield  {journal} {\bibinfo  {journal} {arXiv e-prints}\ ,\ \bibinfo
  {eid} {arXiv:2010.09775}} (\bibinfo {year} {2020})},\ \Eprint
  {http://arxiv.org/abs/2010.09775} {arXiv:2010.09775 [quant-ph]} \BibitemShut
  {NoStop}%
\bibitem [{\citenamefont {{Huang}}\ \emph {et~al.}(2021)\citenamefont
  {{Huang}}, \citenamefont {{Kueng}}, \citenamefont {{Torlai}}, \citenamefont
  {{Albert}},\ and\ \citenamefont {{Preskill}}}]{Huang2021Provably}%
  \BibitemOpen
  \bibfield  {author} {\bibinfo {author} {\bibfnamefont {Hsin-Yuan}\
  \bibnamefont {{Huang}}}, \bibinfo {author} {\bibfnamefont {Richard}\
  \bibnamefont {{Kueng}}}, \bibinfo {author} {\bibfnamefont {Giacomo}\
  \bibnamefont {{Torlai}}}, \bibinfo {author} {\bibfnamefont {Victor~V.}\
  \bibnamefont {{Albert}}}, \ and\ \bibinfo {author} {\bibfnamefont {John}\
  \bibnamefont {{Preskill}}},\ }\bibfield  {title} {\enquote {\bibinfo {title}
  {{Provably efficient machine learning for quantum many-body problems}},}\
  }\href@noop {} {\bibfield  {journal} {\bibinfo  {journal} {arXiv e-prints}\
  ,\ \bibinfo {eid} {arXiv:2106.12627}} (\bibinfo {year} {2021})},\ \Eprint
  {http://arxiv.org/abs/2106.12627} {arXiv:2106.12627 [quant-ph]} \BibitemShut
  {NoStop}%
\bibitem [{\citenamefont {Akhtar}\ and\ \citenamefont
  {You}(2020)}]{PhysRevB.102.134203}%
  \BibitemOpen
  \bibfield  {author} {\bibinfo {author} {\bibfnamefont {A.~A.}\ \bibnamefont
  {Akhtar}}\ and\ \bibinfo {author} {\bibfnamefont {Yi-Zhuang}\ \bibnamefont
  {You}},\ }\bibfield  {title} {\enquote {\bibinfo {title} {Multiregion
  entanglement in locally scrambled quantum dynamics},}\ }\href {\doibase
  10.1103/PhysRevB.102.134203} {\bibfield  {journal} {\bibinfo  {journal}
  {Phys. Rev. B}\ }\textbf {\bibinfo {volume} {102}},\ \bibinfo {pages}
  {134203} (\bibinfo {year} {2020})}\BibitemShut {NoStop}%
\bibitem [{\citenamefont {Grover}\ and\ \citenamefont
  {Fisher}(2015)}]{tarun2015}%
  \BibitemOpen
  \bibfield  {author} {\bibinfo {author} {\bibfnamefont {Tarun}\ \bibnamefont
  {Grover}}\ and\ \bibinfo {author} {\bibfnamefont {Matthew P.~A.}\
  \bibnamefont {Fisher}},\ }\bibfield  {title} {\enquote {\bibinfo {title}
  {Entanglement and the sign structure of quantum states},}\ }\href {\doibase
  10.1103/physreva.92.042308} {\bibfield  {journal} {\bibinfo  {journal}
  {Physical Review A}\ }\textbf {\bibinfo {volume} {92}} (\bibinfo {year}
  {2015}),\ 10.1103/physreva.92.042308}\BibitemShut {NoStop}%
\bibitem [{\citenamefont {Stoudenmire}\ and\ \citenamefont
  {Schwab}(2016)}]{NIPS2016_5314b967}%
  \BibitemOpen
  \bibfield  {author} {\bibinfo {author} {\bibfnamefont {Edwin}\ \bibnamefont
  {Stoudenmire}}\ and\ \bibinfo {author} {\bibfnamefont {David~J}\ \bibnamefont
  {Schwab}},\ }\bibfield  {title} {\enquote {\bibinfo {title} {Supervised
  learning with tensor networks},}\ }in\ \href
  {https://proceedings.neurips.cc/paper/2016/file/5314b9674c86e3f9d1ba25ef9bb32895-Paper.pdf}
  {\emph {\bibinfo {booktitle} {Advances in Neural Information Processing
  Systems}}},\ Vol.~\bibinfo {volume} {29},\ \bibinfo {editor} {edited by\
  \bibinfo {editor} {\bibfnamefont {D.}~\bibnamefont {Lee}}, \bibinfo {editor}
  {\bibfnamefont {M.}~\bibnamefont {Sugiyama}}, \bibinfo {editor}
  {\bibfnamefont {U.}~\bibnamefont {Luxburg}}, \bibinfo {editor} {\bibfnamefont
  {I.}~\bibnamefont {Guyon}}, \ and\ \bibinfo {editor} {\bibfnamefont
  {R.}~\bibnamefont {Garnett}}}\ (\bibinfo  {publisher} {Curran Associates,
  Inc.},\ \bibinfo {year} {2016})\BibitemShut {NoStop}%
\bibitem [{\citenamefont {{Lu}}\ \emph {et~al.}(2021)\citenamefont {{Lu}},
  \citenamefont {{Kan{\'a}sz-Nagy}}, \citenamefont {{Kukuljan}},\ and\
  \citenamefont {{Cirac}}}]{2021arXiv210306872L}%
  \BibitemOpen
  \bibfield  {author} {\bibinfo {author} {\bibfnamefont {Sirui}\ \bibnamefont
  {{Lu}}}, \bibinfo {author} {\bibfnamefont {M{\'a}rton}\ \bibnamefont
  {{Kan{\'a}sz-Nagy}}}, \bibinfo {author} {\bibfnamefont {Ivan}\ \bibnamefont
  {{Kukuljan}}}, \ and\ \bibinfo {author} {\bibfnamefont {J.~Ignacio}\
  \bibnamefont {{Cirac}}},\ }\bibfield  {title} {\enquote {\bibinfo {title}
  {{Tensor networks and efficient descriptions of classical data}},}\
  }\href@noop {} {\bibfield  {journal} {\bibinfo  {journal} {arXiv e-prints}\
  ,\ \bibinfo {eid} {arXiv:2103.06872}} (\bibinfo {year} {2021})},\ \Eprint
  {http://arxiv.org/abs/2103.06872} {arXiv:2103.06872 [quant-ph]} \BibitemShut
  {NoStop}%
\bibitem [{\citenamefont {Reyes}\ and\ \citenamefont
  {Stoudenmire}(2021)}]{Reyes_2021}%
  \BibitemOpen
  \bibfield  {author} {\bibinfo {author} {\bibfnamefont {J~A}\ \bibnamefont
  {Reyes}}\ and\ \bibinfo {author} {\bibfnamefont {E~M}\ \bibnamefont
  {Stoudenmire}},\ }\bibfield  {title} {\enquote {\bibinfo {title} {Multi-scale
  tensor network architecture for machine learning},}\ }\href {\doibase
  10.1088/2632-2153/abffe8} {\bibfield  {journal} {\bibinfo  {journal} {Machine
  Learning: Science and Technology}\ }\textbf {\bibinfo {volume} {2}},\
  \bibinfo {pages} {035036} (\bibinfo {year} {2021})}\BibitemShut {NoStop}%
\bibitem [{\citenamefont {Hauru}\ \emph {et~al.}(2021)\citenamefont {Hauru},
  \citenamefont {Damme},\ and\ \citenamefont
  {Haegeman}}]{10.21468/SciPostPhys.10.2.040}%
  \BibitemOpen
  \bibfield  {author} {\bibinfo {author} {\bibfnamefont {Markus}\ \bibnamefont
  {Hauru}}, \bibinfo {author} {\bibfnamefont {Maarten~Van}\ \bibnamefont
  {Damme}}, \ and\ \bibinfo {author} {\bibfnamefont {Jutho}\ \bibnamefont
  {Haegeman}},\ }\bibfield  {title} {\enquote {\bibinfo {title} {{Riemannian
  optimization of isometric tensor networks}},}\ }\href {\doibase
  10.21468/SciPostPhys.10.2.040} {\bibfield  {journal} {\bibinfo  {journal}
  {SciPost Phys.}\ }\textbf {\bibinfo {volume} {10}},\ \bibinfo {pages} {40}
  (\bibinfo {year} {2021})}\BibitemShut {NoStop}%
\bibitem [{\citenamefont {Geng}\ \emph {et~al.}(2022)\citenamefont {Geng},
  \citenamefont {Hu},\ and\ \citenamefont {Zou}}]{Geng_2022}%
  \BibitemOpen
  \bibfield  {author} {\bibinfo {author} {\bibfnamefont {Chenhua}\ \bibnamefont
  {Geng}}, \bibinfo {author} {\bibfnamefont {Hong-Ye}\ \bibnamefont {Hu}}, \
  and\ \bibinfo {author} {\bibfnamefont {Yijian}\ \bibnamefont {Zou}},\
  }\bibfield  {title} {\enquote {\bibinfo {title} {Differentiable programming
  of isometric tensor networks},}\ }\href {\doibase 10.1088/2632-2153/ac48a2}
  {\bibfield  {journal} {\bibinfo  {journal} {Machine Learning: Science and
  Technology}\ }\textbf {\bibinfo {volume} {3}},\ \bibinfo {pages} {015020}
  (\bibinfo {year} {2022})}\BibitemShut {NoStop}%
\bibitem [{\citenamefont {Haghshenas}(2021)}]{PhysRevResearch.3.023148}%
  \BibitemOpen
  \bibfield  {author} {\bibinfo {author} {\bibfnamefont {Reza}\ \bibnamefont
  {Haghshenas}},\ }\bibfield  {title} {\enquote {\bibinfo {title} {Optimization
  schemes for unitary tensor-network circuit},}\ }\href {\doibase
  10.1103/PhysRevResearch.3.023148} {\bibfield  {journal} {\bibinfo  {journal}
  {Phys. Rev. Research}\ }\textbf {\bibinfo {volume} {3}},\ \bibinfo {pages}
  {023148} (\bibinfo {year} {2021})}\BibitemShut {NoStop}%
\bibitem [{\citenamefont {{Bu}}\ \emph {et~al.}(2022)\citenamefont {{Bu}},
  \citenamefont {{Enshan Koh}}, \citenamefont {{Garcia}},\ and\ \citenamefont
  {{Jaffe}}}]{2022arXiv220203272B}%
  \BibitemOpen
  \bibfield  {author} {\bibinfo {author} {\bibfnamefont {Kaifeng}\ \bibnamefont
  {{Bu}}}, \bibinfo {author} {\bibfnamefont {Dax}\ \bibnamefont {{Enshan
  Koh}}}, \bibinfo {author} {\bibfnamefont {Roy~J.}\ \bibnamefont {{Garcia}}},
  \ and\ \bibinfo {author} {\bibfnamefont {Arthur}\ \bibnamefont {{Jaffe}}},\
  }\bibfield  {title} {\enquote {\bibinfo {title} {{Classical shadows with
  Pauli-invariant unitary ensembles}},}\ }\href@noop {} {\bibfield  {journal}
  {\bibinfo  {journal} {arXiv e-prints}\ ,\ \bibinfo {eid} {arXiv:2202.03272}}
  (\bibinfo {year} {2022})},\ \Eprint {http://arxiv.org/abs/2202.03272}
  {arXiv:2202.03272 [quant-ph]} \BibitemShut {NoStop}%
\end{thebibliography}%

\onecolumngrid
\newpage
\appendix
\section{Entanglement feature and the reconstruction channel\label{app:EF}}
In this appendix, we discuss the details about the derivation on measurement channel $\sigma = \scM[\rho]$ and reconstruction channel $\rho=\scM^{-1}[\sigma]$. With the notation developed in \ref{sec:reconstruction}, the expected classical snapshot $\sigma$ is expressed as
\eq{\sigma=\scM[\rho]=\sum_{\hat{\sigma}\in \mathcal{E}_{\sigma}}\hat{\sigma}\Tr(\hat{\sigma}\rho)d^N.}
By utilizing the assumption that prior ensemble $\scE_{\sigma}$ is locally scrambled, i.e. $P(\hat{\sigma})=P(V^{\dagger}\hat{\sigma}V),\forall V\in\U(d)^N$, we are free to insert local basis transformations $V$, and average it out. By doing so, we have
\eqs{
\sigma&=\dsE_{V\in\U(d)^N}\dsE_{\hat{\sigma}\in \scE_{\sigma}}V^{\dagger}\hat{\sigma}V\Tr(V^{\dagger}\hat{\sigma}V\rho)d^N\\
&=\dia{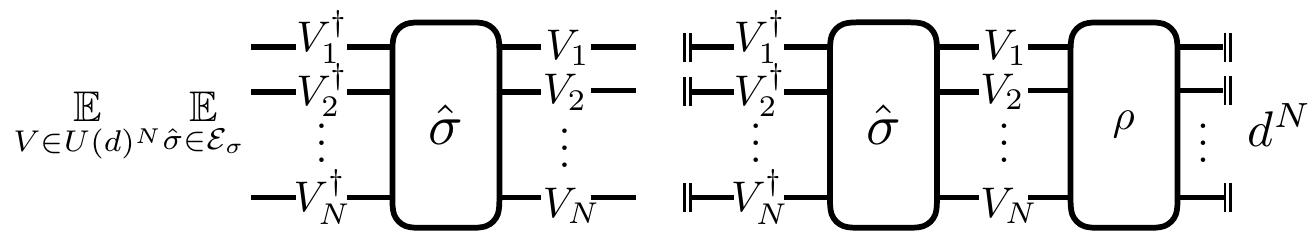}{50}{-25}\\
&=\dia{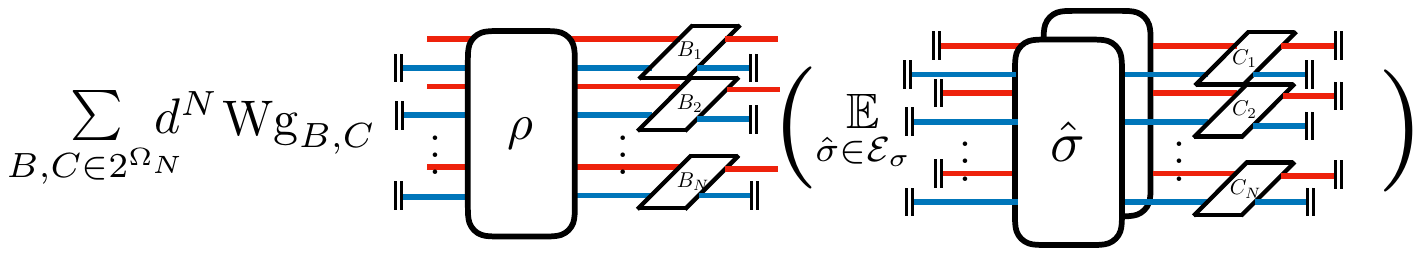}{50}{-25},
\label{eq:tensor_diagram}}
where each $B_i$ and $C_i$ have two choices: swap operator (1) or identity operator (0), and $\Wg_{B,C}=(d^2-1)^{-N}(-1/d)^{|B\ominus C|}$ is the Weingarten function of regions $B$ and $C$, where $B\ominus C=(B\setminus C)\cup(C\setminus B)$ denotes their symmetric difference. In the above tensor diagram, short parallel lines indicate the periodic boundary condition, and the summation of $B$ and $C$ is over all possible subregions of the $N$ qudit system. As we can see, if we choose a subregion $B$ to be the swap operators, then $\rho$ will be traced out on the counter part $\bar{B}$. In addition, the identity operators (red lines) on $\bar{B}$ are inserted. So the first tensor diagram in \eqnref{eq:tensor_diagram} is the reduced density matrix embedded back into the total Hilbert space. We spoil the notation and use $\rho_B d^{\bar{B}}=(\Tr_{\bar{B}}\rho)\otimes \id_{\bar{B}}$ to denote the first tensor diagram, but one should remember the identity operators are supported in region $\bar{B}$. The tensor product $\otimes$ notation indicates that $(\Tr_{\bar{B}}\rho)$ and $\id_{\bar{B}}$ act separately in regions $B$ and $\bar{B}$, which does \emph{not} imply that $B$ should be a consecutive region ``in front of'' $\bar{B}$ (as in the conventional notation). The second tensor diagram in \eqnref{eq:tensor_diagram} is the 2nd \emph{entanglement feature} of the prior POVM $\scE_{\sigma}$,
\eq{
W_{\scE_{\sigma},C}^{(2)}\equiv\E_{\hat{\sigma}\in\scE_{\sigma}}\Tr_{C}(\Tr_{\bar{C}}\hat{\sigma})^2=\E_{\hat{\sigma}\in\scE_{\sigma}}e^{-S_{C}^{(2)}(\hat{\sigma})},}
where $S_C^{(2)}(\hat{\sigma})$ denotes the 2nd R\'enyi entanglement entropy of the state $\hat{\sigma}$ in region $C$. The above tensor diagram representation is equivalent to Eq.\ref{eq:sigma} in the main text.
 
 
\section{Variance estimation and sample complexity\label{app:var}}
In the main text, we relate the sample complexity $M$ with the $\rho$-dependent shadow norm $\norm{O}_{\scE_{\sigma|\rho}}^2$, by
\eq{M\geq \norm{O}_{\scE_{\sigma|\rho}}^2/\epsilon^2\delta.}
However, the $\rho$-dependent shadow norm $\norm{O}_{\scE_{\sigma|\rho}}^2$ is generally complicated to evaluate. If we are not interested in the shadow norm for a specific state $\rho$, but rather the expectation of the shadow norm over an ensemble of states $\{V\rho V^\dagger\}$ that are similar to $\rho$ by local basis transformations $V\in \U(d)^N$, we can actually define a $\rho$-independent shadow norm by averaging over $V$. The result is similar to \eqnref{eq:normO0}
\eqs{\label{eq:normO2}\norm{O}_{\scE_{\sigma}}^2&\equiv\E_{V\in\U(d)^N}\norm{O}_{\scE_{\sigma|{V \rho V^\dagger}}}^2 \\
&=\sum_{g,h\in S_3^{N}}\norm{O}_{g}^2 \Wg_{g,h}W_{\scE_{\sigma},h}^{(3)},}
where $\norm{O}_{g}^2$ is inherited from \eqnref{eq:normO1}
\eqs{\label{eq:normO3}
\norm{O}_{g}^2&\equiv\E_{V\in\U(d)^N}\norm{O}_{V \rho V^\dagger,g}^2\\
&=\Tr\big((\scM^{-1}[O]^{\otimes 2}\otimes \id)\chi_g\big).}
Compared with \eqnref{eq:normO1}, we can see that the ensemble average $\E_{V\in\U(d)^N}$ in \eqnref{eq:normO3} removes the $\rho$ dependence by effectively replacing $\rho$ with $d^{-N}\id$ (the prior density matrix that defines the prior POVM $\scE_{\sigma}$). This explains the consistency in our notation that $\norm{O}_{\scE_{\sigma}}^2=\dsE_{\hat{\sigma}\in\scE_{\sigma}}\hat{o}(\hat{\sigma})^2$ follows from essentially the same definition as in \eqnref{eq:VarO}.  

Note that the reconstruction map $\scM^{-1}$ always commutes with the local basis transformation $V=\prod_{i}V_i$, i.e.~$\scM^{-1}[V^\dagger O V]=V^\dagger \scM^{-1}[O] V$, because $V_i$ acts on each qudit separately and hence does not interfere with the partial trace operation. This indicates that the norm $\norm{O}_{g}^2=\norm{V^\dagger O V}_{g}^2$ is invariant under the transformation $V$. This suggests us to define a locally scrambled ensemble $\scE_O$ (or known as $U(d)^N$-twirling) associated with any given observable $O$
\eq{\scE_O\equiv \{V^\dagger O V|V\in\U(d)^N\},}
such that $\norm{O}_{g}^2$ in \eqnref{eq:normO3} can be redefined as its ensemble average
\eqs{\label{eq:normO4}\norm{O}_{g}^2&=\E_{O\in\scE_O}\norm{O}_{g}^2\\
&=\E_{V\in\U(d)^N}\Tr\big((\scM^{-1}[V^\dagger O V]^{\otimes 2}\otimes \id)\chi_g\big)\\
&=\sum_{{A,B,C,D}\in2^{\Omega_N}}
d^{2N}r_A r_B \Wg_{C,D}W_{\scE_O,D}^{(2)}
\Tr \big(((\chi_C)_{A,B}\otimes\id)\chi_g\big).}
Here $\chi_C$ denotes the swap operator supported in region $C$ that acts between the first two copies of the Hilbert space, and $(\chi_C)_{A,B}$ denotes the reduction of $\chi_C$ in region $A$ and $B$ respectively in the first and the second copies of the Hilbert space, which results in $(\chi_C)_{A,B}=\chi_{A\cap B\cap C}d^{|{A\cap B\cap C}|-|C|}$. The operator entanglement feature $W_{\scE_O,D}^{(2)}=\E_{O\in\scE_O}\Tr_{D}(\Tr_{\bar{D}} O)^2$ follows from the same definition given in \eqnref{eq:WC0}. $r_A,r_B$ are the reconstruction coefficients given by the solution of \eqnref{eq:rA}. Substitute \eqnref{eq:normO4} to \eqnref{eq:normO2}, we can evaluate the summation of $g,h$ given that $\sum_{g,h\in S_3^N}\Tr(\chi_{A\cap B\cap C}\chi_g)\Wg_{g,h}W_{\scE_{\sigma},h}^{(3)}=W_{\scE_{\sigma},{A\cap B\cap C}}^{(2)}$. The reduction of the 3rd entanglement feature to the 2nd entanglement feature is a consequence of the fact that $\rho$ drops out from the tensor product in \eqnref{eq:normO3}, such that only 2-fold Hilbert space is required to define $\norm{O}_{\scE_{\sigma}}^2$.

Thus we finally arrive at the expression for the operator shadow norm purely in terms of the entanglement features of $\scE_{\sigma}$ and $\scE_O$,
\eq{\label{eq:appen_normO5}\norm{O}_{\scE_{\sigma}}^2=\sum_{A,B,C,D\in 2^{\Omega_N}}v_{A,B,C,D}W_{\scE_{\sigma},{A\cap B\cap C}}^{(2)}W_{\scE_O,D}^{(2)},}
where the coefficient $v_{A,B,C,D}$ is given by
\eq{v_{A,B,C,D}=r_A r_B \Big(\frac{d^2}{d^2-1}\Big)^N d^{|{A\cap B\cap C}|-|C|}\Big(-\frac{1}{d}\Big)^{|C\ominus D|}.}

\section{Efficient classical post-processing algorithm with tensor network method\label{app:efficient_alg}}

\subsection{Overview of tensor network based classical post-processing}
\begin{figure}[htbp]
\begin{center}
\includegraphics[width=0.35\columnwidth]{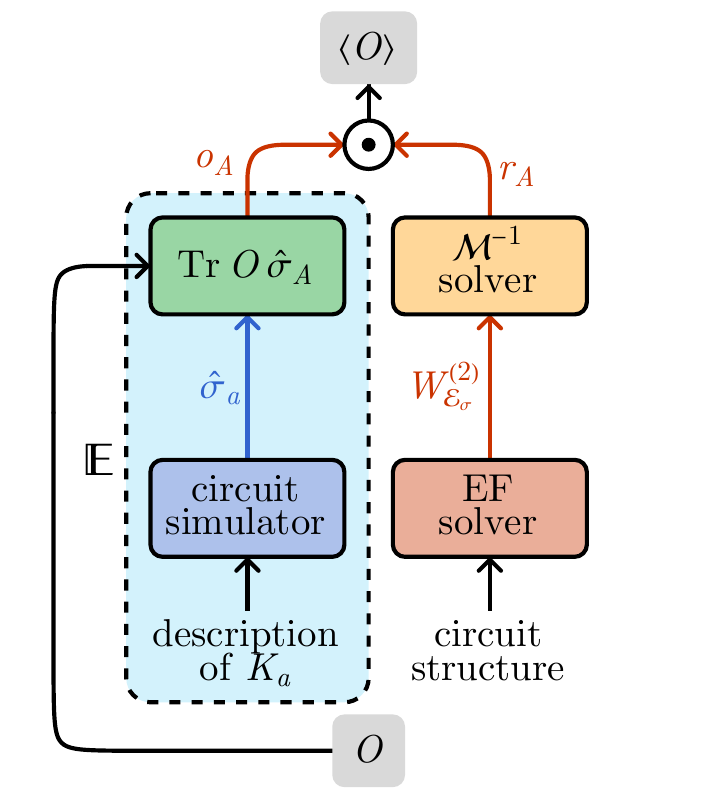}
\caption{Classical post-processing protocol to estimate the operator expectation value and shadow norm.}
\label{appen-fig:ap_flow}
\end{center}
\end{figure}

In the main text, we have derived the following protocol for predicting quantities of states with locally scrambled quantum circuits:
It seems that the number of coefficient $r_A$ scales exponentially with system size, therefore not efficient. Surprisingly, with clever design and the help of tensor network, there indeed exists efficient tensor network method that can achieve efficient classical post-processing. \figref{appen-fig:ap_flow} summarizes the classical post-processing workflow of predicting operator expectations with tensor network methods. 

On the right side of workflow, given the circuit structure, the entanglement feature can be efficiently encoded as a tensor network using ``EF solver". Then the tensor network representation of the entanglement feature is inputted into the $\scM^{-1}$ solver, whose output is a tensor network representation of the reconstruction coefficient $r_A$. The nice thing is that one only need to solve the tensor network representation for $r_A$ once. And this representation can be stored for future usage.

On the left of the workflow, we do experiments on quantum devices and calculate classical shadows $\hat{\sigma}$. And one should notice that our formulation is general enough to include Clifford circuits that do not have group structure. And the classical shadows of those circuits can be calculated efficiently. Then we can combine the classical shadows $\hat{\sigma}$ and tensor network representation of $r_A$ to predict operator expectations $\langle O \rangle$.

A detailed discussion on how to model both $\Tr\left(O\hat{\sigma}_A\right)$ and $r_A$ with tensor network is discussed in the following two subsections.

\subsection{Efficient matrix product state representation of $\vec{o}=\{o_A|\Tr(O\sigma_A)\}$}
In the main text, we have shown the reconstruction channel under local scrambling assumption can be written as
\eqs{
\rho=\scM^{-1}[\sigma]=d^N\sum_{A\in 2^{\Omega_N}}r_A \sigma_A.
}
At first sight, the exponential summation of subregion $A$ seems to be troublesome. However, it can be circumvented by tensor network methods. Here, we will introduce a concrete algorithm. First of all, if the unitaries used in the classical shadow experiment are Clifford gates, then classical shadows $\hat{\sigma}$ are stabilizer states, and each of them can be efficiently stored with $\scO(N^2)$ memory on a classical computer, where $N$ is the system size. 

$\bullet$ \textbf{Proposition I:} Given $O$ is a Pauli observable and $\sigma$ is a stabilizer state, the vector $\vec{o}=\{o_A|o_A = \Tr\left(O \sigma_A\right)\}$ has an efficient matrix product state (MPS) representation with internal bond dimension equals one, where $\sigma_A=\scD_{\bar{A}}[\sigma]$. And $\scD_{A}[\circ]$ is the depolarizing channel acting on region $A$.

\textbf{Proof:} If the circuit is composed of Clifford gates, then classical shadow $\sigma$ is a stabilizer state with stabilizer group generated by
\eqs{
\scS=\langle (-1)^{b_1}U^{\dagger}Z_1 U,\cdots, (-1)^{b_n}U^{\dagger}Z_N U \rangle=\langle \widetilde{Z}_1,\cdots,\widetilde{Z}_N \rangle,
}
\eqs{
\sigma = \prod^{n}_{i=1}\dfrac{\id+\widetilde{Z}_i}{2}=\dfrac{1}{2^N}\sum_{g\in \scS}g.
}
The reduced state $\sigma_A=\scD_{\bar{A}}[\sigma]$ restricted to region $A$ is also a stabilizer state with stabilizer group $\scS_{A}\subseteq \scS$ defined by taking the elements of $\scS$ which have zero support on $\bar{A}$. This is obviously a subgroup of $\scS$ since it is closed under multiplication and inversion. Without loss of generality, we can write 
\eqs{
\sigma_A = \left(\dfrac{1}{2^{|A|}}\sum_{g\in \scS_A}g\right)\otimes\left(\dfrac{\id_2}{2}\right)^{\otimes (N-|A|)}.
}
It is obvious that the expectation $\Tr\left(O\sigma_A\right)=0$ when $\text{supp}(O)\nsubseteq A$. Moreover, the only scenario when $\Tr\left(O\sigma_A\right)$ is non-zero is $\pm O\in \scS_{A}$. Therefore, we have
\eqs{
\Tr\left(O\sigma_A\right)=\begin{cases}
0, &\pm O\notin \scS_A\\
\Tr(O\sigma_A)=\Tr(O\sigma),&\pm O\in \scS_A.
\end{cases}
}
From the above equation, it is clear that for any Pauli observable $O$, $o_A=\Tr\left(O\sigma_A\right)$ can be represented as a trivial MPS with bond dimension $D=1$:
\eqs{
o_A&=\Tr(\sigma O)\Tr\left(o_{1}^{(a_1)}o_{2}^{(a_2)}\cdots o_{N}^{(a_N)}\right)\\ &=\Tr(\sigma O)o_{1}^{(a_1)}o_{2}^{(a_2)}\cdots o_{N}^{(a_N)},
}
where we drop the second trace since the internal bond dimension is one, and each tensor $o_i^{(a_i)}$ on site $i$ with binary physical index $(a_i=0 \text{ or }1)$ is
\eq{
o_i=\begin{cases}
\begin{pmatrix}
1\\ 1
\end{pmatrix},&i\notin \text{supp}(O)\\
\begin{pmatrix}
0\\ 1
\end{pmatrix},&i\in \text{supp}(O)
\end{cases}.\label{eq:MPS_repr}
}
This concludes that even the vector $\vec{o}=\{o_A|o_A = \Tr\left(O \sigma_A\right)\}$ contains exponentially many elements, it has an efficient MPS representation with bond dimension $D=1$. This MPS representation can be easily constructed: given Pauli observable $O$, first calculate $\Tr(O\sigma)$, then construct the MPS using \eqnref{eq:MPS_repr}. $\Tr(O\sigma)$ can computed in $\scO(N^2)$ time, because $O$ is a Pauli observable and $\sigma$ is a stabilizer state. The remaining MPS tensors can be constructed in $\scO(N)$ time by traversing through the Pauli string.

\subsection{Encoding reconstruction coefficient $r_A$ with variational MPS method}
\begin{figure}[htbp]
    \centering
    \includegraphics[width=0.4\linewidth]{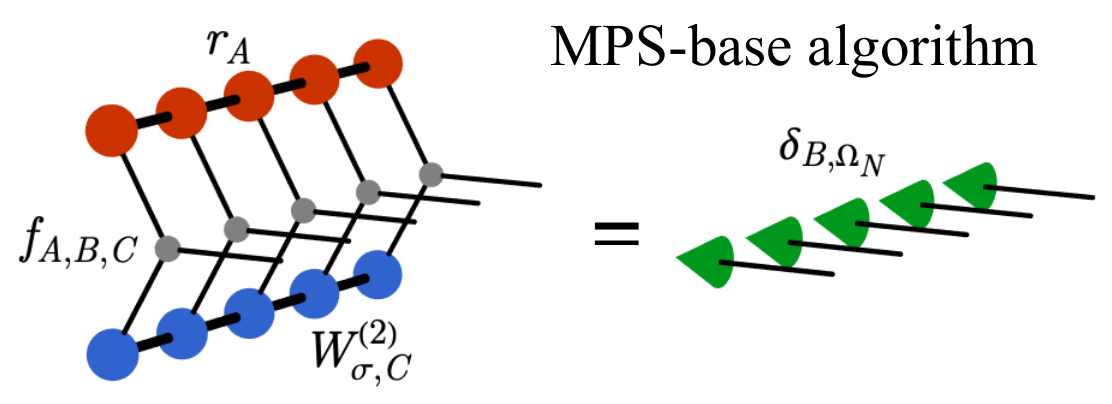}
    \caption{A cartoon illustration of variational solving MPS representation of $r_A$.}
    \label{fig:variational_MPS}
\end{figure}
In the main text, we argued that the vector $r_A$ can be represented as a MPS. Here, we illustrate how to find such a MPS using variational method. First of all, we have shown that the reconstruction coefficient $r_A$ satisfies the linear equation:
\eqs{
\sum_{A,C\in 2^{\Omega_N}}r_A f_{A,B,C}W^{(2)}_{C}=\delta_{B,\Omega_N},\label{eq:linear_equation}
}
where $W^{(2)}_{C}$ is the second entanglement feature vector created by the unitary ensemble. In \refcite{PhysRevB.102.134203}, the authors shows entanglement feature vector $W^{(2)}_{C}$ can be efficiently encoded using MPS representation. The physical intuition behind this efficient representation is that if one views $W^{(2)}_{C}$ as a weight of a quantum state, i.e. $\ket{\psi}=\sum_{C\in 2^{\Omega_N}}W^{(2)}_{C}\ket{C}$, then this state will possess low entanglement \refcite{tarun2015}. Therefore, it can be represented as a MPS with low bond dimension. In \figref{fig:variational_MPS}, the blue nodes indicate the MPS representation of $W^{(2)}_{C}$. For translation invariant circuit structure (such as the brick-wall circuit), the time complexity to construct the MPS representation for $W^{(2)}_{C}$ is $\scO(1)$ (independent of the system size). For general circuit structure, the time complexity is at most $\scO(N)$.

In \eqnref{eq:linear_equation}, the fusion coefficient $f_{A,B,C}$ is
\eqs{
f_{A,B,C}=\Big(\frac{d^3}{d^2-1}\Big)^{N}\sum_{D\in 2^{\Omega_N}}\delta_{B,A\cap D}d^{-|D|}\Big(-\frac{1}{d}\Big)^{|C\ominus D|}.
}
Note that this fusion factor $f_{A,B,C}$ can be factorized to each site as $f_{A,B,C}=\prod_{i}f_{a_i,b_i,c_i}$ where
\eqs{
f_{a_i,b_i,c_i}=\begin{pmatrix}
\begin{pmatrix}
d\\0
\end{pmatrix} & \begin{pmatrix}
0\\ 0
\end{pmatrix}\\
\dfrac{d^2}{d^2-1}\begin{pmatrix}
d \\ -1
\end{pmatrix} & \dfrac{d}{d^2-1}\begin{pmatrix}
-1 \\ d
\end{pmatrix}
\end{pmatrix},
}
as the tensor subscripts $a_i,b_i,c_i=0,1$ enumerates over boolean variables. Therefore, the fusion factor $f_{A,B,C}$ can be represented as the gray tensors in \figref{fig:variational_MPS}.

To find the MPS representation of vector $r_A$, we use the variational method. We can write an MPS ansatz for $r_A$ and try to find the best parameters in the MPS by doing variational optimization. The same idea has been explored in machine learning tensor network optimization \cite{NIPS2016_5314b967,2021arXiv210306872L,Reyes_2021}, and differential programming of tensor networks \cite{10.21468/SciPostPhys.10.2.040,Geng_2022,PhysRevResearch.3.023148}. With differential programming, we can find the best parameters in the MPS ansatz for $r_A$ by minimizing the L1 or L2 loss of the left-hand side tensor and right-hand side tensor of \figref{fig:variational_MPS}. With a fixed bond dimension, the algorithm complexity is $\scO(N)$. In addition, we can utilize the symmetry of the unitary ensemble to minimize the training parameters in the MPS ansatz. In practice, we find that $r_A$ can be represented as a MPS with a low bond dimension using the variational method. A detailed discussion of this new computational method will be in another paper.

In addition, we would like to point out that after the first draft of our paper, our new proposal has caught much attention from both theoretical and experimental sides. Especially, the formal solution of \eqnref{eq:linear_equation} can be solved \cite{2022arXiv220203272B},
\eqs{
r_A = \dfrac{(-1)^{-|A|}}{2^N}\sum_{A\subseteq S}\dfrac{3^{|S|}}{\sum_{B\subseteq S}(-2)^{|B|}W_{B}^{(2)}}\label{eq:solution}.
}
It would be also interesting to directly encode \eqnref{eq:solution} with a MPS without the help of variational optimization. And we leave this to a future study.

\section{Fidelity estimation for mixed state\label{app:mixed_state}}
\begin{figure}[htbp]
    \centering
    \includegraphics[width=0.6\linewidth]{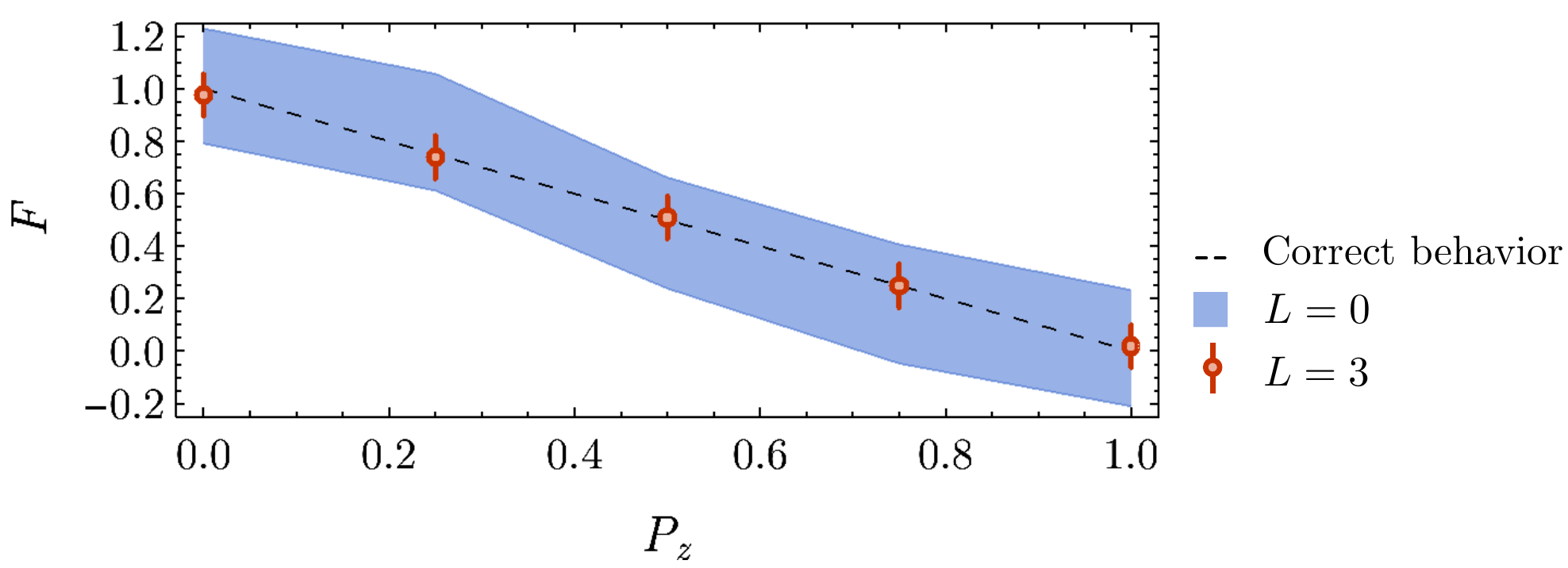}
    \caption{Fidelity estimation between mix state and target state. 5000 experimental classical snapshots are prepared for both random Pauli measurement ($L=0$) and shallow random unitary circuit ($L=3$). Error bar indicates 3 standard deviation.}
    \label{fig:mixstate}
\end{figure}

Our method is not restricted to pure state. In variational quantum state preparation, even the target state is some pure state, noise in the preparation circuit could make the final state in experiments a mixed state. We can use the shallow circuit classical shadow tomography to efficiently estimate the quantum fidelity between final prepared quantum state $\rho_{P}$ and the target quantum state $\rho_{T}$. Fast access to this quantity is crucial for variational quantum state preparation, error mitigation and etc. As an example, we consider the noisy preparation of a perfect GHZ state with $Z$ errors occurs at probability $p$. The prepared state can be expressed as 
\eq{\rho_{P}=(1-p)\ket{\psi_{\text{GHZ}}^{+}}\bra{\psi_{\text{GHZ}}^{+}}+p\ket{\psi_{\text{GHZ}}^{-}}\bra{\psi_{\text{GHZ}}^{-}},}
where $\ket{\psi_{\text{GHZ}}^{\pm}}=\frac{1}{\sqrt{2}}(\ket{0^{\otimes N}}\pm\ket{1^{\otimes N}})$. We compare the performance between random Pauli measurement and shallow circuit shadow tomography with 3 layers of local random unitaries. Experiments are performed on a 9-qubit system and 5000 classical snapshots are collected for both random Pauli shadow tomography and shallow circuits shadow tomography. The result is shown in \figref{fig:mixstate}. As we can see, for 5000 experiments, the quantum fidelity estimated using random Pauli measurement has huge error bar, indicated by the blue shaded region. However, same amount of data collected after shallow circuit evolution can give accurate estimation of quantum fidelity, and error bar is almost four times smaller. Practically, this makes the usage of shallow circuits more appealing.

\section{Approximated unitary ensemble and purification\label{app:purification}}
\begin{figure}[htbp]
    \centering
    \includegraphics[width=0.45\linewidth]{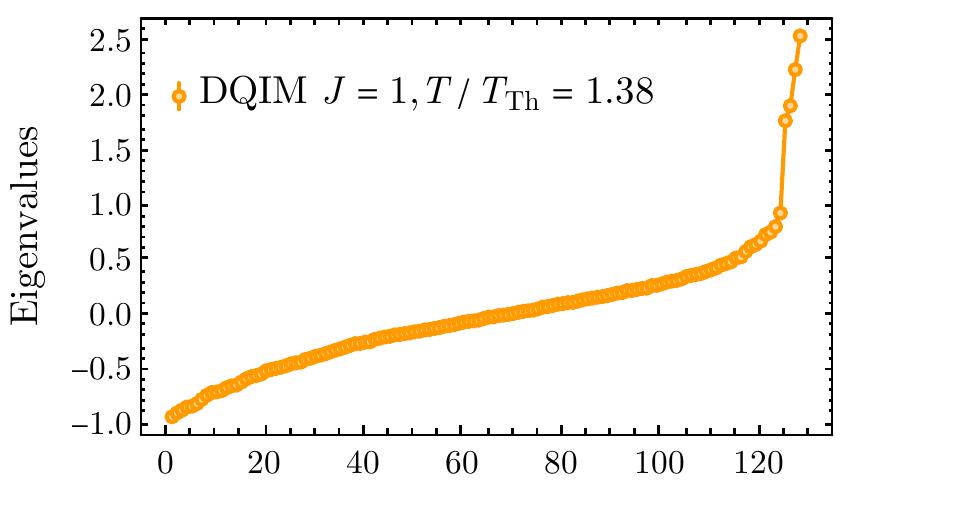}
    \caption{Eigenvalues of reconstructed density matrix $\tilde{\rho}$ of 7-qubit GHZ state using mismatched channels. The unitary ensemble is single instance of DQIM ensemble with $J=1$, and $T/T_{\text{Th}}=1.38$. Under this condition, the unitary ensemble is not locally scrambled.}
    \label{fig:eigenvalues}
\end{figure}

In the main text, we have seen when the measurement channel $\scM$ in data acquisition and the reconstruction channel $\scM^{-1}$ in classical post-processing mismatch, the reconstructed density matrix $\frac{1}{M}\sum_{\hat{\sigma}\in\scE_{\sigma|\rho}}\scM^{-1}[\hat{\sigma}]$ may not be positive-definite. And it results in biased prediction of physical quantities. In \figref{fig:bias} (a) and \figref{fig:practical_fidelity}, we have seen the biased prediction of fidelity that is larger than one. In \figref{fig:eigenvalues}, we plot the eigenvalues of reconstructed density matrix of 7-qubit GHZ state using DQIM ensemble with $T/T_{\text{Th}}=1.38$. In the main text, we have seen the DQIM ensemble with one period of evolutional time or $T/T_{\text{Th}}=1.38$ is not sufficient to achieve the local scrambling assumption, such that there is mismatch between data acquisition channel $\scM$ and reconstruction channel $\scM^{-1}$. We see the spectrum of density matrix contains some negative eigenvalues.

\begin{figure}[htbp]
    \centering
    \includegraphics[width=\linewidth]{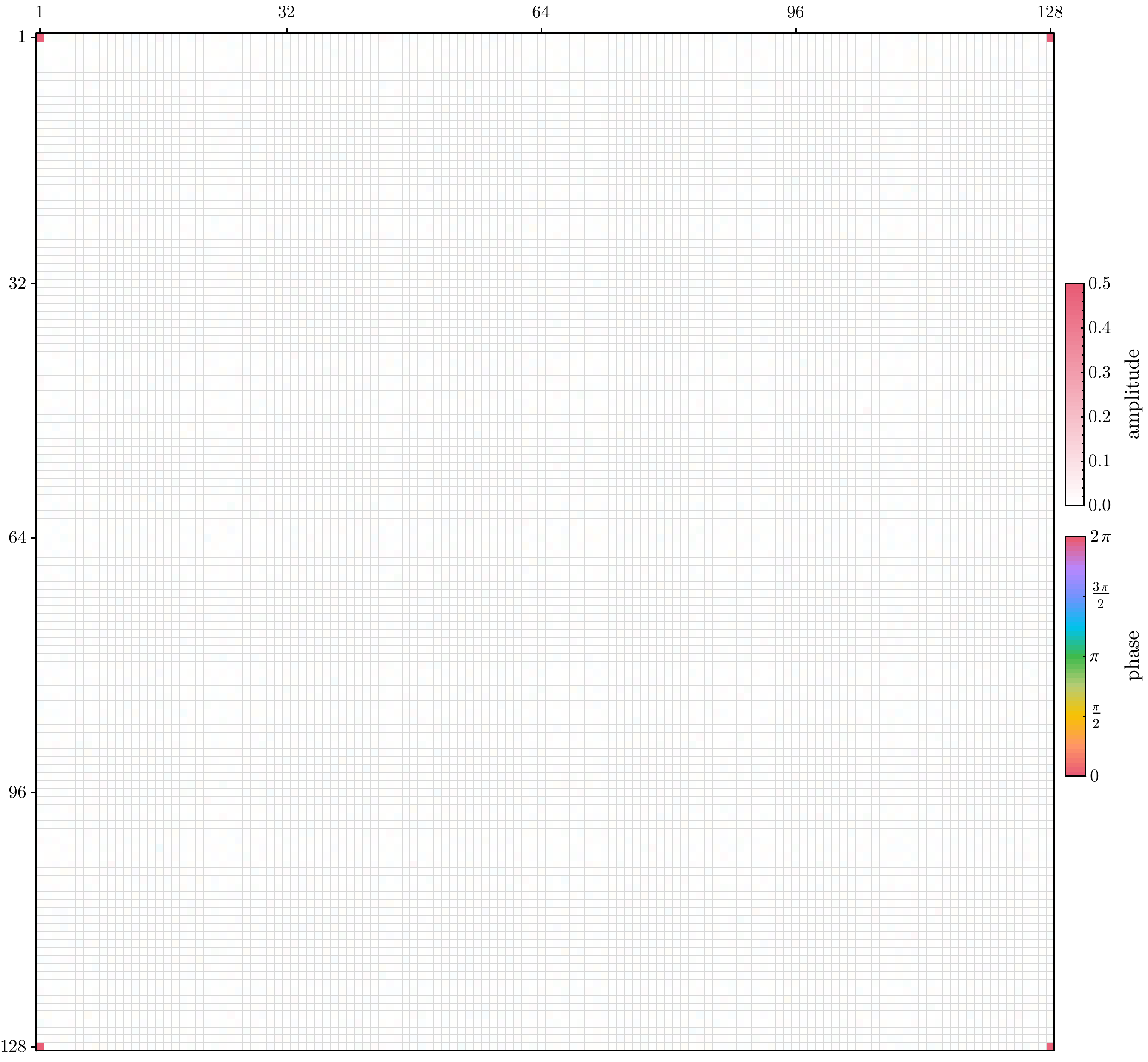}
    \caption{Unbiased reconstruction of a 7-qubit GHZ density matrix, using a single instance of Hamiltonian in the DQIM ensemble at $T/T_{\text{Th}}=25.3$ (after the local scrambling condition is achieved).}
    \label{fig:J2_2}
\end{figure}

\begin{figure}[htbp]
    \centering
    \includegraphics[width=\linewidth]{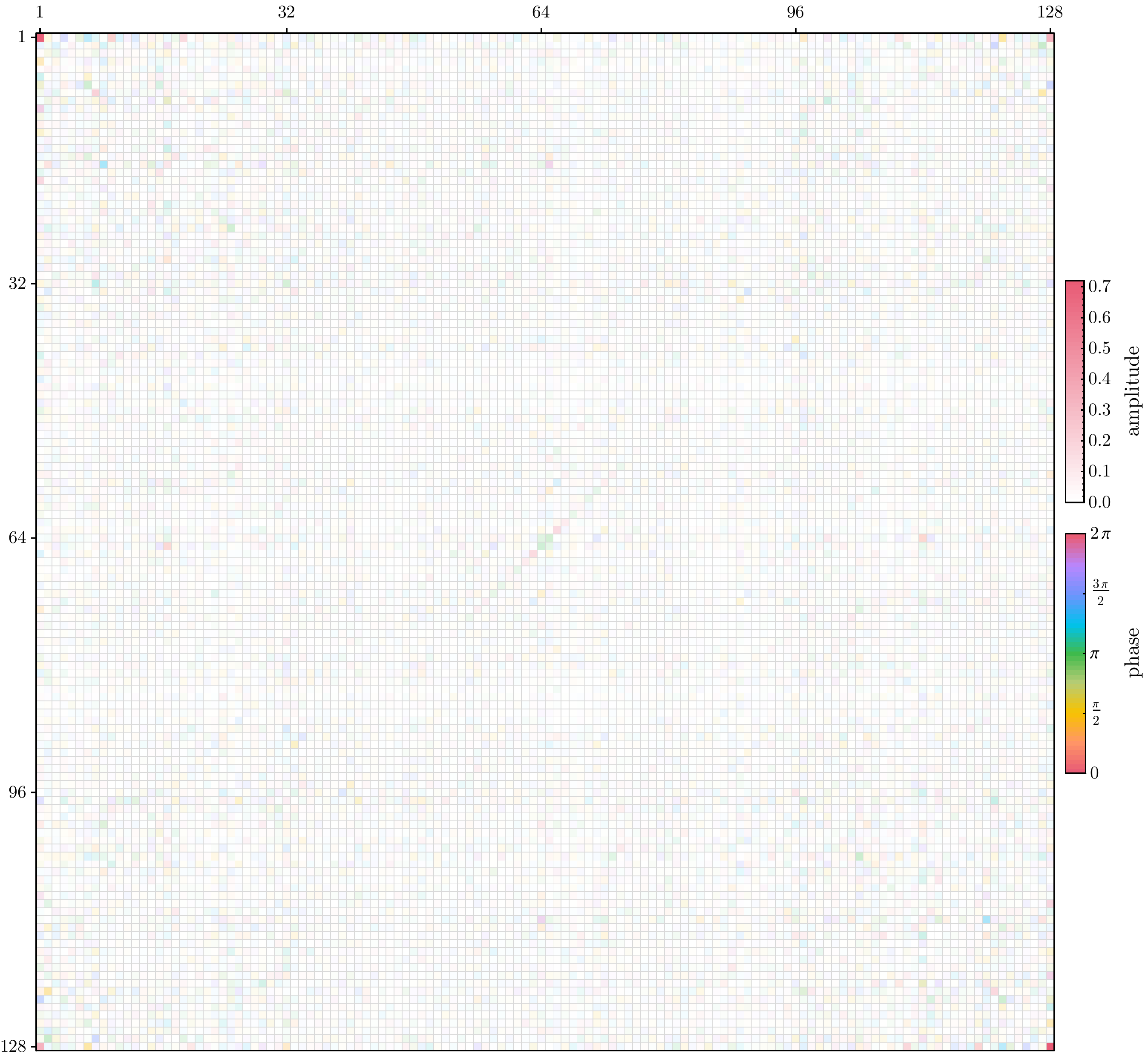}
    \caption{Biased reconstruction of a 7-qubit GHZ density matrix, using a single instance of Hamiltonian in the DQIM ensemble at $T/T_{\text{Th}}=1.95$ (before the local scrambling condition is achieved).}
    \label{fig:J2_1}
\end{figure}

In addition, the approximate shadow tomography based on locally scrambling Hamiltonian evolution, such as DQIM ensemble or GUE2 ensemble, is approximately unbiased when local scrambling is approximately satisfied or frame potential gap is vanishingly small. In \figref{fig:practical_fidelity}, we have seen they all can give unbiased prediction of quantum fidelity when $T\geq 10T_{\text{Th}}$. We directly visualize the reconstructed density matrix using approximated DQIM ensemble in \figref{fig:J2_2} and \figref{fig:J2_1}. As we see in \figref{fig:J2_1}, at $T/T_{\text{Th}}=1.95$, the locally scrambling assumption is not satisfied, and reconstructed density matrix is biased. In contrast, at $T/T_{\text{Th}}=25.3$ (\figref{fig:J2_2}), the reconstructed density matrix using a single instance of DQIM Hamiltonian is perfect, justifying the validity of our approach when the locally scrambling assumption is approximated satisfied.



\begin{figure}[htbp]
    \centering
    \includegraphics[width=0.42\linewidth]{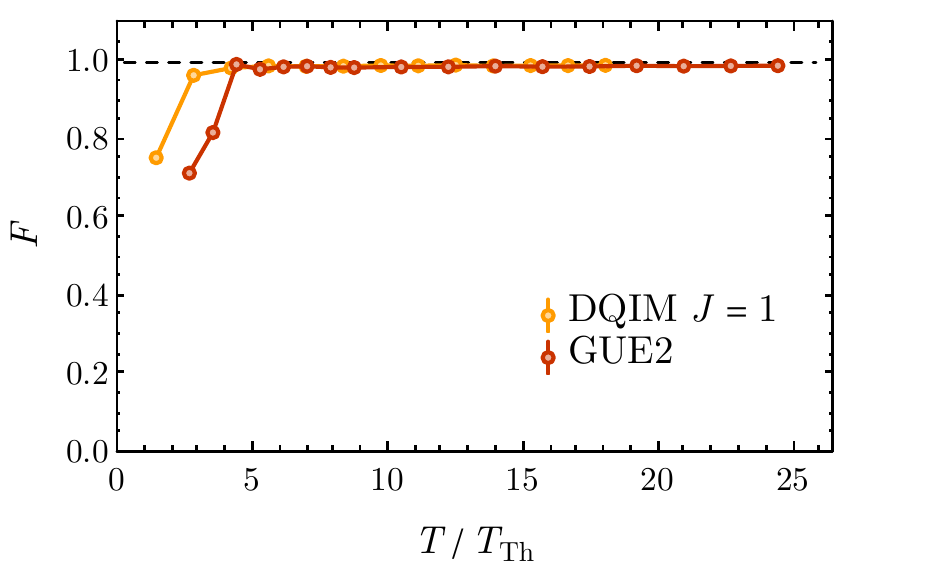}
    \caption{Fidelity estimation of approximated unitary ensemble after purification. After around $T\sim 10 T_{\text{Th}}$, the fidelity is around 0.99. Same data are used as \figref{fig:practical_fidelity}}
    \label{fig:purification}
\end{figure}

Further more, for biased reconstruction, in order to make it positive definite, we can nonlinear project the reconstructed $\rho$ to the convex set of physical states $C=\{\rho|\rho\succeq 1,\Tr(\rho)=1\}$ by minimizing \eq{\Pi_C(\sigma)=\arg \min_{\rho\in C}\Tr((\rho-\sigma)^2),}
which is the method mentioned in \refcite{Acharya2021Informationally}. If we have more prior knowledge about the quantum state, such as it is a pure state, then we can further impose those assumptions into the projection. Here, as an illustration, we utilize the knowledge that the target quantum state is pure, and we project the reconstructed $\rho$ to a pure state $\tilde{\rho}$ in $C$ by choosing the eigenstate of $\rho$ with the largest eigenvalue. As shown in \figref{fig:purification}, for the approximated ensembles, the GUE2 and DQIM are biased in the short time region, and projected state $\tilde{\rho}$ has a fidelity less than one. And when locally scrambling assumption is approximately satisfied, the projected $\tilde{\rho}$ will have fidelity that is approximately 0.99. With these checks: 
\begin{itemize}
    \item unbiased prediction of physical quantities, see \figref{fig:practical_fidelity}
    \item high fidelity of reconstructed density matrix projected back to physical space, see \figref{fig:purification}
\end{itemize}
we confirm the approximated shadow tomography can perform unbiased reconstruction.

\end{document}